\documentclass{aa}

\usepackage{fixltx2e}
\usepackage{graphicx}
\usepackage{txfonts}
\usepackage{rotating}
\usepackage{subcaption}
\usepackage{makecell}

\begin{document}

\title{Polarization properties of methanol masers}
\titlerunning{Methanol masers polarization landscape}
\author{D. Dall'Olio\inst{1}
         \and
          W. H. T. Vlemmings\inst{1}
         \and
         B. Lankhaar \inst{1}
         \and
         G. Surcis \inst{2}
            }

\institute{Department of Space, Earth and Environment, Chalmers University of Technology,
            Onsala Space Observatory,\\ Observatoriev\"agen 90, 43992 Onsala, Sweden;
            \email{daria.dallolio@chalmers.se}
            \and
             INAF--Osservatorio Astronomico di Cagliari, Via della Scienza 5, 09047 Selargius, Italy
               }

           \date{Received giorno mese anno; accepted giorno mese
             anno}

           \abstract{Astronomical masers have been effective tools to
             study magnetic fields for many years. Observing the
             linear and circular polarization of different maser
             species allows the determination of magnetic field
             properties, such as morphology and strength. In
             particular, methanol can be used to probe different parts
             of protostars such as accretion discs and outflows, since
             it produces one of the strongest and the most commonly
             observed masers in massive star-forming regions.}  {We
             investigate the polarization properties of selected
             methanol maser transitions in light of newly calculated
             methanol Land{\'e} g-factors and considering hyperfine
             components. We compare our results with previous
             observations and we evaluate the effect of preferred
             hyperfine pumping and non-Zeeman effects.}  {We run
             simulations using the radiative transfer code CHAMP for
             different magnetic field values, hyperfine components and
             pumping efficiencies.}  {We find a dependence of the
             linear polarization fraction on the magnetic field
             strength and on hyperfine transitions. The circular
             polarization fraction also shows a dependence on the
             hyperfine transitions. Preferred hyperfine pumping can
             explain some high levels of linear and circular
             polarization and some of the peculiar features seen in
             the S-shape of observed V-profiles. By comparing some
             methanol maser observations taken from the literature
             with our simulations, we find that the observed methanol
             masers are not significantly affected by non-Zeeman
             effects related to the competition between stimulated
             emission rates and Zeeman rates, like the rotation of the
             symmetry axis. We consider also the relevance of
               other non-Zeeman effects that are likely active for
               modest saturation levels, like the effect of magnetic
               field changes along the maser path and anisotropic
               resonant scattering.}
           {Our models show that for methanol maser emission,
               both the linear and circular polarization percentages
               depend on which hyperfine transition is masing and the
               degree to which it is being pumped. Since non-Zeeman
               effects become more relevant at high values of
               brightness temperatures, it is important to obtain good
               estimates of these quantities and on maser beaming
               angles.  Better constraints on the brightness
               temperature will help in understand about the extent
               to which non-Zeeman effects contribute to the observed
               polarization percentages.
             In order to detect separate hyperfine
             components, an intrinsic thermal line width significantly
             smaller than the hyperfine separation is required.}


   \maketitle

\section{Introduction}
The role of magnetic field during star formation has been a topic of great
debate for years. Many observations have been performed trying to
detect magnetic field morphology and strength towards star forming
regions. Several works have already demonstrated that astrophysical
masers are powerful tools to investigate magnetic field properties in
young protostars \citep[recently reviewed by][and references
therein]{Crutcher2019rev}.  Through the study of linearly and
circularly polarized maser emission, it is possible to obtain
information about the magnetic field, such as direction and strength,
over spatial scales of 10-100 au \citep[e.g.][]{Vlemmings2010,
  Surcis2013}. Moreover, different maser species and transitions
probe different regions of the protostar, providing a unique picture
of the physical conditions of the material where star formation
processes are ongoing \citep{Surcis2011w75n}. Masers can also help to
investigate the link between the gas properties and the magnetic
field: for instance, maser polarization observations were used to
infer a relationship between the density of the gas and the magnetic
field acting in the region \citep{Fish2006a, Vlemmings2008}.
Maser polarization observations have also been used to infer
properties of the large scale Galactic magnetic field \citep[e.g.][]{Green2012}.

Circular polarization has been detected in the majority of the maser species such as
hydroxyl, water and methanol \citep{Etoka2005,
  Vlemmings2006, Sarma2008, Surcis2011ngc, Caswell2011, Hunter2018}, and
in particular methanol masers have emerged as excellent tools to probe
magnetic fields during star formation \citep[e.g.][and references
therein]{Vlemmings2008, Surcis2019, Momjian2019, Sarma2020}.  Although the circularly
polarized emission of methanol maser has been regularly detected, no
exact estimates of the magnetic field strength were possible due to
unknown  Land{\'e} g-factors \citep{Vlemmings2011}. Thanks to the recent detailed calculations of
the Land{\'e} g-factors for all methanol transitions and the associated
hyperfine components \citep{Lankhaar2016, Lankhaar2018}, it is now
possible to obtain a complete interpretation of the methanol maser
polarization properties and infer the magnetic field characteristics.

In this paper we investigate the methanol maser polarization
properties using the maser polarization radiative transfer code
CHAMP \citep{Lankhaar2019}, performing several simulations of
different methanol masers transitions, as described in
Sect.~\ref{sec:methods}.  We report our results for the different
masers transitions in Sect.~\ref{sec:results}, with a more detailed
view on the 6.7 GHz one given the large amount of observations in the
literature, and with a more general description of the most important
features observed at other frequencies. Then, in
Sect.~\ref{sec:discussion} we compare the results of our
simulations with previous observations and we discuss the
importance of hyperfine preferred pumping, its effect on the
polarization fraction and the V spectra, and the presence of non-Zeeman
effects. In Sect.~\ref{sec:conclusions} we present our conclusions
and future perspectives, considering that this work can be the
starting point for a larger and more detailed study of magnetic fields,
and that further high-resolution observations will help to better
understand the action of preferred hyperfine pumping.

\subsection{Origin of circular polarization and non-Zeeman effects}
\label{sec:Non-zeeman}

One of the major sources of circular polarization of molecular lines
is the Zeeman effect \citep{Zeeman1897, FiebigGuesten1989, Sarma2001}.
According to the theory, under the action of a magnetic field $B$, the
emission from a molecule is separated in several components due to the
magnetic sub-levels. The shift between these components is named
Zeeman splitting and can be used to derive the amount of circular
polarization, which is proportional to the magnetic field
strength. Studying circular polarization in maser emissions is
therefore fundamental to infer the magnetic field strength of the
masing region.

In general, the saturation level and the nature of the masing
molecule (paramagnetic or non-paramagnetic) are the main factors responsible
for the maser polarization properties \citep[e.g.][]{Watson2008,
  Dinh-v-Trung2009_II}.
In addition, maser polarization is also affected by the ratio
between the Zeeman frequency $g\Omega$, the rate of stimulated
emission $R$ and the decay rate of the molecular state $\Gamma$
\citep{WW1984}.

The rate of stimulated emission can be obtained from
\begin{equation}
  R \simeq \frac{A k T_{B}\Delta\Omega}{4\pi h\nu} \quad,
  \label{eq:R}
\end{equation}
where $A$ is the Einstein coefficient which depends on the hyperfine transition,  $k$
and $h$ are the Boltzmann and Planck constants respectively,  $\nu$ is the maser
frequency. $T_{B}$ and $\Delta\Omega$ are the maser brightness temperature and
beaming solid angle \citep[see also][]{Vlemmings2011}.

The Zeeman frequency is defined as
\begin{equation}
  g\Omega={{2g\mu B}\over{\hbar}} \qquad,
  \label{eq:gomega}
\end{equation}
where $B$ is the magnetic
field in
G,
$\hbar$ is the reduced Planck constant and $g$ is the Land{\'e}
g-factor \citep[see e.g.][]{NW1990b}. In case of paramagnetic molecule (e.g.\ OH) $\mu$ is the
Bohr magneton ($\mu_B=e\hbar/2m_ec$) while in case of non-paramagnetic
molecule (e.g. H$_2$O or CH$_3$OH) it is the nuclear magneton
($\mu_N=e\hbar/2m_nc$),
where $e$ is the electron charge, $m_e$ and $m_n$ are the electron
and nucleon mass respectively.

The magnitude of the Zeeman effect depends on $g\Omega$ and it is
different between paramagnetic and non-paramagnetic molecules. Since $\mu_B/\mu_N\sim10^3$,
paramagnetic molecules can show a split three orders of magnitude
larger than the non-paramagnetic ones \citep{Vlemmings2007}.
Moreover, for paramagnetic molecules, the Zeeman frequency $g\Omega$
is usually larger than the intrinsic line width; thus the Zeeman
components are separated and resolved and from the Zeeman splitting it
is possible to obtain the magnetic field along the line of sight
without ambiguity, $B_{\parallel}=B\cos\theta$, with $\theta$ defined
as the angle between the magnetic field and the line of
sight.
On the contrary, for non-paramagnetic molecules like methanol,
$g\Omega$ is smaller than the line width, and the Zeeman components
overlap. In this case it is more complicated to infer circular
polarization measurements, because it also depends on the saturation
level and non-Zeeman effects might arise, compromising the measurement
of the regular Zeeman splitting.  A maser is defined saturated when
the rate of stimulated emission $R$ exceeds the decay rate of the
involved molecular state $\Gamma$.

\subsubsection{Rotation of the symmetry axis}
\label{Sec:nonZ1}
When $g\Omega>R$, the magnetic field direction is the quantization
axis, but when the maser brightness increases or for weak $B$,
$R$ can become much larger than $g\Omega$ and a rotation of the symmetry
axis can occur. This change of quantization axis can generate an
intensity-dependent circular polarization similar to the regular
Zeeman splitting \citep{NW1990}.  Therefore, it is important to know
when this effect might occur, by estimating $R$ and $g\Omega$.  The
ratio between the Zeeman splitting rate and the stimulated emission
rate is
\begin{equation}
\frac{R}{g\Omega}\simeq 1.7{{1}\over{g}} {{[{\rm mG}]}\over{B}}
{{T_b}\over{[10^{10} {\rm K}]}}{{\Delta\Omega}\over{[10^{-2} {\rm
      sr}]}} {{[{\rm GHz}]}\over{\nu}} {{A}\over{[10^{-7}{\rm s}^{-1}]}}.
\end{equation}
The rotation of the symmetry axis might occur when
${{R}\over{g\Omega}} >1$. In the case of the 6.7 GHz methanol maser,
considering the typical values for the beaming solid angle
$\Delta\Omega=10^{-2}$ K sr, a magnetic field $B=10$ mG,
$A=1.074\times10^{-9}$~s$^{-1}$ and an average Land{\'e} factor
calculated using all hyperfine components $g_a=0.236$ \citep{Lankhaar2016}, 
this effect might arise only when the maser is deeply saturated, with
$T_B\ge10^{12}$ K. This result also holds when considering the largest
and the smallest g-factor of the 6.7 GHz methanol maser hyperfine transition
$3\rightarrow 4$A and $5\rightarrow 6$B respectively, 
leading to $T_B\ge10^{10}$ K.

\subsubsection{Effect of magnetic field changes}
\label{Sec:nonZ2}

There are also other non-Zeeman mechanisms that can generate high
levels of polarization at lower $T_B$. One of these
effects is due to magnetic field changes along the maser path, for
example a rotation \citep{Wiebe1998}. This rotation converts the
linear polarization fraction $P_L$ to circular polarization fraction
$P_V$. Considering a rotation of 1 rad in the magnetic field
direction along the maser path, the fractional circular polarization
generated by this mechanism is $P_L^2/4$. According to previous
observations of 6.7 GHz methanol masers, this mechanism only produces
minor contributions \citep{Vlemmings2011}.  For example, the
polarization observed in high angular resolution observations for 6.7
GHz masers ranges usually from 1\% to 4\%. Therefore the change of
magnetic field direction along the maser propagation direction can
contribute at most $\sim0.04$\%, that is only a fraction of the
observed values of $P_V$ \citep[e.g.][]{Surcis2009, Surcis2012,
  Surcis2019}. But this effect can be important for very high levels
of linear polarization $P_L\sim10\%$, causing an extra
$P_V\sim0.25\%$. Recent methanol maser observations by \citet{Breen2019}
have registered a level of linear polarization $\sim7.5\%$, and we will
discuss them in Sect.~\ref{sec:discussion}.

\subsubsection{Anisotropic resonant scattering}
\label{sec:ArScattering}
Resonant scattering is a higher-order radiation matter interaction
that describes the absorption and subsequent immediate emission of a
photon by a molecule or atom. \citet{Houde2013} point out that forward
resonant scattering is coherent in nature, and that a large ensemble
of molecules, collectively scatter incoming radiation. In the presence
of a magnetic field, such collective resonant scattering leads to a
significant phase shift between the left and right-circularly
polarized radiation modes. This process is called anisotropic resonant
scattering (ARS) and leads to the conversion of linear polarization
(by way of the Stokes U component) to circular polarization.

\citet{Houde2014} propose that the circular polarization of SiO masers
might be generated through the anisotropic resonant scattering of
linearly polarized maser radiation by a foreground cloud between the
maser and the observer. The foreground cloud may be out of the
velocity range of the maser. Alternatively, the collective resonant
scattering of radiation might be a feature of the radiative transfer
of maser radiation. CHAMP only accounts for first-order radiative
interactions and neglects resonant scattering. Proper estimates of the
probability of ARS in relation to maser amplification have to be
developed before the importance of ARS to maser radiative transfer can
be evaluated.
 
Circular polarization profiles generated through ARS do not
necessarily lead to the antisymmetric S-shaped profile that
characterizes Zeeman circular polarization. Only under special
conditions are antisymmetric Stokes-V profiles generated by ARS. The
Stokes-V profiles of methanol masers are generally characterized by 
an antisymmetric spectrum.

\subsubsection{Other non-Zeeman effects}
\label{sec:ArScattering}

Other non-Zeeman effects might be due to instrumental effects and to
the presence of a velocity gradient across an extended source.
Potential instrumental effects causing extra circular polarization
depend on the instrument characteristics, and are generally reported
in the literature.  A velocity gradient could originate an S-shaped V
spectrum like the one produced by the Zeeman effect, but this effect
is rare with masers since they are point-like sources producing narrow
spectral lines developing in a narrow velocity range (e.g. \citealt{Sarma2020}).

\section{Methods}
\label{sec:methods}

In order to investigate the polarization of methanol masers by a
magnetic field and its effects on the hyperfine structure, we ran
simulations using the CHAMP code \citep{Lankhaar2019}. CHAMP is a
maser polarization radiative transfer code that takes into account all
dominant hyperfine components of a molecule and their individual Land{\'e}
factors. CHAMP implements both the methods in \citet[][hereafter
N\&W92]{NW92}, which do not treat non-Zeeman effects, and those in
\citet[][hereafter N\&W94]{NW94}, which do consider non-Zeeman effects. The
user can choose which method to use.  In addition to a combined
treatment of all hyperfines according to their transition
probabilities, CHAMP includes the possibility \textit{i)} to change the pumping
efficiency for different hyperfine components, and \textit{ii)} of anisotropic
pumping.

The linear polarization degree is defined as
\begin{equation}
 P_L = \frac{\sqrt{Q^2 + U^2}}{I} \quad,
\end{equation}
the polarization angle is
\begin{equation}
\psi = \frac{1}{2}\, \mathrm{atan}\left(\frac{U}{Q}\right) \quad,  
\end{equation}
and the circular polarization degree
is
\begin{equation}
P_V = \frac{(V_{\mathrm{max}} - V_{\mathrm{min}})}{I}  \quad,
\end{equation}
where I, Q, U and V are the Stokes parameters.

The polarization angle is relative to the projection of the magnetic
field direction onto the plane of the sky, that we take to
be North-South (unless along the maser beam when $\theta=0^\circ$).

In Table \ref{tab:detection} we report the maser transitions that we
used in our simulations.  Since the size of the masing region or the maser beaming angle
$\Delta\Omega$ are often unknown, for comparison with our models we give the
value of the brightness temperature $T_B$ assuming a maser spot size
of 3 mas. We also give a lower limit for $T_B$ based on the
observations from \citet{Breen2019}, \citet{ Sarma2020},
\citet{Momjian2019},  \citet{Surcis2009} and \citet{Surcis2019}. In the case of the 12.2
GHz maser, we are not modelling a specific source but we are giving a
typical value considering the previous work by \citet{Moscadelli2003}.
For known spectral features, $T_B$ can be derived using the equation
\begin{equation}
 \frac{ T_B}{[\mathrm{K}]} = \frac{S(\nu)}{[\mathrm{Jy}]}\left(\frac{\Sigma^2}{[\mathrm{mas}^2]}\right)^{-1}\zeta_\nu\qquad,
  \label{eq:tb}
  \end{equation}
where $S(\nu)$ is the detected flux density, $\Sigma$ is the maser
angular size and $\zeta_\nu$ is a constant that includes a
proportionality factor obtained for a Gaussian shape \citep{Burns1979} and scales with
the frequency according to the relation
\begin{equation}
\zeta_\nu\simeq 6.1305\times10^{11}\left(\frac{\nu}{[\mathrm{GHz}]}\right)^{-2}\frac{\mathrm{mas^2}}{\mathrm{Jy}}\mathrm{K}\qquad.
  \label{eq:zeta}
\end{equation}

In Table \ref{tab:detection} we also report which method between
N\&W92 and N\&W94 we used.  We performed simulations for different
methanol maser transitions at different magnetic field strengths,
angular momentum transitions, and propagation angles $\theta$.  The
molecular parameters used in the simulation are presented in
Tables~\ref{tab:meth6.2}--~\ref{tab:meth46} and are taken from
\citealt{Lankhaar2018} or, for the newly discovered transitions
observed by \citealt{Breen2019}, computed for this paper. In these
Tables, the quantities marked as ``g-factors'' are defined as
$\mu_N g / \hbar$. We ran calculations assuming an intrinsic thermal
velocity width $v_{th}=1$ km~s$^{-1}$.  For all the transitions we
used a pumping efficiency $\delta=0.02$, a decay rate of the upper and
lower level $\Gamma=1$ s$^{-1}$ \citep{Vlemmings2010, Lankhaar2019}
and, in case of anisotropic pumping, the anisotropy degree
$\epsilon=0.01$. The parameters are described in
\citealt{Lankhaar2019}.

\begin{sidewaystable*}
  \caption{Methanol maser transitions considered in our simulations and observational parameters}
  \begin{center}
  \begin{tabular}{lllcllllcccccl} \hline
Spectral line                   & \!\!\!\!\!\!\!Frequency &
                                                            {T$_{B}$~Lower
                                                            Limit}
                                                            \tablefootmark{a}
    &{T$_{B}$} \tablefootmark{a}      & {Obs. linear } &
                                                         Obs. circular & Source and & code\\ 
                                & GHz                                                 &{K} & {K}     & {pol. \%}       & {pol. \%}&references &       \\
\hline                                                                                                   
                                                                                                        
CH$_3$OH 17$_{-2}$ $\rightarrow$ 18$_{-3}$ E (v$_t$=1)         & 6.18    &  2.2$\times10^{6}$            &   5.2$\times10^{11}$   & 7.0    & $<$0.5    & G358.931--0.030 \tablefootmark{b}  & N\&W92       \\                                                                                
\noalign{\vskip4pt}
CH$_3$OH 5$_1$ $\rightarrow$ 6$_0$ A$^{+}$ (v$_t$=0)          & 6.68     &\makecell[l]{$7.6\times10^{6}$\\$2.8\times10^{13}$\\$1.3\times10^{12}$}            &  \makecell[l]{$1.5\times10^{12}$\\$8.7\times10^{13}$\\$6.3\times10^{12}$}   & \makecell[l]{7.5\\8.1\\4.5}   &  \makecell[l]{0.5\\2.1\\0.2}     & \makecell[l]{G358.931--0.030 \tablefootmark{b},\\W3(OH) \tablefootmark{c},\\W75N\tablefootmark{c}  }  & N\&W94       \\ 
\noalign{\vskip4pt}
CH$_3$OH 12$_4$ $\rightarrow$ 13$_3$ A$^-$ (v$_t$=0)         & 7.68      & 4.5$\times10^{6}$            &    6.5$\times10^{11}$ & 3.5    & $<$0.5    & G358.931--0.030 \tablefootmark{b}  & N\&W92        \\
\noalign{\vskip4pt}
CH$_3$OH 12$_4$ $\rightarrow$ 13$_3$ A$^+$ (v$_t$=0)         & 7.83      & 4.5$\times10^{6}$             &    6.3$\times10^{11}$ & 3.5    & $<$0.5    & G358.931--0.030 \tablefootmark{b}  & N\&W92        \\
\noalign{\vskip4pt}
CH$_3$OH 2$_0$ $\rightarrow$ 3$_1$ E (v$_t$=0)               & 12.2       &  $10^{9}$ -- $10^{12}$   & --      & --       &  --     &    W3(OH) \tablefootmark{d}         & N\&W94        \\
\noalign{\vskip4pt}
CH$_3$OH 17$_6$ $\rightarrow$ 18$_5$ E (v$_t$=0)               & 20.3       &  $4.5\times10^4$   & $1.7\times10^{10}$      & 4.0       &  2.0     &   G358.931--0.030 \tablefootmark{e}         & N\&W92        \\
\noalign{\vskip4pt}
CH$_3$OH 10$_1$ $\rightarrow$ 11$_2$ A$^+$ (v$_t$=1)          & 20.9       & 3.6$\times10^{4}$                               &     1.5$\times10^{11}$ & 7.0    & $<$0.5   & G358.931--0.030 \tablefootmark{b}   & N\&W92      \\
\noalign{\vskip4pt}
CH$_3$OH 5$_2$ $\rightarrow$ 5$_1$ E (v$_t$=0)                & 25.0      & 9.1$\times10^{3}$                                 & 7.2$\times10^{9}$     &   --   & 0.3   & OMC--1 \tablefootmark{f}              & N\&W94       \\
\noalign{\vskip4pt}
CH$_3$OH 4$_{-1}$ $\rightarrow$ 3$_0$ E                       & 36.2      & \makecell[l]{$1.7\times10^1$\\ $3.7\times10^{4}$}             &   \makecell[l]{$2.6\times10^{7}$ \\ $3.4\times10^{9}$}  & \makecell[l]{$<$0.5\\--} & \makecell[l]{$<$0.5\\0.1}   &\makecell[l]{G358.931--0.030 \tablefootmark{b},\\  M8E \tablefootmark{g}}  & N\&W94       \\
\noalign{\vskip4pt}
CH$_3$OH 7$_{-2}$ $\rightarrow$ 8$_{-1}$ E (v$_t$=0)          & 37.7      & 8.7$\times10^{3}$                                  &    1.2$\times10^{10}$  & 3.5    & 0.5    & G358.931--0.030 \tablefootmark{b}   & N\&W92         \\
\noalign{\vskip4pt}
CH$_3$OH 7$_0$ $\rightarrow$ 6$_1$ A$^+$                     & 44.1       & \makecell[l]{$1.2\times10^{2}$\\ $5.0\times10^{4}$ }          &   \makecell[l]{ $3.15\times10^{8}$\\$1.3\times10^{10}$} &\makecell[l]{$<$0.5\\--}    & \makecell[l]{$<$0.5\\0.05}   & \makecell[l]{G358.931--0.030 \tablefootmark{b},\\DR21W \tablefootmark{h}}   & N\&W94         \\
\noalign{\vskip4pt}
CH$_3$OH 2$_0$ $\rightarrow$ 3$_1$ E (v$_t$=1)               & 44.9        & 1.9$\times10^{4}$                                &   1.7$\times10^{10}$& 2.5    & $<$0.5   & G358.931--0.030 \tablefootmark{b}    & N\&W94         \\
\noalign{\vskip4pt}
CH$_3$OH 9$_3$ $\rightarrow$ 10$_2$ E (v$_t$=0)              & 45.8        & 1.4$\times10^{4}$                                &    1.3$\times10^{10}$&  7.0    & 1.5      & G358.931--0.030 \tablefootmark{b}    & N\&W92          \\
\hline
  \end{tabular}
\end{center}
\tablefoot{
    \tablefoottext{a}{Brightness temperature is estimated using a maser
      spot size of 3 mas using Eq.~(\ref{eq:tb}) and Eq.~(\ref{eq:zeta}), while the lower
     limit is based on the observational details 
     given in the source references.}
   \tablefoottext{b}{From \citet{Breen2019}.}
    \tablefoottext{c}{For W3(OH) we refer to the feature W3OH.22 from \citet{Surcis2019}, and for W75N to \citet{Surcis2009}.}
    \tablefoottext{d}{We do not model any specific maser but we only report values observed by \citet{Moscadelli2003}.}
    \tablefoottext{e}{From \citet{MacLeod2019}.}
    \tablefoottext{f}{We used the component 2 of the maser observed by \citet{Sarma2020}.}
    \tablefoottext{g}{From \citet{Sarma2009}.}
    \tablefoottext{h}{We used the component 1 of the maser observed by \citet{Momjian2019}.}
    
}
\label{tab:detection}
\end{sidewaystable*}

\section{Results}
\label{sec:results}
\subsection{6.7 GHz methanol maser}
\label{sec:results_6.7}

We report on our simulations of the 6.7 GHz methanol maser in a range
of different magnetic field strengths (1, 3, 10, 20, 30 and 100 mG),
with varying luminosity and magnetic field angles. The results
  of these simulations are shown in several figures in the body of the
  paper (Fig.~\ref{fig:PQPV_TB}--~\ref{fig:profile_iso1}) and in the
  appendix (Fig.~\ref{fig:PQPV_TB_altriB}--~\ref{Fig:spectra6.2}).
The magnetic field strength, the thermal velocity width v$_{th}$, the
propagation angle $\theta$ and the transition angular momenta are
indicated in each figure. In
  Fig.~\ref{fig:PQPV_TB}--~\ref{fig:profile_iso1},~\ref{fig:PQPV_TB_altriB},
  and \ref{fig:PL_cont_altriB}--~\ref{fig:pa_profile_iso_altriB}, the
  panel at the top left labelled ``baseline'' indicates a maser
emission where all the eight hyperfine transitions contribute equally,
while all other panels assume a preferred pumping for the indicated
$i\rightarrow j$ transition. The preferred pumping rate is ten times
larger than the other transitions' rate.

\begin{figure}
 \centering
 \includegraphics[width=\columnwidth, clip]{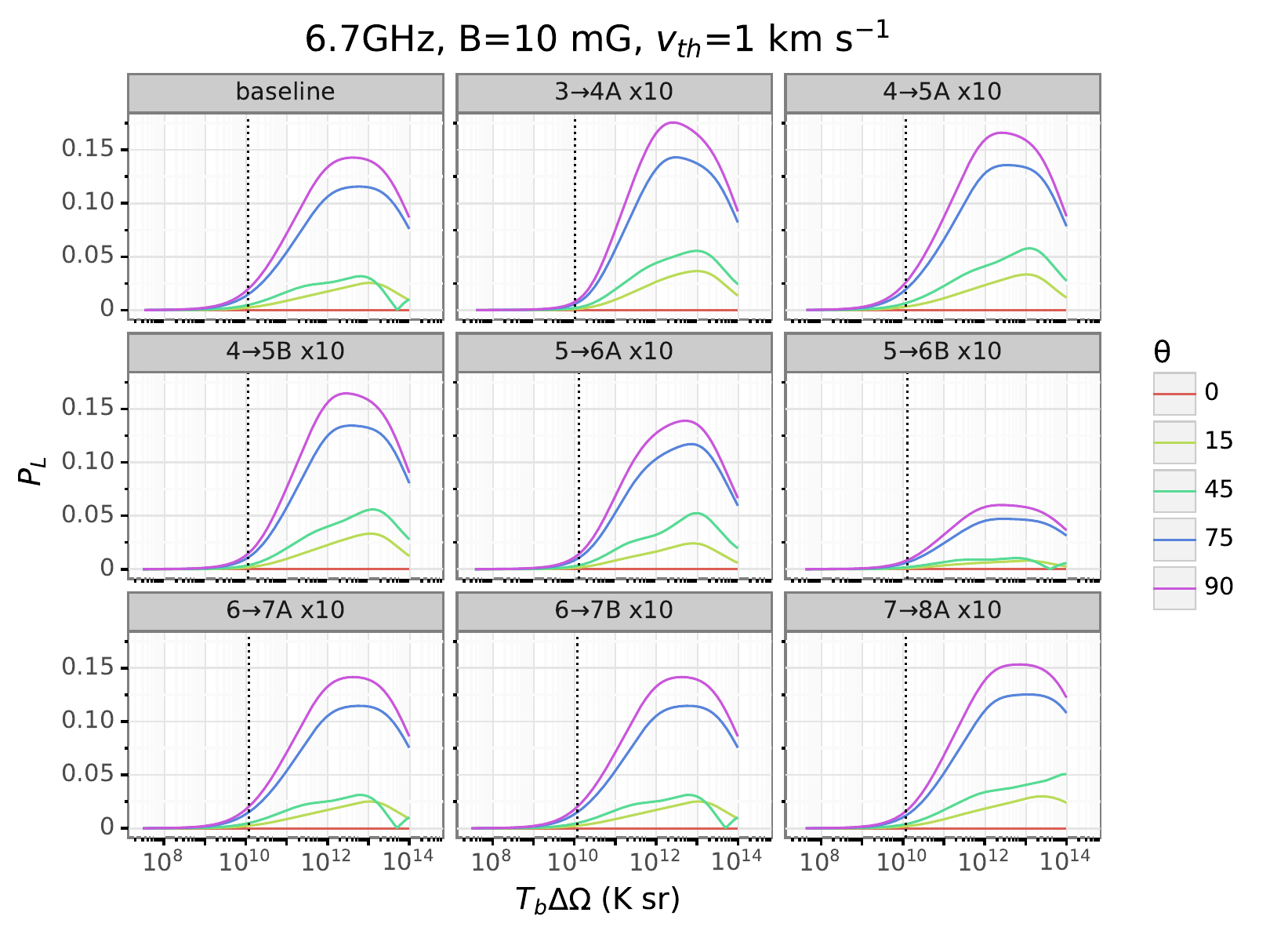}
  \includegraphics[width=\columnwidth, clip]{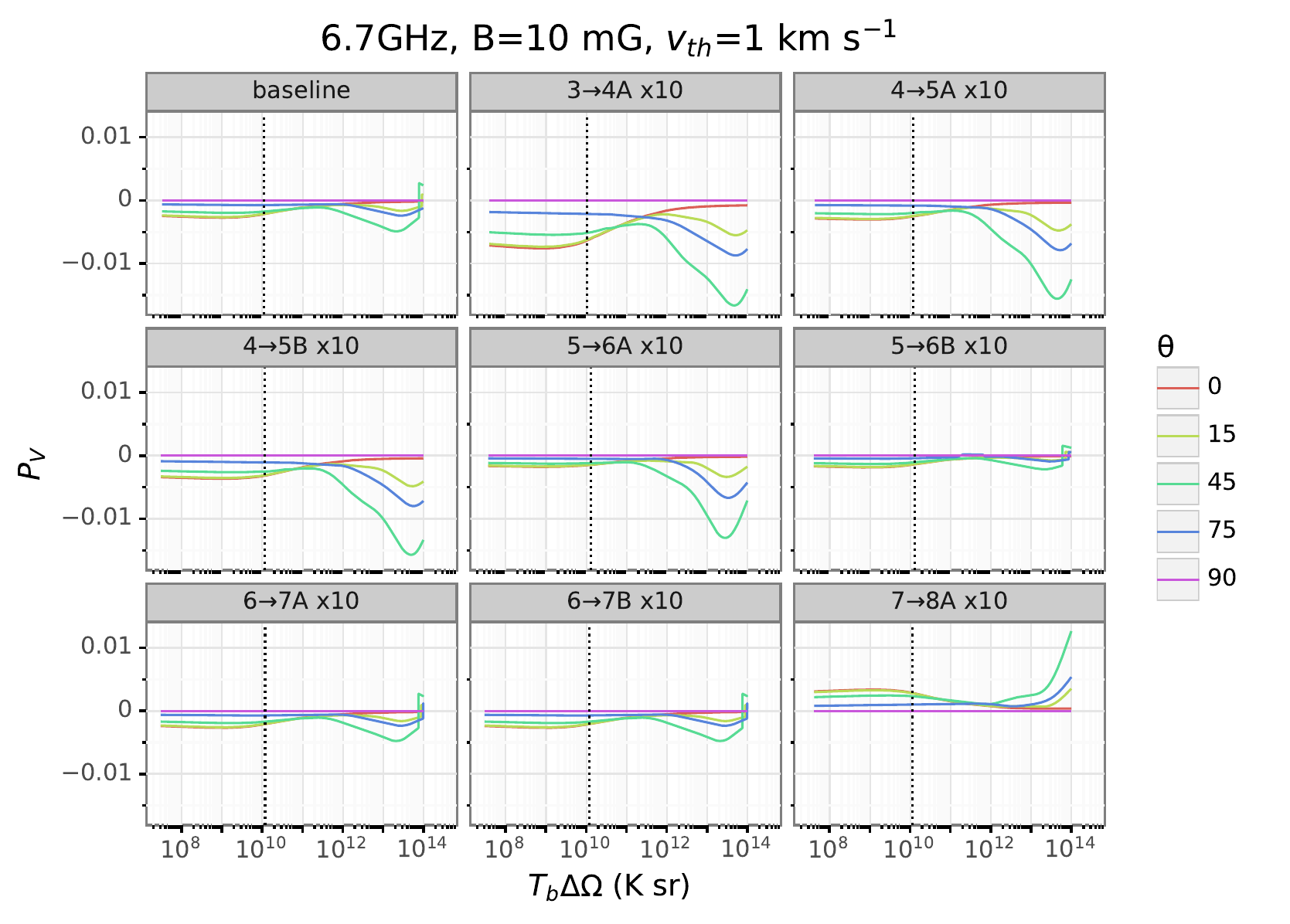}
  \caption{6.7 GHz methanol maser linear and circular
    polarization fraction as a function of the maser luminosity for
    five different $\theta$. The vertical line marks $T_B\Delta\Omega$
    where $g\Omega=10R$. The magnetic field strength is 10 mG, the
    thermal velocity width is 1 km s$^{-1}$, and the preferred hyperfine
    transitions are indicated in each panel. The panel at the top left
    labelled ``baseline'' indicates a fixed pumping rate equal for all
    the hyperfine transitions, while all others assume a $10\times$
    preferred pumping for the indicated $i\rightarrow j$ transition.
    The results of the simulation performed with other magnetic field
    strengths are given in Appendix~\ref{sec:appendix}.}
 \label{fig:PQPV_TB}
\end{figure}
We present in Fig.~\ref{fig:PQPV_TB} the linear and circular
polarization fraction $P_L$ and $P_V$ as a function of the maser
luminosity for five different $\theta$ values. We mark with a vertical
line the locus $g\Omega=10R$, that for 6.7 GHz methanol maser falls at
$T_B\Delta\Omega\sim10^{10}$K sr. For $T_B\Delta\Omega\lesssim10^{10}$ K sr
the Zeeman frequency $g\Omega$ is much higher than the
stimulated emission rate $R$. On the contrary for
$T_B\Delta\Omega\gtrsim10^{10}$ K sr, the Zeeman frequency is similar to or lower than
the rate of stimulated emission. As already mentioned in
Sect.~\ref{sec:Non-zeeman}, masers can be affected by several
non-Zeeman processes that can intensify or generate linear and
circular polarization. Usually $g\Omega=10R$ has been used in the
literature (e.g.\ by \citealt{Perez-sanches2013}) to mark the region
where non-Zeeman effects become relevant, and therefore inferring
magnetic field properties becomes more challenging, because the
magnetic field is not directly related to $P_L$ and $P_V$.  One of the
most prominent non-Zeeman effects is the rotation of the symmetry axis
that can occur when $g\Omega<10R$ \citep{Perez-sanches2013, Wiebe1998},
thus we decided to indicate $g\Omega=10R$ as an upper limit for having
reliable $P_L$ and $P_V$.

$P_L$ and $P_V$ show a clear dependence on the hyperfine transitions,
on the magnetic field strength and on the angle $\theta$. From these
plots it is quite evident that each single hyperfine component is
affected by the magnetic field in a different way, and therefore, in
presence of preferred hyperfine pumping, we observe a specific level
of polarization fraction. At a first glance, these
  dependencies appear most prominent above $T_B\Delta\Omega$
  corresponding to $g\Omega=10R$, however noticeable effects also
  occur when $g\Omega>10R$ for several transitions, such as
  $3\rightarrow 4 $~A and $4\rightarrow 5 $~A. For example the
  dependence on $\theta$ shown in Fig.~\ref{fig:msersPV_TB} is
  different from the usual cos$\theta$ dependence
  \citep{WatsonWyld2001} also for $T_B\Delta\Omega$ below
  $g\Omega=10R$. Also these dependencies seem weaker for $P_V$
  compared to that for $P_L$ , unless the magnetic field is very high,
  $\sim$100 mG (Fig.~\ref{fig:PQPV_TB_altriB}), or the hyperfine is
  preferentially pumped by a large degree,$\sim100$ times
  (Fig.~\ref{Fig:trans43}).

We will now consider the $P_L$ and $P_V$ values at $g\Omega=10R$. When
$\theta=90^o$, and $B=10$ mG, the maximum linear polarization fraction
is found to be of the order of 3\% for the hyperfine transition
$4\rightarrow 5 $~A. When $\theta=0^o$ and $B=10$ mG, we obtain a
maximum circular polarization fraction of the order of 1\% for the
hyperfine transition $3\rightarrow 4 $~A. As the magnetic field
strength increases, the position of $g\Omega=10R$ falls at higher
$T_B\Delta\Omega$. Thus, for $B=100$ mG, $P_L$ can reach 10\% in the
hyperfine transitions $4\rightarrow 5 $~A at $\theta=90^o$, while
$P_V$ can rise up to 8\% when the hyperfine transition
$3\rightarrow 4 $~A is preferentially pumped, and $\theta=0^o$.

We also investigated different degrees of preferred pumping for the
two strongest hyperfine transitions $3\rightarrow 4$~A and
$4\rightarrow 5$~A. We observed that both the linear and circular
polarization fractions increase with the preferred pumping.
These results are shown in Fig.~\ref{Fig:trans43} and
Fig.~\ref{Fig:trans54}, where the linear and circular polarization
fractions are plotted as a function of the maser luminosity. Different
degrees of preferred pumping are shown in different colours for the
two hyperfine transitions, assuming a magnetic field of 10 mG. The
vertical dotted lines indicate when
$g\Omega=10R$.

In Tables \ref{tab:maxPL} and
\ref{tab:maxPV}, we report the $P_L$ and $P_V$ values taken at
$g\Omega=10R$, for all the masers and for a magnetic field $B=10$
mG. The results of the simulation performed with other magnetic field
strengths (1 and 100 mG) are given in Appendix~\ref{sec:appendix} in
Fig.~\ref{fig:PQPV_TB_altriB}.

\begin{figure}[h!]
 \centering
 \includegraphics[width=0.9\columnwidth,clip]{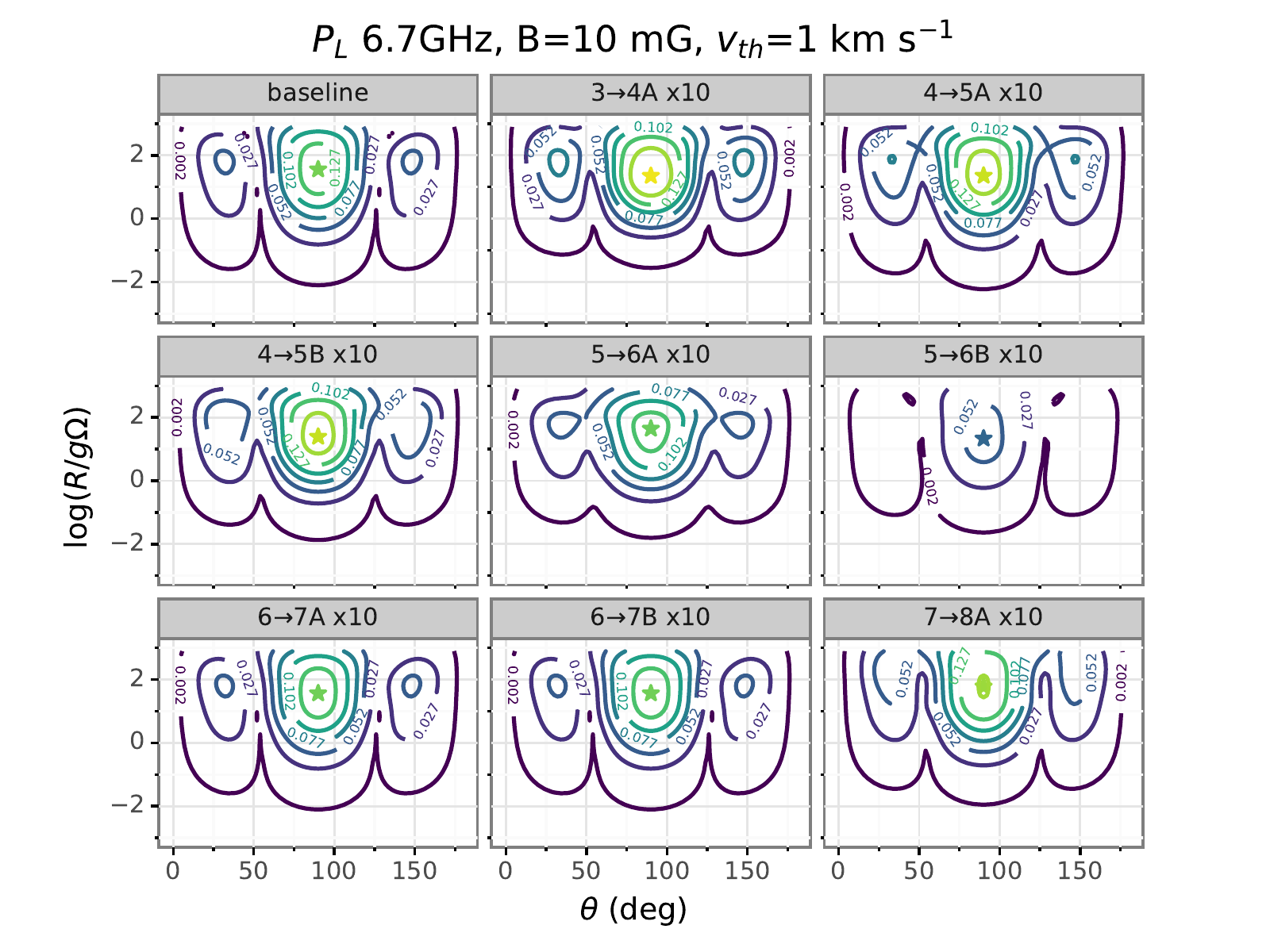}\\
 \includegraphics[width=0.9\columnwidth,clip]{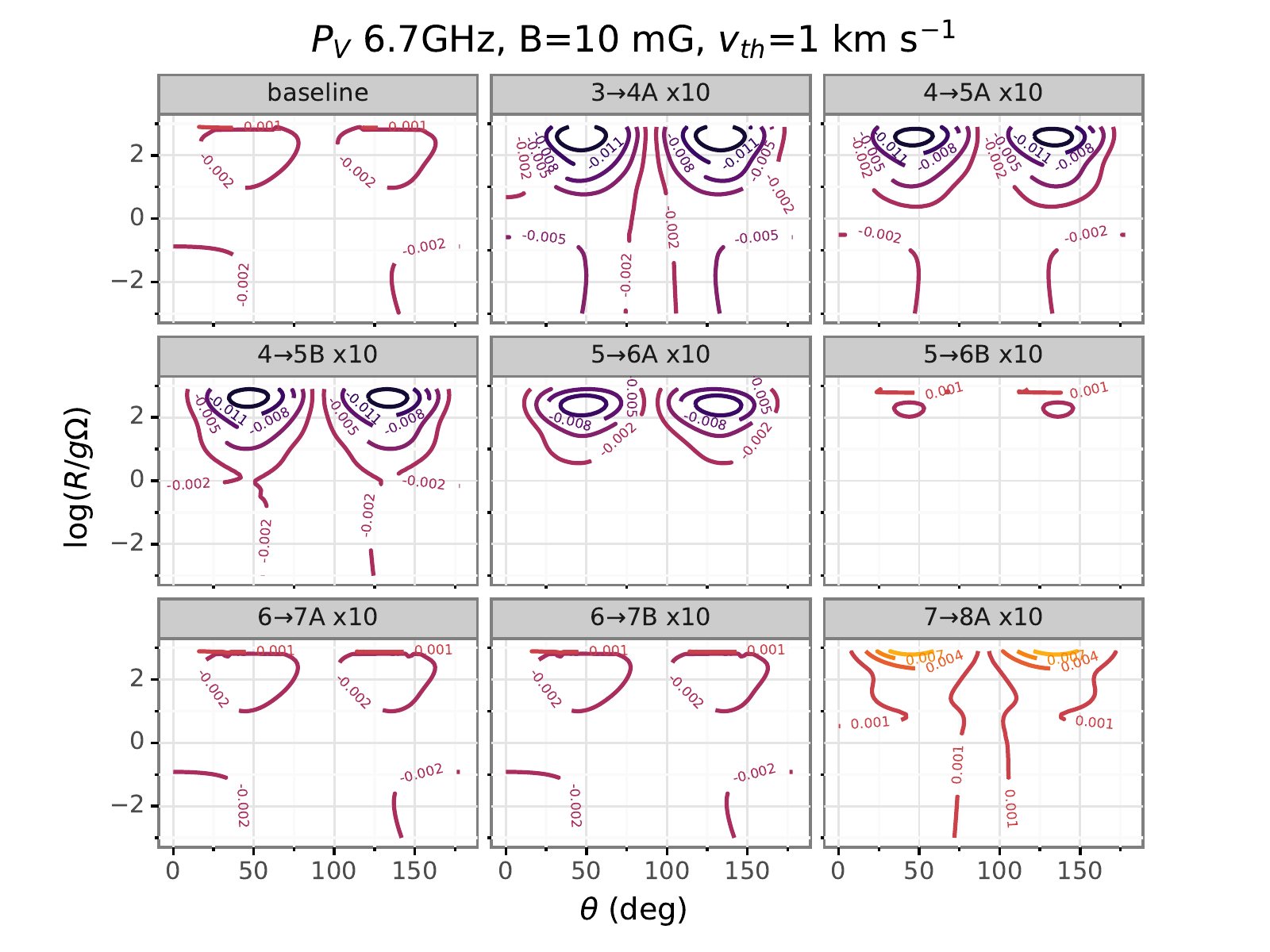}\\
 \includegraphics[width=0.9\columnwidth, clip]{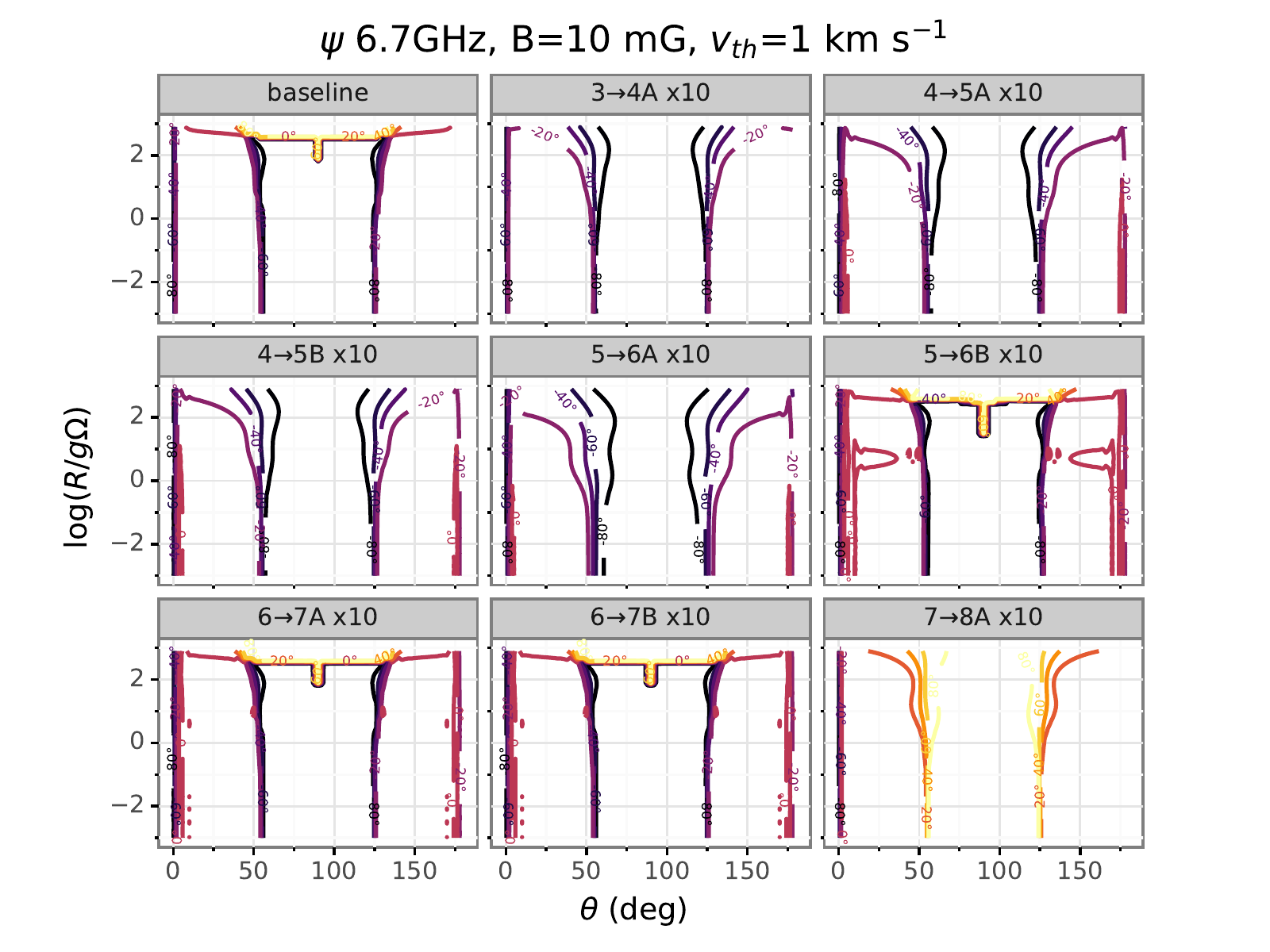}

 \caption{6.7 GHz methanol maser linear polarization fraction
   P$_L$ (top), circular polarization fraction P$_V$ (middle) and
   Linear polarization angle $\psi$ (bottom) plotted as a function of
   the propagation angle $\theta$ and the rate of stimulated
   emission. Stars indicate the position of the peak of
   P$_L$. Magnetic field strength is 10 mG, thermal velocity width is
   1 km s$^{-1}$ and the hyperfine transitions are indicated in each
   panel. The panel at the top left labelled ``baseline'' indicates a
   fixed pumping rate equal for all the hyperfine transitions, while
   all others assume a $10\times$ preferred pumping for the indicated
   $i\rightarrow j$ transition. The results of the simulation
   performed with other magnetic field strengths are given in
   Appendix~\ref{sec:appendix}.}
 \label{fig:contour_iso1}
\end{figure}

\begin{figure}[h!]
 \centering

 \includegraphics[width=\columnwidth,clip]{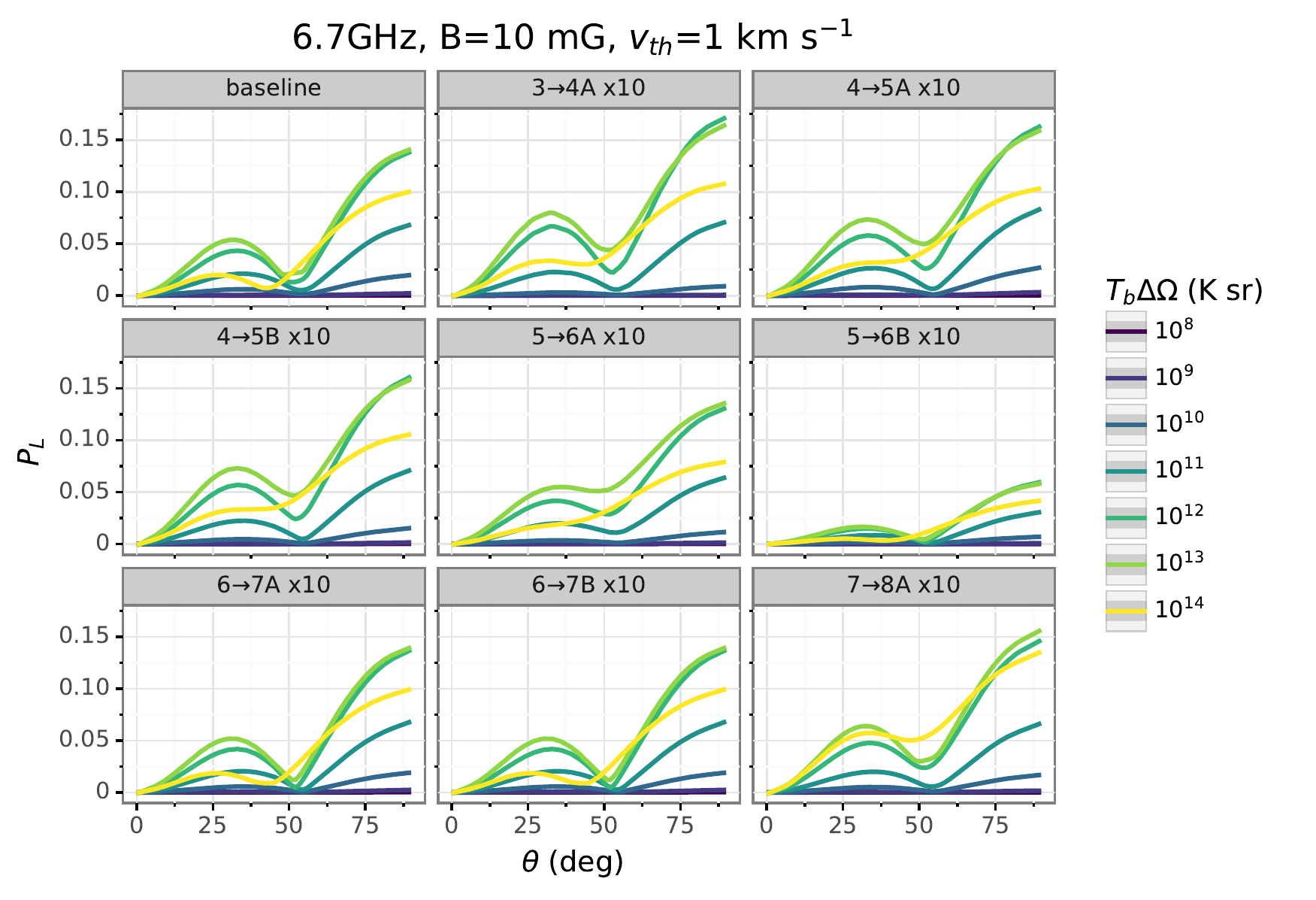}\\
 \includegraphics[width=\columnwidth,clip]{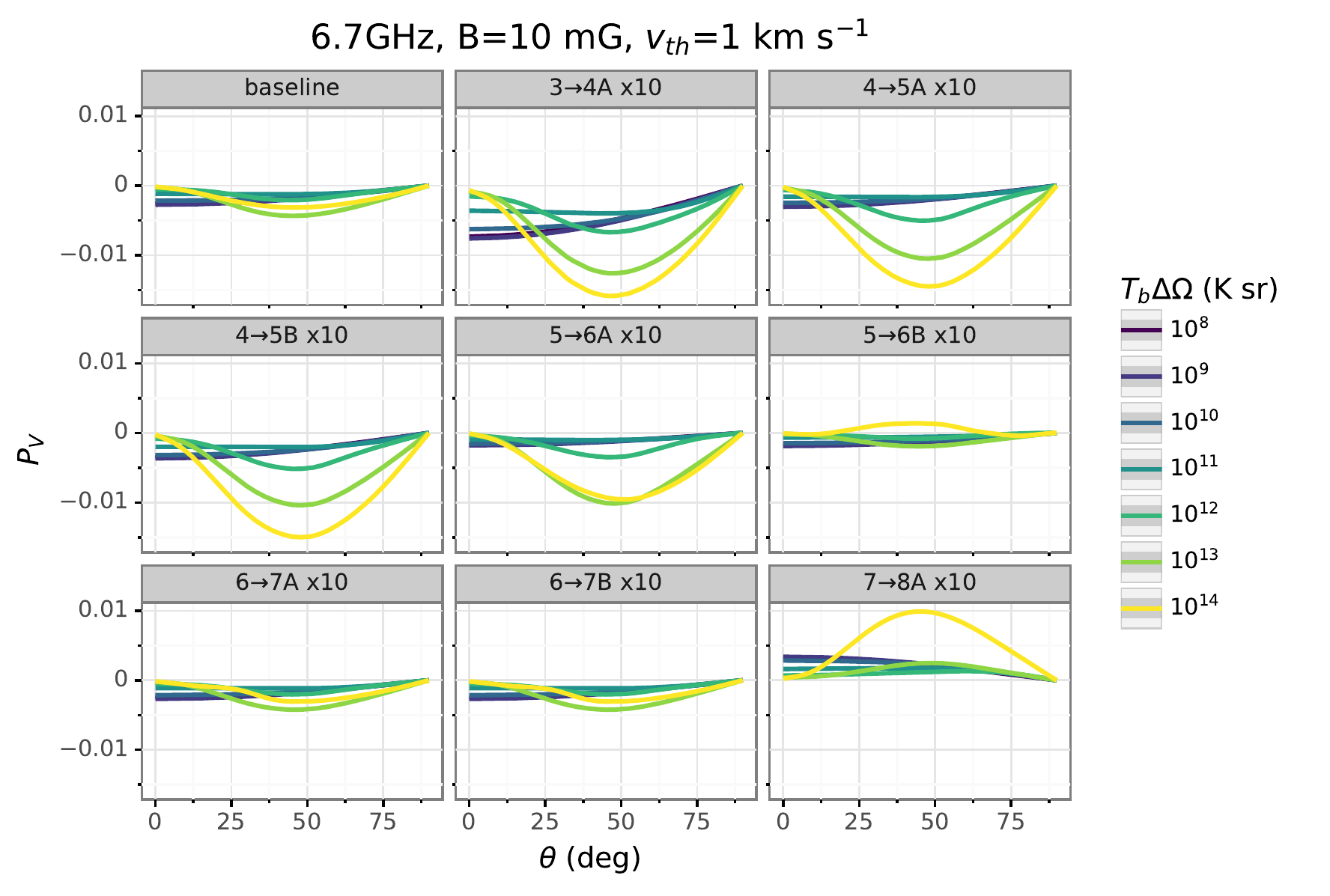}\\
 \includegraphics[width=\columnwidth,clip]{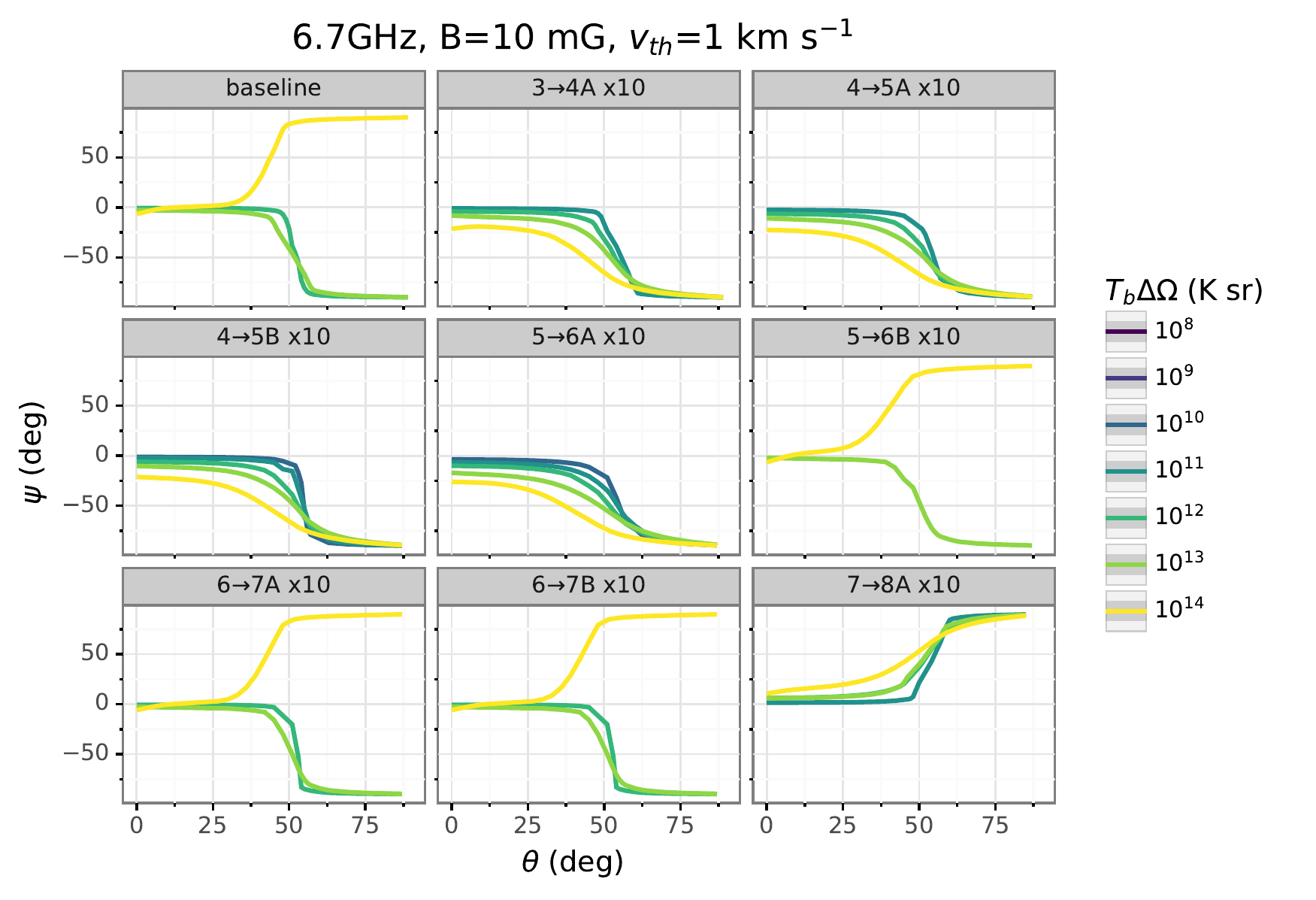}
 
  \caption{6.7 GHz methanol maser linear polarization
    fraction P$_L$ (top), circular polarization 
    fraction P$_V$(middle) and polarization angle
    $\psi$ (bottom) plotted as a function of the propagation angle
    $\theta$ for different brightness temperatures. Panels as in
    Fig.~\ref{fig:contour_iso1}. }
 \label{fig:profile_iso1}
\end{figure}

In Fig.~\ref{fig:contour_iso1} and from top to bottom,
respectively, we plot the linear polarization fraction
$P_L$, the circular polarization fraction $P_V$ and the linear
polarization angle $\psi$ as a function of the
propagation angle $\theta$ and of the rate of stimulated emission for
a magnetic field of 10 mG. These plots are symmetric along the line $\theta=90^\circ$.
The results of the
simulation performed with other magnetic field strengths are given in
Appendix~\ref{sec:appendix} in
Figs.~\ref{fig:PL_cont_altriB}--~\ref{fig:Pa_cont_altriB}.

Also in this case, $P_L$ shows a dependence on the magnetic field
strength and on preferred hyperfine pumping. $P_L$ presents a sharp
drop around the Van Vleck angle $\theta=54^\circ$.  $P_L$ peaks around
$\theta=90^\circ$ and $\mathrm{log}(R / g\Omega)=1$ (the peak is
marked with a star in Fig.~\ref{fig:contour_iso1}, top), with a
secondary peak around $\theta=30^\circ$ and
$\mathrm{log}(R / g\Omega)=2$. This morphology is consistent across
different magnetic field strengths and angular momenta of the
transitions.  We observe that also $P_V$ contours show a dependence on
magnetic field strength and hyperfine transitions
(Fig.~\ref{fig:contour_iso1}, middle). The contour morphology shows
two peaks (plus their symmetric counterparts at $\theta>90^\circ$)
located approximately at $\mathrm{log}(R / g\Omega)\sim2.5$ and
$\theta\sim50^\circ$, and at $\mathrm{log}(R / g\Omega)\sim-2$ and
$\theta\sim0^\circ$. The first one is dominant at lower $B$ strengths
(1, 3, 10 mG), while the second becomes more important at higher $B$
(20, 30, 100 mG).
As the $B$ strength increases, some hyperfine transitions present emerging
features in the $\psi$ contours, where the polarization angles
flip. For instance, in the case of $B=10$ mG
(Fig.~\ref{fig:contour_iso1}, bottom), the transition 5$\rightarrow$6~B
presents two emerging regions located at
$\mathrm{log}(R / g\Omega)\sim0.5$ and $\theta\sim25^\circ$. Similar
marks appear also for magnetic fields of 30 and 100 mG in the
hyperfine transition 3$\rightarrow$4 A. We notice a $90^\circ$ flip of
the polarization angle $\psi$ happening around the Van Vleck angle,
and this trait becomes sharper with increasing magnetic field
strength. This characteristic has been observed also for SiO and
H$_2$O masers \citep{Vlemmings2006, Tobin2019}.  Generally, when the saturation level is low
($\mathrm{log}(R/g\Omega)<0$), the linear polarization angle $\psi$
appears quite stable in the range $0^\circ<\theta<54^\circ$, 
while variations of $20^\circ$--$40^\circ$ are visible for high level of
saturation. Around $\theta\sim90^\circ$, $\psi\sim90^\circ$ for almost all
the saturation values.

In these plots we do not take into account the limit $g\Omega=10R$ and
therefore the strongest peak in $P_L\sim$ 0.2 and $P_V\sim$ 0.02,
registered for the hyperfine transition 3$\rightarrow$4 A, are due to
non-Zeeman effects.

In Fig.~\ref{fig:profile_iso1} the linear and circular polarization
fraction and the polarization angle are plotted as functions of the
propagation angle $\theta$ for different brightness temperatures in a
magnetic field $B=10$ mG.  From Fig.~\ref{fig:PQPV_TB}, we know that for
the 6.7 GHz methanol maser $g\Omega=10R$ happens around
$T_B\Delta\Omega\sim10^{10}$~K~sr.  When $T_B\Delta\Omega>10^{10}$ K
sr and $g\Omega<<10R$, the maser starts to saturate and some
non-Zeeman effects (such as the rotation of the symmetry axis) can
arise. $P_L$ decreases approaching the Van Vleck angle from both smaller
and greater angles, and at the Van Vleck angle $P_L$ has a minimum. This
behaviour is maintained across different levels of saturation. We
note that $P_L$ reaches its highest values when $\theta$ is greater than
Van Vleck angle and for $T_B\Delta\Omega\sim10^{12}-10^{13}$ K sr.

Looking at the circular polarization profile, when
$T_B\Delta\Omega<10^{10}$ K sr, the maser is not saturated and $P_V$
decreases for increasing $\theta$, following the cosine profile
described by \citet{WatsonWyld2001}. In this regime $P_V$ can be
directly related to the magnetic field strength.  When the maser saturates
and $T_B\Delta\Omega>10^{10}$ K sr, a rotation of the symmetry axis
might arise, and $P_V$ is not linked anymore to magnetic field
strength. In Fig.~\ref{fig:msersPV_TB} we plot the cosine (dotted
line) for different $T_B\Delta\Omega$ (solid line) to show this
behaviour. We also observe an inversion of the linear polarization
angle corresponding to the Van Vleck angle, which is sharp for
$T_B\Delta\Omega<10^{10}$ K sr and it becomes smoother with increasing
brightness temperatures.

\begin{table}
  \centering
  \caption{$P_L$ upper limits at  $g\Omega=10R$, for $B=10$ mG}
\begin{tabular}{rrrl}
\hline
  Maser       & $P_L$  baseline\tablefootmark{a} & \multicolumn{2}{l}{$P_L$ with one hyperfine} \\
              &   \%              & \multicolumn{2}{l}{transition $\times$10}\tablefootmark{b}    \\
  \hline
       6.2GHz & 1.9           & 2.2 & (16$\rightarrow$17A)                           \\
       6.7GHz & 2.0           & 2.6 & (4$\rightarrow$5A)                             \\
       7.7GHz & 2.5           & 4.1 & (10$\rightarrow$11A)                           \\
       7.8GHz & 2.5           & 4.1 & (10$\rightarrow$11A)                           \\
       12GHz  & 2.8           & 3.6 & (1$\rightarrow$2B)                             \\
      20.3GHz & 1.7           & 2.1 & (16$\rightarrow$17A)                           \\
        21GHz & 3.2           & 5.6 & (8$\rightarrow$9A)                             \\
        25GHz & 9.5           & 11  & (4$\rightarrow$4A)                             \\
        36GHz & 2.9           & 3.7 & (3$\rightarrow$2A)                             \\
      37.7GHz & 4.6           & 6.1 & (6$\rightarrow$7B)                             \\
        44GHz & 2.2           & 2.8 & (5$\rightarrow$4A)                              \\
        45GHz & 4.1           & 6.2 & (1$\rightarrow$2A)                             \\
        46GHz & 3.5           & 4.5 & (8$\rightarrow$9A)                             \\
  \hline
  \label{tab:maxPL}
\end{tabular}
\tablefoot{
\tablefoottext{a}{``Baseline'' indicates a
  fixed pumping rate equal for all the hyperfine transitions.}
\tablefoottext{b}{$10\times$ preferred pumping for the 
   $i\rightarrow j$ hyperfine transition indicated between parentheses.}}
\end{table}

\begin{table}
  \caption{$P_V$ upper limits at  $g\Omega=10R$, for $B=10$ mG}
  \centering
\begin{tabular}{rrrl}
\hline
  Maser        & $P_V$  baseline\tablefootmark{a} & \multicolumn{2}{l}{$P_V$ with one hyperfine} \\
               &   \%             & \multicolumn{2}{l}{transition $\times$10}    \\
  \hline
       6.2GHz  &4.2          &4.9       & (16$\rightarrow$17A)                  \\
       6.7GHz  &0.22         &0.63      & (3$\rightarrow$4A)                    \\
       7.7GHz  &5.4          &8.3       & (10$\rightarrow$11A)                  \\
       7.8GHz  &5.4          &8.3       & (10$\rightarrow$11A)                   \\
        12GHz  &0.095        &0.34       & (1$\rightarrow$2A)                    \\
       20.3GHz &3.9          &4.7        & (16$\rightarrow$17A)                \\
        21GHz  &6.0          &10       & (8$\rightarrow$9A)                    \\
        25GHz  &0.014        &0.098       & (4$\rightarrow$4B)\tablefootmark{c}                    \\
        36GHz  &0.011        &0.05      & (3$\rightarrow$2A)                    \\
      37.7GHz  &8.0          &10       & (6$\rightarrow$7B)                    \\
        44GHz  &0.011        &0.056       & (5$\rightarrow$4A)                     \\
        45GHz  &0.011        &0.055       & (1$\rightarrow$2B)                    \\
        46GHz  &6.7          &8.3       & (8$\rightarrow$9A)                    \\
  \hline
  \label{tab:maxPV}
\end{tabular}
\tablefoot{
  \tablefoottext{a}{``Baseline'' indicates a
   fixed pumping rate equal for all the hyperfine transitions.}
 \tablefoottext{b}{$10\times$ preferred pumping for the 
   $i\rightarrow j$ hyperfine transition indicated between parentheses.}
    \tablefoottext{c}{The same level of $P_V$ has been registered also
    for the 4$\rightarrow$4A and 6$\rightarrow$6B. }}
\end{table}

We also produced spectra of the 6.7 GHz methanol maser for different
levels of saturation. We observe that the line broadens as the maser
starts to saturate. Also the Stokes $Q$, $U$ and $V$
  intensify for high saturation levels. It is also interesting to
observe that, when preferred pumping is acting, a flip of the S-shape
profile of the circular polarization can be observed between different
hyperfine transitions.
The S-shaped Stokes V profile in
  Fig.~\ref{Fig:spectra6.7_1} for the 3$\rightarrow$4~A transition has
  the same shape as for the baseline profile in
  Fig.~\ref{Fig:spectra6.7_baseline}, but the profile reverses
  for the 8$\rightarrow$7~A transition in
  Fig.~\ref{Fig:spectra6.7_8}.  All Stokes parameters are plotted
and several propagation angles $\theta$ are taken into account,
assuming a magnetic field of 10 mG and v=1 km~s$^{-1}$.

In addition, we investigated anisotropic pumping and found, as already
noticed for SiO masers \citep{Lankhaar2019}, an increase of linear
polarization fraction up to 100\%.  In light of methanol maser
pumping, the amount of anisotropy that can occur is low, and since
such a high level of linear polarization has never been observed, we
expect that anisotropic pumping can be considered less important for
methanol masers than for SiO masers.
Anisotropic pumping in SiO masers occurs primarily through the
  absorption of IR radiation from a nearby central stellar
  object. Methanol masers are either pumped collisionally (class I) or
  radiatively (class II). Collisional pumping generally does not
  introduce any anisotropy in the molecular states. The radiative
  pumping of class II masers can be anisotropic, albeit less so than
  compared to the case of SiO masers, since it occurs through
  co-spatial dust. Additionally, the high density of class II masers
  de-polarizes the molecular states \citep{Lankhaar2020}, resulting in
  generally low degrees of anisotropic pumping.
We also investigated the effect
of amplification of a background polarized emission and we found that it
does not contribute significantly to the linear polarization fraction
unless that background emission itself is already polarized at a high
enough level.

\subsection{Other methanol maser transitions}
\label{sec:results_others}

For the other methanol maser transitions we investigated $P_L$ and
$P_V$ as a function of the maser luminosity for five different
$\theta$ values and magnetic fields of 10, 20 and 30 mG.
In case of 25 GHz methanol maser we investigated also a
  magnetic field of 100 mG. All the well-known maser
  transitions are modelled using the N\&W94 code that can treat
  non-Zeeman effects (see Tab.~\ref{tab:detection}), and only the
  lesser known transitions like 6.2, 7.7, 7.8, 20.3, 20.9, 37.7 and 46
  GHz, have been investigated with the N\&W92 code. N\&W92 code can
  not deal with non-Zeeman effects, and therefore the simulations stop
  at $g\Omega=10R$. Within this limit they behave like the other
  masers modelled with the N\&W94 code.

In general, also at these other frequencies, $P_L$ and $P_V$ show a
dependence on the hyperfine transitions, on the magnetic field
strength and on the angle $\theta$, although less so for $P_V$
compared to $P_L$ unless the magnetic field is very high ($\sim$100
mG) or a hyperfine is preferentially pumped by a large degree
($\sim$100 times), consistent with our results for 6.7 GHz masers
reported in Sec.~\ref{sec:results_6.7}. We confirm also for these
transitions that hyperfine preferred pumping can influence the
polarization fraction. In Tables \ref{tab:maxPL} and \ref{tab:maxPV}
we report $P_L$ and $P_V$ for these maser transition, in case of
$B=10$ mG and taken at $g\Omega=10R$. Usually the maximum value of
$P_L$ and $P_V$ is reached for $\theta=90^{\circ}$ and
$\theta=0^{\circ}$, respectively.
In the case of the 25GHz maser, the maximum in $P_V$ is registered for
the hyperfines 4$\rightarrow$4~A and B and 6$\rightarrow$6~B, that are
the transitions where the g-factors are highest.
The plots with the results of the
simulations for the 6.2 GHz baseline in a magnetic field $B=10$ mG
are given in Appendix~\ref{sec:appendix} in Fig.~\ref{fig:6.2PQPV_TB}.
 
Contour plots were produced only for simulations made with the N\&W94
code. The 12, 25, 36, 44, 45 GHz masers show contours similar to the
6.7 GHz. $P_L$ and $P_V$ typically present the same morphology
reported for the 6.7 GHz with a dependence on the magnetic field
strength and hyperfine transitions.  The $P_L$ peaks are located in
the same positions registered for the 6.7 GHz, around
$\theta=90^\circ$ and $\mathrm{log}(R / g\Omega)=1$ around
$\theta=30^\circ$ and $\mathrm{log}(R / g\Omega)=2$. The $P_V$ peaks
are situated approximately at $\mathrm{log}(R / g\Omega)\sim2.5$ and
$\theta\sim50^\circ$, and at $\mathrm{log}(R/g\Omega)\sim-2$ and
$\theta\sim0^\circ$. Also for these masers, the peaks become more
evident as the magnetic field strength increases. The polarization
angle shows a flip of $90^\circ$ around the Van Vleck
angle. Generally, $\psi$ presents a region of stability in a range
$0^{\circ}<\theta<54^{\circ}$ around $-2<\mathrm{log}(R /
g\Omega)<0$. This region in the 25 GHz maser is limited only to the
area around $\mathrm{log}(R/g\Omega)\sim -2$.

Also for these transitions we observe high levels of $P_L$ and $P_V$
due to non-Zeeman effects.  When the maser is not saturated, P$_V$
decreases for increasing $\theta$, as predicted by
\citet{WatsonWyld2001}, while for saturated masers the curve do not
follow the cosines. We show this effect in Fig.~\ref{fig:msersPV_TB}
for the baselines of the 6.2, 6.7, 25 and 45 GHz
masers. In the case of the 6.2 GHz maser we report the values until
$R/g\Omega=10$, and we see that all $T_B$ are following the cosine
curve and non-Zeeman effects are not present. On the contrary, for the
three other masers, we can see that the $P_V$ curves for high
brightness temperatures and high levels of saturation do not decrease
with increasing $\theta$, instead they peak and show non-Zeeman
effects that can lead to an overestimate of the magnetic field.

We identify also in these masers an inversion of the linear
polarization angle close to the Van Vleck angle, which is sharp when
the masers start to saturate and becomes smoother with increasing
brightness temperatures.

Looking at the circular polarization fraction of 25, 36, 44 and 45 GHz
they appear stable also for $T_B\Delta\Omega$ above
$R/g\Omega=10$. As an example, in the case of the 25 GHz maser, we
obtain at $R/g\Omega=10$ a $T_B\Delta\Omega\sim10^9$K sr for B=10 mG. Here
the $P_L$ starts to increase while $P_V$ keeps its stability until
$T_B\Delta\Omega\sim10^{10}$ K sr . Therefore, it might be that for $P_V$,
non-Zeeman effects become dominant only for $T_B$ higher than those
where $P_L$ is sensible. However, from Fig~\ref{fig:msersPV_TB}, the 25
GHz maser already presents at $T_B\Delta\Omega\sim10^{10}$ K sr a curve that
slightly deviates from the cosine, indicating that some non-Zeeman
effects are already acting. Also for the other masers we observe
similar deviations.

\subsection{6.2 GHz methanol maser}
\label{sec:6.2splitting}
Of the masers studied here, the 6.2 GHz maser recently discovered by
\citet{Breen2019} presents the largest hyperfine split of $\sim1$
km~s$^{-1}$. The two peaks appear quite similar in intensity. In order
to investigate the two hyperfine components, we ran the models for
thermal line widths $v_{th}=1$, 1.25 and 1.50 km~s$^{-1}$ and we give
the results of the simulations in Fig.~\ref{Fig:spectra6.2}. We see
that for $v_{th}=$1 km~s$^{-1}$ the two components are separated and
they start to blend for $v_{th}=$1.25 km~s$^{-1}$. At $v_{th}=$1.5
km~s$^{-1}$ the two hyperfines appear totally blended in one single
line.

This case illustrates why it is complicated to detect the different
hyperfine components of methanol masers at other frequencies. To
observe the hyperfine components it is necessary to have an
intrinsic thermal line width that is less that the hyperfine split
($v_{th}<$1.25 km~s$^{-1}$ in this case), otherwise we can only detect one single
blended component.   However, when the split is large enough with
respect to the intrinsic thermal line width, our simulations predict
that multiple components can be observed.

\section{Discussion}
\label{sec:discussion}

\subsection{Comparison between models and observations}
\label{sec:modelVSObs}

Within the limit $g\Omega=10R$, our model predicts linear and
circular polarization fractions that can be considered upper limits,
and any polarization above them will likely include some non-Zeeman
contributions.  We compared $P_L$ and $P_V$ observed in methanol
masers (Table~\ref{tab:detection}) with the results of our simulations
(Tables~\ref{tab:maxPL} and ~\ref{tab:maxPV} for $B=10$ mG).  In general
we noticed that linear polarization fraction predictions are mostly
lower than the observed values, even for magnetic fields higher than 10
mG, while in the case of circular polarization fractions we often
observe predictions higher than the observed values. In the following
part we will discuss each transition, comparing observations and
models and taking into account brightness temperatures, magnetic fields
and the presence of non-Zeeman effects.

\subsubsection{6.7 GHz methanol masers}
\label{sec:mv6.7}

We compared the results of our simulations with 6.7 GHz maser
observations from \citet{Breen2019}, \citet{Surcis2019} and
\citet{Surcis2009} (these masers are reported in Table
~\ref{tab:detection}).  These works reported high $P_L$ of the order of
7.5\%, 8.1\% and 4.5\% respectively. We selected some extreme
measurements of $P_L$ from the works of Surcis et al.\ and 6.7 GHz
methanol masers typically present lower values ranging from
1\% to 4\% \citep[][and references therein]{Surcis2019}.

We now discuss the possible action of non-Zeeman effects on
  the three selected 6.7 GHz methanol masers and reported in Table
~\ref{tab:detection}. The observed brightness
temperatures range between $10^{11}$--$10^{13}$ K and, assuming
$\Delta\Omega\sim10^{-2}$ sr, the $T_B\Delta\Omega$ obtained from
observations vary between $10^{9}$--$10^{11}$ K~sr.  Our 6.7 GHz
methanol maser model gives $T_B\Delta\Omega\sim10^{10}$ K~sr at
$g\Omega=10R$, for a magnetic field of 10 mG, but when the magnetic
field strength reaches $B=100$ mG, the limit increases by one order of
magnitude becoming $T_B\Delta\Omega\sim10^{11}$ K~sr.  This indicates
that the selected masers are not affected by the rotation of the
symmetry axis, but might be influenced by other non-Zeeman
contributions (e.g.\ a rotating magnetic field along the maser
direction) that can occur even at modest T$_B$.  Thus, if we consider
$g\Omega=10R$ as an upper limit, we can estimate that the maximum
brightness temperature reachable before being severely affected by
non-Zeeman effects is $\sim10^{12}$ K (for $B=10$ mG) or $\sim10^{13}$
K (for $B=100$ mG).

For the 6.7 GHz maser transition observed
by \citet{Breen2019}, we do not have information regarding the
magnetic field strength, but $T_B$ inferred from
observations is within the limit $g\Omega=10R$ for both $B=10$ mG and $B=100$ mG.
According to our model, such a high level of $P_L$ can be generated by
a high $B$ or by a specific hyperfine preferred pumping or by a
combination of both. Indeed for $B=100$ mG, it is possible to reach a
$P_L\sim 10\%$ (see Fig.~\ref{fig:PQPV_TB_altriB}, e.g.\ with the
transitions $4\rightarrow5$A and B). In addition, under the action of
different preferred pumping, from 3 to 100 times over the baseline
level, the amount of linear and circular polarization fraction
increases and can reach, depending on the angle, up to 7.5\%, even in
$B=10$ mG (see Fig.~\ref{Fig:trans43} and Fig.~\ref{Fig:trans54}).
\citet{Breen2019} also reported a $P_V=0.5$\% that is within the
values reported by our model for a $B=10$ mG
(Table~\ref{tab:maxPV}). However $P_V=0.5$\% is also consistent with
the results of our model considering a $B=100$ mG and an hyperfine
preferred pumping on the transition $4\rightarrow5$A.  In any case,
given the high $P_L$ observed, we can not exclude the influence of
some non-Zeeman effects. Since $T_B$ inferred from observations is
within the $g\Omega=10R$ limit, we can rule out a rotation of the
symmetry axis, but it could be possible to have a contribution in
$P_V$ of $\sim0.14\%$ due to magnetic field changes along the maser
path (as described in Sect.~\ref{Sec:nonZ2}).

\citet{Surcis2019} reported $B>182$ mG, $P_L=8.1\%$ and $P_V=2.1\%$ for
the maser feature WH3(OH).22 and the $T_B$ inferred from their
observations is $\sim10^{13}$ K. These values are in agreement with
our models for a magnetic field $B=100$ mG and within the $g\Omega=10R$
limit.
However, since $P_L$ is high, we cannot exclude minor non-Zeeman
contributions affecting $P_V$, that could count $\sim$0.16\% due to
magnetic field changes along the maser path (Sect.~\ref{Sec:nonZ2}).

In the case of W75N, \citet{Surcis2009} estimated a
$T_B\Delta\Omega<10^{9}$ K~sr and, given $\Delta\Omega\sim10^{-1}$ sr and a
maser angular size of 7 mas, they obtained $\theta=70^{\circ}$ and
B=50 mG.  They also registered a $P_L=4.5$\% and a $P_V$=0.2\%. 
As for W3(OH), this value of $P_L$ and $P_V$ can be generated under the action
of different degrees of hyperfine preferred pumping and $B=10$--100 mG.
If we exclude a different pointing direction of the magnetic field,
hyperfine preferred pumping could be probable in the case of W75N: indeed, if
we compare the S-shape profile of the observed V spectrum with the
synthetic one, we see that the emission detected by \citet{Surcis2009}
is compatible with a preferred pumping on hyperfine transitions
$7\rightarrow8$A, or $6\rightarrow7$A.  It is notable that these are
not the same hyperfines of Tab.~\ref{tab:maxPL} and
Tab.~\ref{tab:maxPV}.

\subsubsection{25 GHz methanol masers}
\label{sec:mv25}
\citet{Sarma2020} observed 25 GHz methanol masers in two different
epochs in the OMC--1 region. They reported for the two detections a
level of $P_V\sim0.3$\% and $P_V\sim0.4$\%, and, based on the low
$T_B\Delta\Omega$, conclude that non-Zeeman effect likely contributes
little to this percentage. From our models in case of B=100 mG and for
$T_B\Delta\Omega\sim10^{6}$ K sr, we confirm that non-Zeeman effects
are neglegible and we found a $P_V\sim1$\%, for the strongest
hyperfine transitions, also considered by \citet{Sarma2020}.  While
this value is still higher than the measured values, we note that our
simulations were run for $v_{th}=1$ km~s$^{-1}$ . At
$T_B\Delta\Omega\sim10^{6}$ K sr, our spectra have a FWHM line-width
of slightly less than half that of the observed spectrum (0.24
km~s$^{-1}$ vs. 0.53 km~s$^{-1}$). Since in the normal Zeeman
interpretation, the fractional polarization is inversely proportional
to the line-width this indicates that the observed polarization in
OMC-1 can indeed be the result of a magnetic field of $\sim$100 mG.

From our models in case of B=30 mG we derive at the $g\Omega=10R$
limit a value of $T_B\Delta\Omega\sim10^9$ K sr and $P_V\sim$0.2\% for
the same hyperfine transitions considered by \citet{Sarma2020}. While
this indicates that a level of circular polarization can be reached
that is similar to the observed level for a smaller magnetic field
strength, this only occurs for values of $T_B\Delta\Omega$ that are
more than three orders of magnitude larger than estimated from the
observations. In that case, part of $P_V$ originates from non-Zeeman
effects, which do not contribute at the lower observed
$T_B\Delta\Omega$. Alternatively, if the maser spot size is
overestimated, and thus $T_B$ underestimated, and/or $\Delta\Omega$ is
significantly higher, non-Zeeman effects might after all be present in
the observed signal. However, currently, there are no observational
indications that either is the case, leading us to conclude that the
normal Zeeman analysis performed in \citet{Sarma2020} can be used and
that the 25 GHz methanol masers trace a shock enhanced magnetic field.

\subsubsection{36 GHz methanol masers}
\label{sec:mv36}
\citet{Sarma2009} detected 36 GHz methanol masers across the M8E region
with a $P_V\sim0.1$\% and estimated a magnetic field
$B_{LOS}\sim20$--30 mG. Also \citet{Breen2019} reported a very low
level in both linear and circular polarization fractions, lower than
0.5\%. From our simulations, we confirm a low $P_V$
for this maser and we found an agreement with the observations, since
the $3\rightarrow2$~A hyperfine can produce a $P_V\sim0.1$\% for a magnetic
field $B_{LOS}\sim20$--30 mG. In addition, comparing the $V$ spectrum
observed by \citet{Sarma2009} with the one produced by our models, we
find that also in this case preferred pumping might have acted on
the hyperfine transition $3\rightarrow2$~A; this transition indeed
generates an evident S-shape profile --- one that appears much less prominent
in the baseline case.

For this maser at the $g\Omega=10R$, we obtain a
$T_B\Delta\Omega\sim10^9$ K sr, and assuming the same
$\Delta\Omega\sim10^{-2}$ sr, this leads to a limit in brightness
temperature of $\sim10^{11}$ K, which is above the range constrained
from observations. Given the low level of $P_L$ observed by
\citet{Breen2019}, we exclude a significant non-Zeeman contribution
for this maser emission.

\subsubsection{44 GHz methanol masers}
\label{sec:mv44}
\citet{Momjian2019} observed 44.1 GHz methanol masers in the DR21W
star forming region, detecting a $P_V\sim0.05$\%. They considered the
hyperfine transition $5\rightarrow4$~A and constrained a lower limit
for the magnetic field $B_{LOS}\sim25$ mG. In our model we also obtain
$P_V\sim0.056$\% when this hyperfine transition is preferably pumped
and for a magnetic field of 10 mG. For a $B=30$ mG the resulting
$P_V\sim0.1$\%. In both the case, $B$ is in the same order of magnitude
of the one suggested by \citet{Momjian2019}. However the observed
spectra show the presence of multiple components and therefore we
cannot totally exclude other combinations of hyperfines producing this
emission. Even so, the synthetic $V$ spectra of the $5\rightarrow4$~A
present a S-shaped profile that is in agreement with the observed
one.  \citet{Breen2019} observed $P_L$ and $P_V$ less than 0.5\%, but
from our simulations the predicted maximum $P_L\sim3\%$. Given the low
level of $P_L$ observed by \citet{Breen2019}, we can rule out any
contribution due to magnetic field changes along the maser path. By comparing $T_B$ from
observations and model, we can exclude also a rotation of the
symmetry axis. From our model $T_B\Delta\Omega\sim10^9$ K sr at
$g\Omega=10R$ limit for $B=10$ mG. Therefore if
$\Delta\Omega\sim10^{-2}$ sr, the maximum brightness temperature
reachable to exclude the rotation of the symmetric axis is
$\sim10^{11}$ K which is higher than the observed ones.

\subsubsection{45 GHz methanol masers}
\label{sec:mv45}
From our simulations we obtain $T_B\Delta\Omega\sim10^9$ K sr at the $g\Omega=10R$ limit
for $B=10$ mG; thus we estimate a maximum brightness temperature reachable
to exclude a rotation of the symmetric axis of $T_B\sim10^{11}$ K, which
is several orders of magnitude higher than that observed by
\citet{Breen2019}.  We also note that the level of $P_L$ and $P_V$
predicted by our model for $B=10$ mG are higher than the one detected
by \citet{Breen2019}.
\subsubsection{Other methanol masers}
\label{sec:mv6.7}
Other methanol masers transitions presenting high level of $P_L$ were
detected at 6.2, 7.7, 7.8, 21, 37.7, 46 GHz and at 20.3 GHz by
\citet{Breen2019} and \citet{MacLeod2019} respectively. We could not
model non-Zeeman effects because the simulations for these transitions
were performed with the N\&W92 code, limited to $g\Omega=10R$. However
looking at the $T_B\Delta\Omega$ inferred from our models and
comparing them with the ones estimated from observations, for all
these masers we can exclude a rotation of the symmetry axis.

For the 6.2 GHz maser, the observed $P_L\sim7$\% has not been predicted by
our simulations, while, the predicted $P_V$ is instead much higher
than the one detected by \citet{Breen2019}. In this case we cannot
exclude the contribution of non-Zeeman effects enhancing linear
polarization. Looking at 7.7 GHz and 7.8 GHz masers, the detected
$P_L$ and $P_V$ are lower than the predicted ones. For the 21 GHz, the
observed $P_L$ is $\sim7$\% and our models show $P_L\sim6$\% in case
of $B=30$ mG and for the hyperfine transitions 8$\rightarrow$9~A pumped
10 times more than the baseline. 37.7 GHz observations shows high
$P_L$ values which are almost in agreement with our model for a $B=10$
mG, while 46 GHz presents $P_L$ higher than the predicted one. We can
not exclude the action of non-Zeeman effects or a combination of
hyperfine preferred pumping at higher $B$ strength.

Since these masers are likely arising from the same region of the
protostar \citep{Breen2019, MacLeod2019}, one possible explanation for
these high observed levels of $P_L$ could be due to the action of a
magnetic field $\gtrsim 100$ mG or to
specific hyperfine pumping or to the synergy of both. More
observations will probably help in understanding which process in
ongoing in this region, by measuring $P_V$ and by inferring the $B$ strength.

\subsection{Reversed profile in circular polarization spectra}
\label{sec:reversed}
\citet{Daria2017_IRAS18089} observed a reversed profile in 6.7 GHz
methanol maser I and V spectra detected between two different epochs
in the high mass protostar IRAS\,18089-1732. In these observations the
V spectra presented two S-shape profiles, one being the opposite of
the other. One of the possible explanations given by these authors to
interpret the reversed I profiles and opposite circular polarization
was the presence of two hyperfine components of the 6.7~GHz methanol
transition $5_{1}\rightarrow~6_{0}$. Under this hypothesis, one
hyperfine transition would be preferred over the other in one epoch,
and vice versa in the following epoch. Our simulations shows that this
possibility can occur if the two components of the spectra are given
by the hyperfine transitions $3\rightarrow~4$~A and
$7\rightarrow~8$~A, as shown in Figs.~\ref{Fig:spectra6.7_baseline},
\ref{Fig:spectra6.7_1}, and \ref{Fig:spectra6.7_8}. It can also occur
when the two hyperfine transitions $6\rightarrow~7$~A and B are
simultaneously preferentially pumped.
However, since the same effect
could be also due either to a flip in the magnetic field strength
intrinsic to the source \citep{Vlemmings2009} or to different and
blended masers, originating in two places lying along the same line of
sight (e.g. \citealt{Momjian2017}), further methanol masers
observations are needed to help in understanding the magnetic field
action and its link on preferred hyperfine pumping.

\subsection{6.2 GHz methanol maser and its hyperfine splitting}
\label{sec:6.2hyperfine}
As reported in Sect.~\ref{sec:6.2splitting}, our simulations show that
it is possible to observe the hyperfine splitting of two components
for the 6.2 GHz methanol maser, when the thermal line width
$v_{th}\lesssim 1.25$ km~s$^{-1}$. Therefore, we showed that a splitting
significantly larger than the intrinsic thermal line width is required
to actually observe the split. In addition, the first detection of this
maser was performed by \citet{Breen2019}, but from their single dish
observations it is impossible to discern if the features in the
spectra are clear signs of hyperfine splitting or not.  Further high
angular resolution observations might help to observe the splitting
between these two hyperfine components.

\section{Conclusions}
\label{sec:conclusions}
Masers can be considered useful tools to infer magnetic field
properties in star forming regions, and by observing linear and
circular polarization we can study the magnetic field morphology and
strength in different parts of the protostar.

We ran simulations using the radiative transfer CHAMP code, for several
magnetic field strengths, hyperfine components, and pumping
efficiencies. We  explored the polarization properties of some observed methanol
maser transitions, considering newly calculated methanol Land{\'e}
factors and preferred hyperfine pumping.  We also investigated the
action of non-Zeeman effects, that can contaminate magnetic field estimates.

We noticed a dependence of the linear polarization fraction on the
magnetic field strength and on the hyperfine transitions. Also the
circular polarization fraction presents a dependence on the hyperfine
transitions. We found that distinct hyperfine components react to the
magnetic field differently. Thus, in case of preferred hyperfine
pumping, high levels of linear and circular polarization can be
produced, explaining some of the high $P_L$ and $P_V$ observed. We
discussed some of the peculiar features seen in the S-shape of the
observed V-profiles. Comparing $T_B$ obtained from our models with the
observations, we argue that methanol masers do not appear affected by
a rotation of the symmetry axis.  However other non-Zeeman effects
might arise also at modest $T_B$ and they need to be considered in the
study of magnetic fields.

A possible way to constrain these effects is represented by
  observing polarized emission from other maser transitions expected
  to arise from the same region. Observations of only $P_L$ would be
  used to determine the direction of the magnetic field that can be
  parallel or perpendicular to the linear polarization vector;
  however, the observed $P_L$ values will be consistent with several
  models of preferred hyperfine pumping and it will be impossible to
  discern which hyperfine preferred pumping might be occurring. $P_L$
  will allow to achieve estimate on saturation level and the possible
  effect of magnetic field rotation on measured $P_V$. Single
  observations of masers presenting only $P_V$, can be used to derive
  lower limits of magnetic field strength and can be helpful to
  determine and discuss the most probable hyperfine transitions
  acting. By simultaneous observations of $P_L$ and $P_V$, if possible also in
  different transitions, one can confidently rule out the presence of
  non-Zeeman contributions or estimate their actual action. Also a
  more accurate knowledge of the maser beaming angle $\Delta\Omega$
  and thus more precise measurement of the brightness temperature
  $T_B$ will help to understand the amount of non-Zeeman contributions
  and on which hyperfine transitions they are more relevant.

We showed an hyperfine splitting occurring
between two components of the 6.2 GHz methanol maser, when
$v_{th}<1.25$ km~s$^{-1}$. We have seen that a splitting significantly
larger than the intrinsic thermal line width is required to detect the
splitting between the hyperfine components.

Therefore further high
angular resolution observations of methanol masers are necessary to
understand how hyperfine preferred pumping works. The comparison
between these observational details will be fundamental to better
understand the simultaneous action of magnetic fields and preferred
pumping.

\begin{acknowledgements}

  We thank the referee for the insightful comments and accurate
  suggestions that have contributed to the improvement of this
  work. Simulations were performed on resources at the Chalmers Centre
  for Computational Science and Engineering (C3SE) provided by the
  Swedish National Infrastructure for Computing (SNIC).

\end{acknowledgements}

\begin{appendix}
  \section{Appendix}
  \label{sec:appendix}

% 6.2 GHz
\begin{table}[t]
\centering
\caption{Characteristics of hyperfine transitions for 6.2 GHz methanol maser}
\begin{tabular}{l c c c }
  \hline
  Hyperfine    & Hyperfine   & g-factor       & A$_{i-j}$ \\
  transition    & shift (Hz) &(s$^{-1}$mG$^{-1}$)&  s$^{-1}$ \\
\hline
$16\rightarrow17$ A &-11180  &-0.8094  &$9.247\times 10^{-10}$  \\
$17\rightarrow18$ A &11720  &-0.0335  &$4.624\times 10^{-10}$  \\
$17\rightarrow18$ B &-12410  &0.0048  &$4.624\times 10^{-10}$  \\
$18\rightarrow19$ A &10830  &0.6945  &$9.226\times 10^{-10}$  \\
$16\rightarrow17$ B &-11730  &-0.8094  &$9.247\times 10^{-10}$  \\
$17\rightarrow18$ C &12240  &-0.0335  &$4.624\times 10^{-10}$  \\
$17\rightarrow18$ D &-12810  &0.0048  &$4.634\times 10^{-10}$  \\
$18\rightarrow19$ B &11210  &0.6945  &$9.226\times 10^{-10}$  \\
\hline
\end{tabular}
\label{tab:meth6.2}
\end{table}

% 6.7GHz
\begin{table}[t]
\centering
\caption{Characteristics of hyperfine transitions for 6.7 GHz methanol maser}
\begin{tabular}{l c c c }
  \hline
  Hyperfine    & Hyperfine   & g-factor       & A$_{i-j}$ \\
  transition    & shift (Hz) &(s$^{-1}$mG$^{-1}$)&  s$^{-1}$ \\
\hline
$7\rightarrow8$ A &2500  &2.9712  &$1.03\times 10^{-9}$  \\
$6\rightarrow7$ A &-4397  &1.6416  &$1.04\times 10^{-9}$  \\
$6\rightarrow7$ B &3541  &1.4112  &$1.02\times 10^{-9}$  \\
$5\rightarrow6$ A &-2889  &-0.8016  &$1.01\times 10^{-9}$  \\
$5\rightarrow6$ B &3015  &0.0096  &$1.00\times 10^{-9}$  \\
$4\rightarrow5$ A &1240  &-2.9376  &$1.06\times 10^{-9}$  \\
$4\rightarrow5$ B &-2835  &-3.2496  &$1.03\times 10^{-9}$  \\
$3\rightarrow4$ A &-4417  &-7.1472  &$1.08\times 10^{-9}$  \\
\hline
\end{tabular}
\label{tab:meth6.7}
\end{table}

% 7.7 GHz
\begin{table}[t]
\centering
\caption{Characteristics of hyperfine transitions for 7.7 GHz methanol maser}
\begin{tabular}{l c c c }
  \hline
  Hyperfine    & Hyperfine   & g-factor       & A$_{i-j}$ \\
  transition    & shift (Hz) &(s$^{-1}$mG$^{-1}$)&  s$^{-1}$ \\
\hline
$10\rightarrow11$ A &10970  &-2.9311  &$7.063\times 10^{-10}$  \\
$11\rightarrow12$ A &-3300  &-1.5661  &$4.395\times 10^{-10}$  \\
$11\rightarrow12$ B &7710  &-1.6428  &$4.355\times 10^{-10}$  \\
$12\rightarrow13$ A &-7390  &-0.4215  &$3.483\times 10^{-10}$  \\
$12\rightarrow13$ B &6680  &-0.5460  &$3.483\times 10^{-10}$  \\
$13\rightarrow14$ A &-9370  &0.5460  &$4.305\times 10^{-10}$  \\
$13\rightarrow14$ B &8230  &0.4071  &$4.345\times 10^{-10}$  \\
$14\rightarrow15$ A &-8830  &1.3075  &$6.982\times 10^{-10}$  \\
\hline
\end{tabular}
\label{tab:meth7.7}
\end{table}

% 7.8 GHz
\begin{table}[t]
\centering
\caption{Characteristics of hyperfine transitions for 7.8 GHz methanol maser}
\begin{tabular}{l c c c }
  \hline
  Hyperfine    & Hyperfine   & g-factor       & A$_{i-j}$ \\
  transition    & shift (Hz) &(s$^{-1}$mG$^{-1}$)&  s$^{-1}$ \\
\hline
$10\rightarrow11$ A &10870  &-2.9263  &$7.464\times 10^{-10}$  \\
$11\rightarrow12$ A &-3230  &-1.5613  &$4.656\times 10^{-10}$  \\
$11\rightarrow12$ B &7620  &-1.6380  &$4.613\times 10^{-10}$  \\
$12\rightarrow13$ A &-7320  &-0.4167  &$3.690\times 10^{-10}$  \\
$12\rightarrow13$ B &6600  &-0.5412  &$3.690\times 10^{-10}$  \\
$13\rightarrow14$ A &-9290  &0.5508  &$4.560\times 10^{-10}$  \\
$13\rightarrow14$ B &8160  &0.4119  &$4.603\times 10^{-10}$  \\
$14\rightarrow15$ A &-8750  &1.3123  &$7.396\times 10^{-10}$  \\
\hline
\end{tabular}
\label{tab:meth7.8}
\end{table}

% 12 GHz
\begin{table}[t]
\centering
\caption{Characteristics of hyperfine transitions for 12 GHz methanol maser}
\begin{tabular}{l c c c }
  \hline
  Hyperfine    & Hyperfine   & g-factor       & A$_{i-j}$ \\
  transition    & shift (Hz) &(s$^{-1}$mG$^{-1}$)&  s$^{-1}$ \\
\hline
$3\rightarrow4$ A &-4965  &3.4180  &$0.737\times 10^{-8}$  \\
$2\rightarrow3$ A &-4802  &0.6911  &$0.752\times 10^{-8}$  \\
$2\rightarrow3$ B &-1475  &-0.5592  &$0.732\times 10^{-8}$  \\
$1\rightarrow2$ A &7978  &-6.8670  &$0.803\times 10^{-8}$  \\
$3\rightarrow4$ B &-6106  &2.9400  &$0.737\times 10^{-8}$  \\
$2\rightarrow3$ C &5847  &1.8470  &$0.610\times 10^{-8}$  \\
$2\rightarrow3$ D &-1475  &-1.0430  &$0.571\times 10^{-8}$  \\
$1\rightarrow2$ B &-3263  &-6.6920  &$0.803\times 10^{-8}$  \\
\hline
\end{tabular}
\label{tab:meth12}
\end{table}

% 20.3 GHz
\begin{table}[t]
\centering
\caption{Characteristics of hyperfine transitions for 20.3 GHz methanol maser}
\begin{tabular}{l c c c }
  \hline
  Hyperfine    & Hyperfine   & g-factor       & A$_{i-j}$ \\
  transition    & shift (Hz) &(s$^{-1}$mG$^{-1}$)&  s$^{-1}$ \\
\hline
$16\rightarrow17$ &10020  &-1.2884  &$1.197\times 10^{-8}$  \\
$17\rightarrow18$ &-8010  &-0.4933  &$5.984\times 10^{-9}$  \\
$17\rightarrow18$ &9100  &-0.5029  &$5.984\times 10^{-9}$  \\
$18\rightarrow19$ &-9710  &0.2059  &$1.194\times 10^{-8}$  \\
$16\rightarrow17$ &10020  &-1.2884  &$1.197\times 10^{-8}$  \\
$17\rightarrow18$ &-8010  &-0.4933  &$5.984\times 10^{-9}$  \\
$17\rightarrow18$ &9100  &-0.5029  &$5.984\times 10^{-9}$  \\
$18\rightarrow19$ &-9710  &0.2059  &$1.194\times 10^{-8}$  \\
\hline
\end{tabular}
\label{tab:meth20.3}
\end{table}

% 21 GHz
\begin{table}[t]
\centering
\caption{Characteristics of hyperfine transitions for 21 GHz methanol maser}
\begin{tabular}{l c c c }
  \hline
  Hyperfine    & Hyperfine   & g-factor       & A$_{i-j}$ \\
  transition    & shift (Hz) &(s$^{-1}$mG$^{-1}$)&  s$^{-1}$ \\
\hline
$8\rightarrow9$ A &-21710  &-2.5719  &$3.373\times 10^{-8}$  \\
$9\rightarrow10$ A &-1730  &-1.0297  &$2.018\times 10^{-8}$  \\
$9\rightarrow10$ B &-13560  &-0.8238  &$1.991\times 10^{-8}$  \\
$10\rightarrow11$ A &6920  &0.1581  &$1.660\times 10^{-8}$  \\
$10\rightarrow11$ B &-7210  &0.6801  &$1.663\times 10^{-8}$  \\
$11\rightarrow12$ A &5150  &2.3803  &$1.445\times 10^{-8}$  \\
$11\rightarrow12$ B &4990  &0.7040  &$1.449\times 10^{-8}$  \\
$12\rightarrow13$ A &18650  &2.4953  &$3.327\times 10^{-8}$  \\
\hline
\end{tabular}
\label{tab:meth21}
\end{table}

% 25 GHz
\begin{table}[t]
\centering
\caption{Characteristics of hyperfine transitions for 25 GHz methanol maser}
\begin{tabular}{l c c c }
  \hline
  Hyperfine    & Hyperfine   & g-factor       & A$_{i-j}$ \\
  transition    & shift (Hz) &(s$^{-1}$mG$^{-1}$)&  s$^{-1}$ \\
\hline
$6\rightarrow6$ A &-26  &4.4480  &$5.570\times 10^{-8}$  \\
$5\rightarrow5$ A &4223  &0.2830  &$5.538\times 10^{-8}$  \\
$5\rightarrow5$ B &-1070  &0.3896  &$5.533\times 10^{-8}$  \\
$4\rightarrow4$ A &1462  &-5.4850  &$5.500\times 10^{-8}$  \\
$6\rightarrow6$ B &3754  &4.5990  &$5.570\times 10^{-8}$  \\
$5\rightarrow5$ C &-271  &0.1131  &$5.537\times 10^{-8}$  \\
$5\rightarrow5$ D &-1070  &0.3393  &$5.532\times 10^{-8}$  \\
$4\rightarrow4$ B &-3784  &-5.4290  &$5.500\times 10^{-8}$  \\
\hline
\end{tabular}
\label{tab:meth25}
\end{table}

% 36 GHz
\begin{table}[t]
\centering
\caption{Characteristics of hyperfine transitions for 36 GHz methanol maser}
\begin{tabular}{l c c c }
  \hline
  Hyperfine    & Hyperfine   & g-factor       & A$_{i-j}$ \\
  transition    & shift (Hz) &(s$^{-1}$mG$^{-1}$)&  s$^{-1}$ \\
\hline
$5\rightarrow4$ A &5402.  &3.054  &$15.238\times 10^{-8}$  \\
$4\rightarrow3$ A &-4791.  &1.093  &$13.009\times 10^{-8}$  \\
$4\rightarrow3$ B &-3395.  &-1.722  &$12.615\times 10^{-8}$  \\
$3\rightarrow2$ A &-3032.  &-4.524  &$13.994\times 10^{-8}$  \\
$5\rightarrow4$ B &4042.  &2.664  &$15.238\times 10^{-8}$  \\
$4\rightarrow3$ C &4303.  &0.3519  &$14.595\times 10^{-8}$  \\
$4\rightarrow3$ D &-3395.  &-0.4712  &$14.469\times 10^{-8}$  \\
$3\rightarrow2$ B &-6802.  &-4.423  &$13.995\times 10^{-8}$  \\
\hline
\end{tabular}
\label{tab:meth36}
\end{table}

% 37.7 GHz
\begin{table}[t]
\centering
\caption{Characteristics of hyperfine transitions for 37.7 GHz methanol maser}
\begin{tabular}{l c c c }
  \hline
  Hyperfine    & Hyperfine   & g-factor       & A$_{i-j}$ \\
  transition    & shift (Hz) &(s$^{-1}$mG$^{-1}$)&  s$^{-1}$ \\
\hline
$6\rightarrow7$ A &8220  &-2.2989  &$9.638\times 10^{-8}$  \\
$7\rightarrow8$ A &-6150  &-0.1341  &$4.831\times 10^{-8}$  \\
$7\rightarrow8$ B &8110  &-0.6178  &$4.875\times 10^{-8}$  \\
$8\rightarrow9$ A &-7750  &1.1207  &$9.594\times 10^{-8}$  \\
$6\rightarrow7$ B &1550  &-2.2989  &$9.638\times 10^{-8}$  \\
$7\rightarrow8$ C &-350  &-0.3448  &$4.842\times 10^{-8}$  \\
$7\rightarrow8$ D &1090  &-0.4023  &$4.842\times 10^{-8}$  \\
$8\rightarrow9$ B &-1380  &1.1207  &$9.594\times 10^{-8}$  \\
\hline
\end{tabular}
\label{tab:meth37.7}
\end{table}

% 44 GHz
\begin{table}[t]
\centering
\caption{Characteristics of hyperfine transitions for 44 GHz methanol maser}
\begin{tabular}{l c c c }
  \hline
  Hyperfine    & Hyperfine   & g-factor       & A$_{i-j}$ \\
  transition    & shift (Hz) &(s$^{-1}$mG$^{-1}$)&  s$^{-1}$ \\
\hline
$9\rightarrow8$ A &-2444  &2.5950  &$27.449\times 10^{-8}$  \\
$8\rightarrow7$ A &3934  &1.2755  &$27.116\times 10^{-8}$  \\
$8\rightarrow7$ B &-3216  &1.3006  &$26.796\times 10^{-8}$  \\
$7\rightarrow6$ A &2897  &-0.6786  &$26.285\times 10^{-8}$  \\
$7\rightarrow6$ B &-2897  &-0.1005  &$26.078\times 10^{-8}$  \\
$6\rightarrow5$ A &-1684  &-2.5321  &$26.407\times 10^{-8}$  \\
$6\rightarrow5$ B &2863  &-2.7395  &$25.874\times 10^{-8}$  \\
$5\rightarrow4$ A &3898  &-5.7805  &$25.913\times 10^{-8}$  \\
\hline
\end{tabular}
\label{tab:meth44}
\end{table}

% 45 GHz
\begin{table}[t]
\centering
\caption{Characteristics of hyperfine transitions for 45 GHz methanol maser}
\begin{tabular}{l c c c }
  \hline
  Hyperfine    & Hyperfine   & g-factor       & A$_{i-j}$ \\
  transition    & shift (Hz) &(s$^{-1}$mG$^{-1}$)&  s$^{-1}$ \\
\hline
$1\rightarrow2$ A &-3850  &-6.7387  &$2.512\times 10^{-7}$  \\
$2\rightarrow3$ A &540  &-1.8104  &$1.352\times 10^{-7}$  \\
$2\rightarrow3$ B &-4250  &1.8008  &$1.679\times 10^{-7}$  \\
$3\rightarrow4$ A &5870  &3.3622  &$2.884\times 10^{-7}$  \\
$1\rightarrow2$ B &-5600  &-6.7387  &$2.512\times 10^{-7}$  \\
$2\rightarrow3$ C &1940  &0.1868  &$1.556\times 10^{-7}$  \\
$2\rightarrow3$ D &-2030  &-0.2012  &$1.455\times 10^{-7}$  \\
$3\rightarrow4$ B &3780  &3.3622  &$2.884\times 10^{-7}$  \\
\hline
\end{tabular}
\label{tab:meth45}
\end{table}

% 46 GHz
\begin{table}[t]
\centering
\caption{Characteristics of hyperfine transitions for 46 GHz methanol maser}
\begin{tabular}{l c c c }
  \hline
  Hyperfine    & Hyperfine   & g-factor       & A$_{i-j}$ \\
  transition    & shift (Hz) &(s$^{-1}$mG$^{-1}$)&  s$^{-1}$ \\
\hline
$8\rightarrow9$ A &7930  &-1.9828  &$1.531\times 10^{-7}$  \\
$9\rightarrow10$ A &-5650  &-0.4454  &$7.674\times 10^{-8}$  \\
$9\rightarrow10$ B &7300  &-0.5268  &$7.674\times 10^{-8}$  \\
$10\rightarrow11$ A &-7570  &0.7328  &$1.524\times 10^{-7}$  \\
$8\rightarrow9$ B &8930  &-1.9828  &$1.531\times 10^{-7}$  \\
$9\rightarrow10$ C &-6560  &-0.4454  &$7.674\times 10^{-8}$  \\
$9\rightarrow10$ D &8110  &-0.5268  &$7.674\times 10^{-8}$  \\
$10\rightarrow11$ B &-8300  &0.7328  &$1.524\times 10^{-7}$  \\
\hline
\end{tabular}
\label{tab:meth46}
\end{table}

% \begin{table}[t]
% \centering
% \caption{Characteristics of hyperfine transitions for 46 GHz methanol maser}
% \begin{tabular}{rcrl c c c }
%   \hline
%   \multicolumn{4}{l}{Hyperfine}    & Hyperfine   & g-factor       & A$_{i-j}$ \\
%     \multicolumn{4}{l}{transition}    & shift (Hz) &(s$^{-1}$mG$^{-1}$)&  s$^{-1}$ \\
% \hline
% 8&$\rightarrow$&9 &A &7930  &-1.9828  &$1.531\times 10^{-7}$  \\
% 9&$\rightarrow$&10 &A &-5650  &-0.4454  &$7.674\times 10^{-8}$  \\
% 9&$\rightarrow$&10 &B &7300  &-0.5268  &$7.674\times 10^{-8}$  \\
% 10&$\rightarrow$&11& A &-7570  &0.7328  &$1.524\times 10^{-7}$  \\
% 8&$\rightarrow$&9 &B &8930  &-1.9828  &$1.531\times 10^{-7}$  \\
% 9&$\rightarrow$&10 &C &-6560  &-0.4454  &$7.674\times 10^{-8}$  \\
% 9&$\rightarrow$&10 &D &8110  &-0.5268  &$7.674\times 10^{-8}$  \\
% 10&$\rightarrow$&11& B &-8300  &0.7328  &$1.524\times 10^{-7}$  \\
% \hline
% \end{tabular}\
% label{tab:meth46bis}
% \end{table}

\begin{figure*}
 \centering
 \includegraphics[width=\columnwidth, clip]{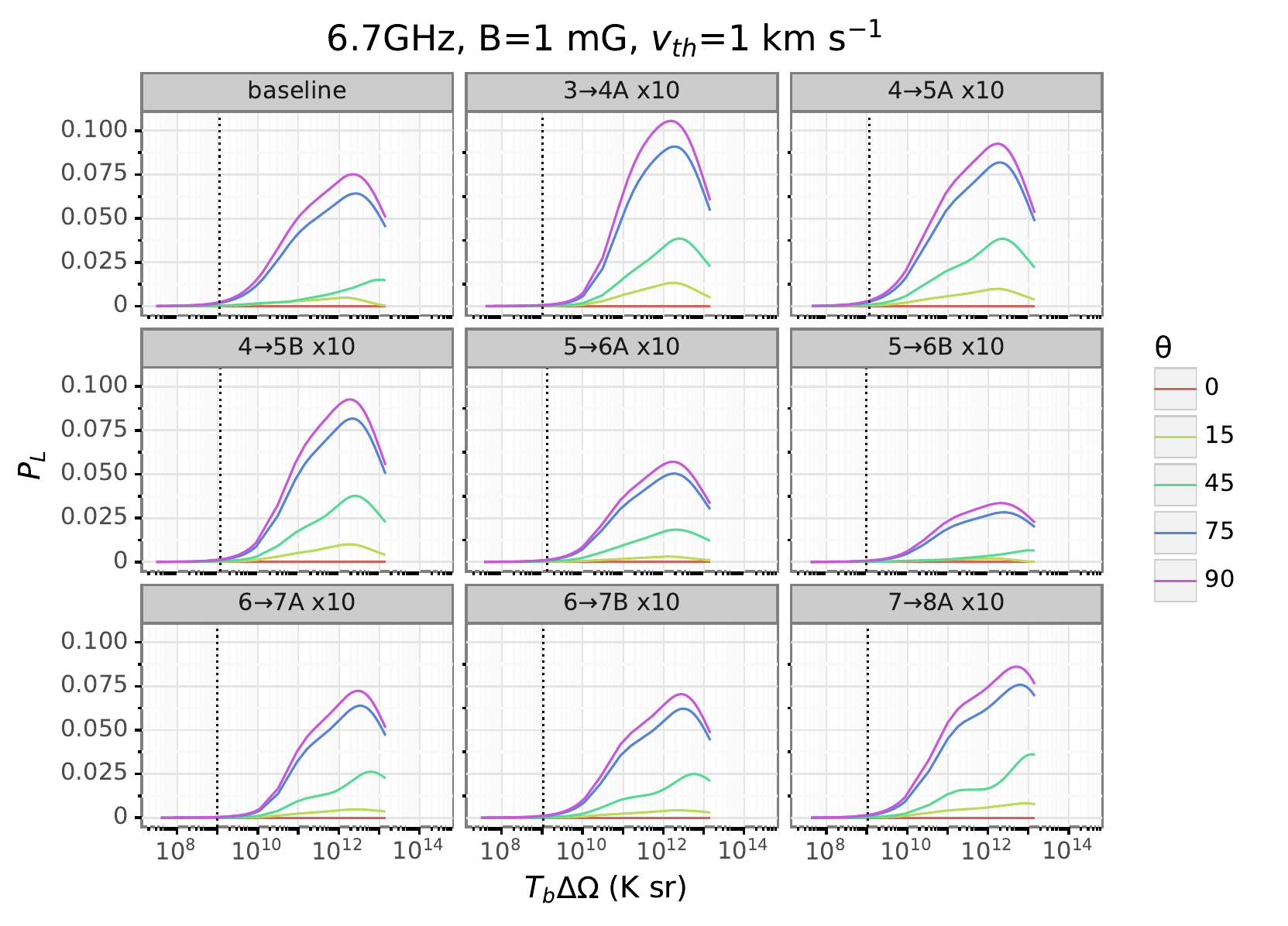}
  \includegraphics[width=\columnwidth, clip]{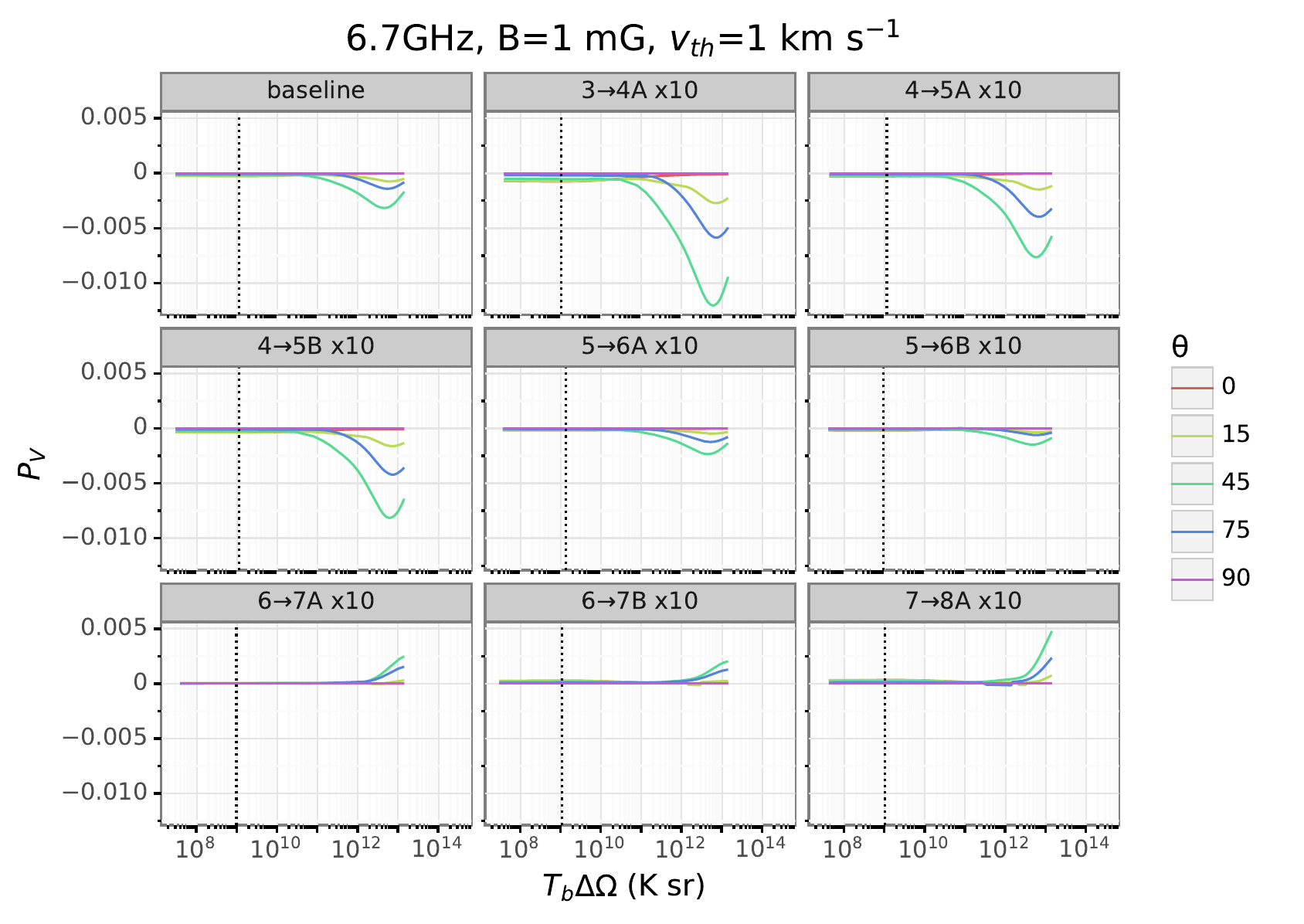}\\
  \includegraphics[width=\columnwidth, clip]{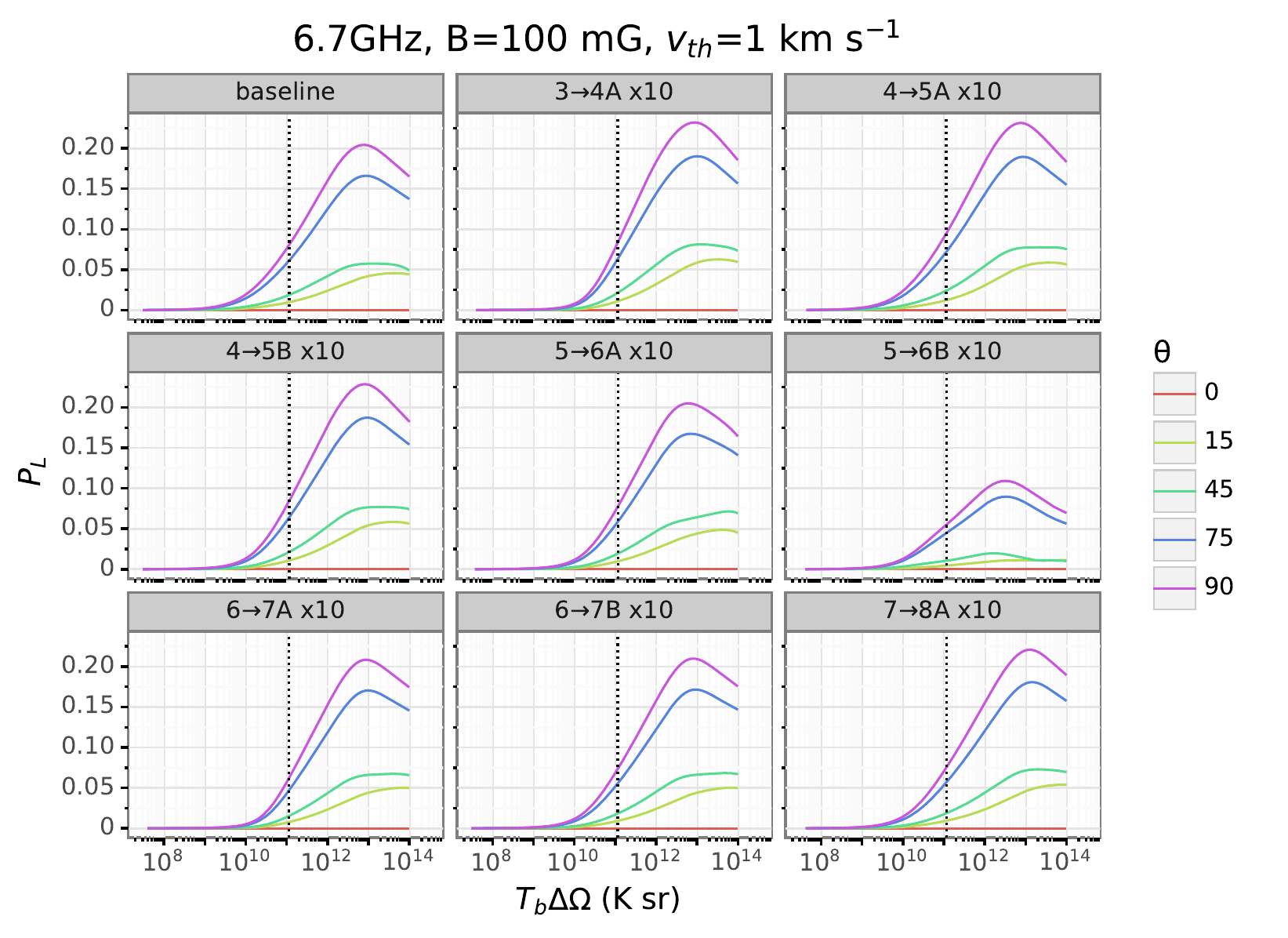}
  \includegraphics[width=\columnwidth, clip]{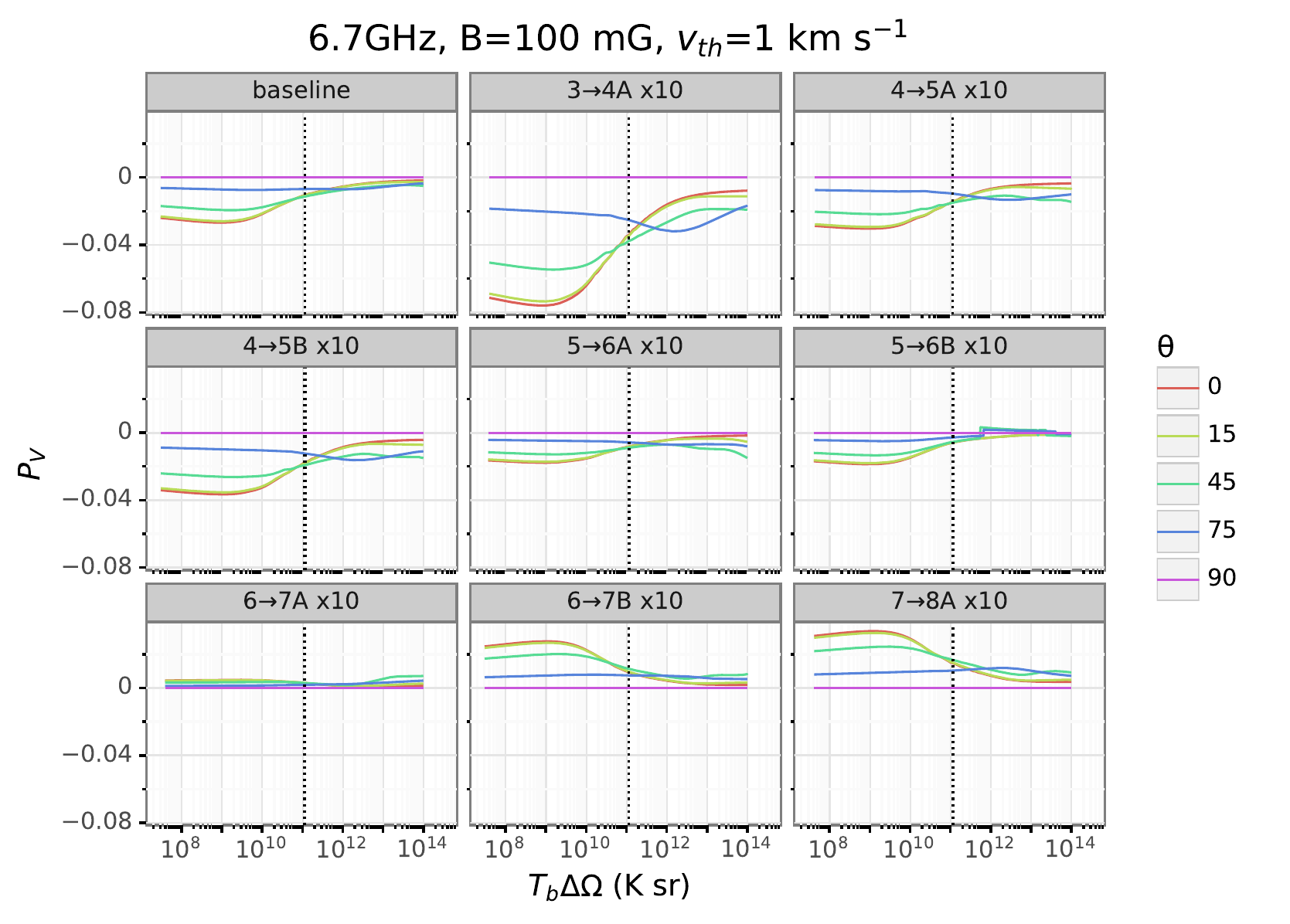}
  \caption{6.7 GHz methanol maser linear and circular
    polarization fraction as a function of the maser luminosity for
    five different $\theta$. The vertical line marks $T_B\Delta\Omega$
    where $g\Omega=10R$. Magnetic field strength is 1 and 100 mG, thermal velocity width is
   1 km s$^{-1}$ and the hyperfine transitions are indicated in each
   panel. The panel at the top left labelled ``baseline'' indicates a
   fixed pumping rate equal for all the hyperfine transitions, while
   all others assume a $10\times$ preferred pumping for the indicated
   $i\rightarrow j$ transition.}
 \label{fig:PQPV_TB_altriB}
\end{figure*}

\begin{figure*}[h!]
 \centering
 \includegraphics[width=\columnwidth, clip]{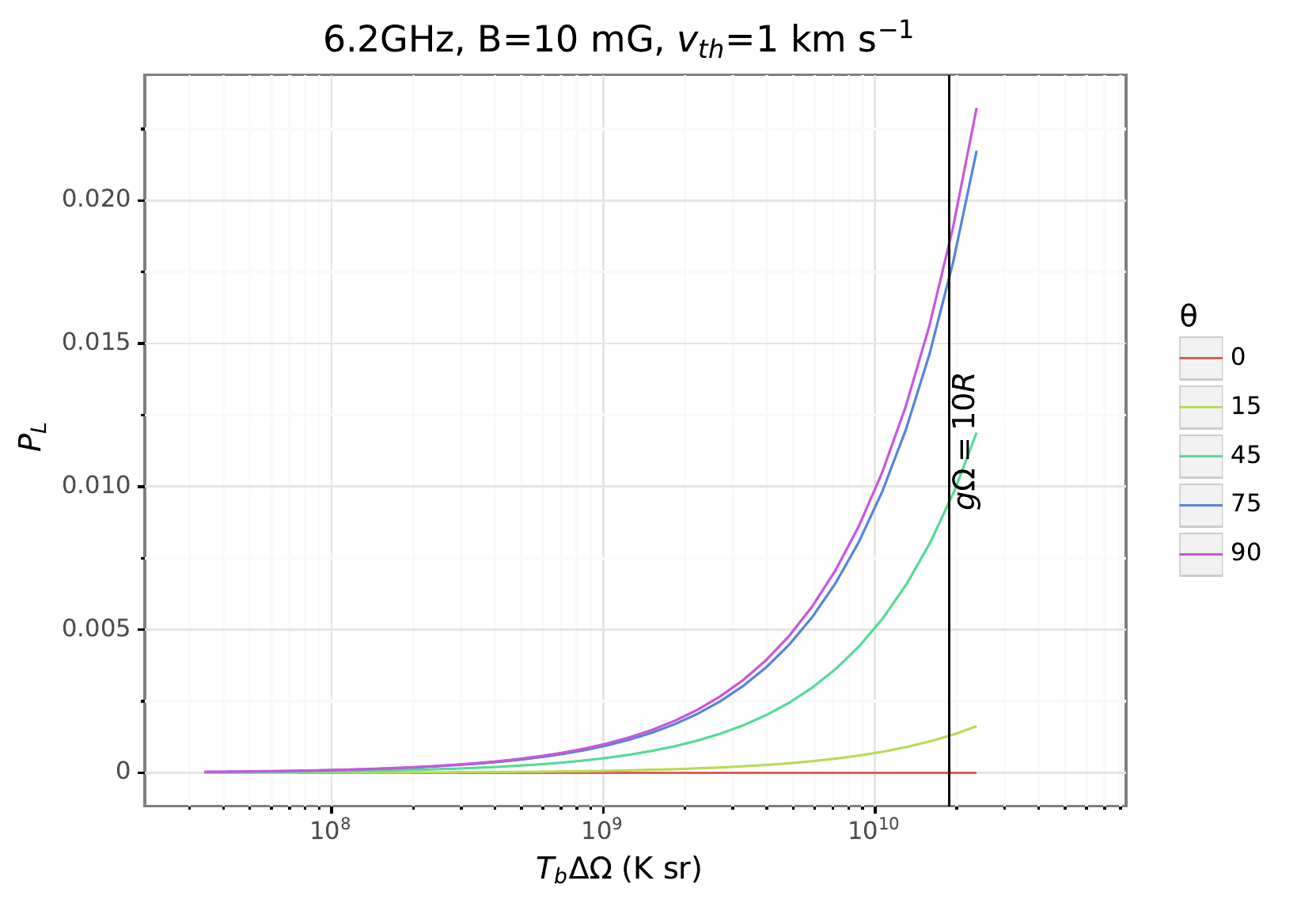}
  \includegraphics[width=\columnwidth, clip]{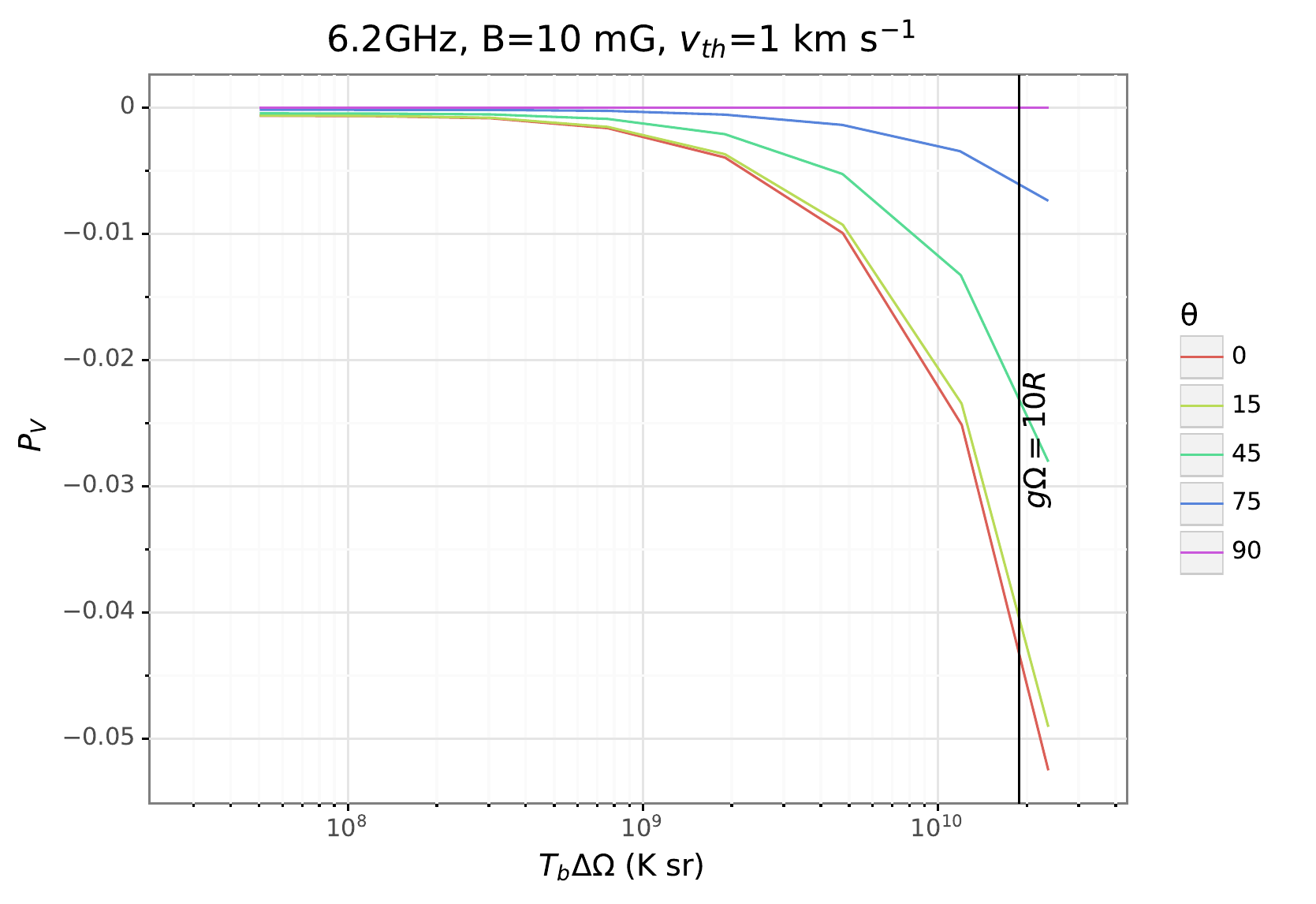}
  \caption{6.2 GHz methanol maser baseline. Linear and circular
    polarization fraction as a function of the maser luminosity for
    five different $\theta$. The vertical line marks $T_B\Delta\Omega$
    where $g\Omega=10R$. Magnetic field strength is 10 mG, thermal velocity width is
   1 km s$^{-1}$. }
 \label{fig:6.2PQPV_TB}
\end{figure*}

\begin{figure}
 \centering
  \includegraphics[width=\columnwidth, clip]{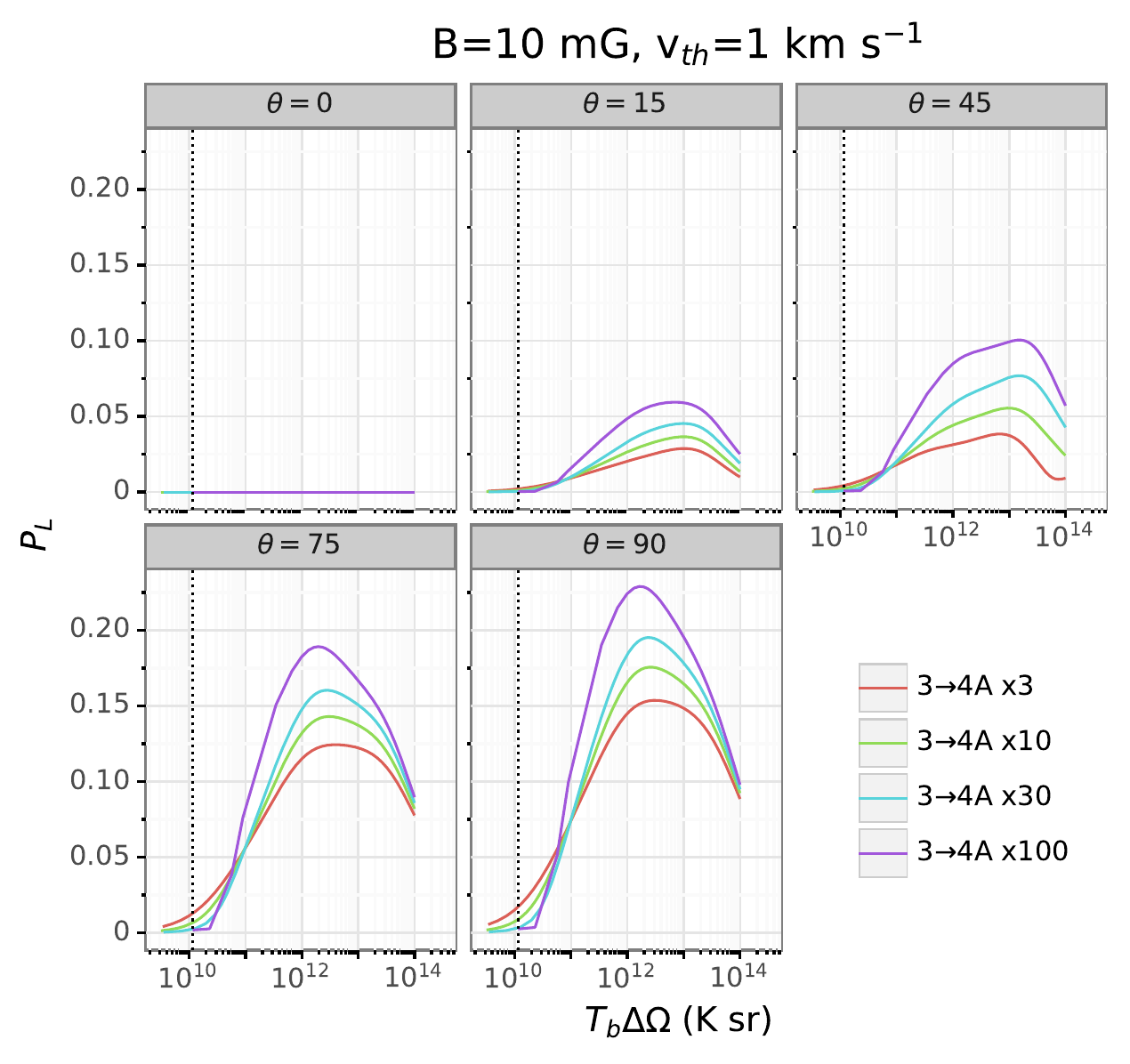}
  \includegraphics[width=\columnwidth, clip]{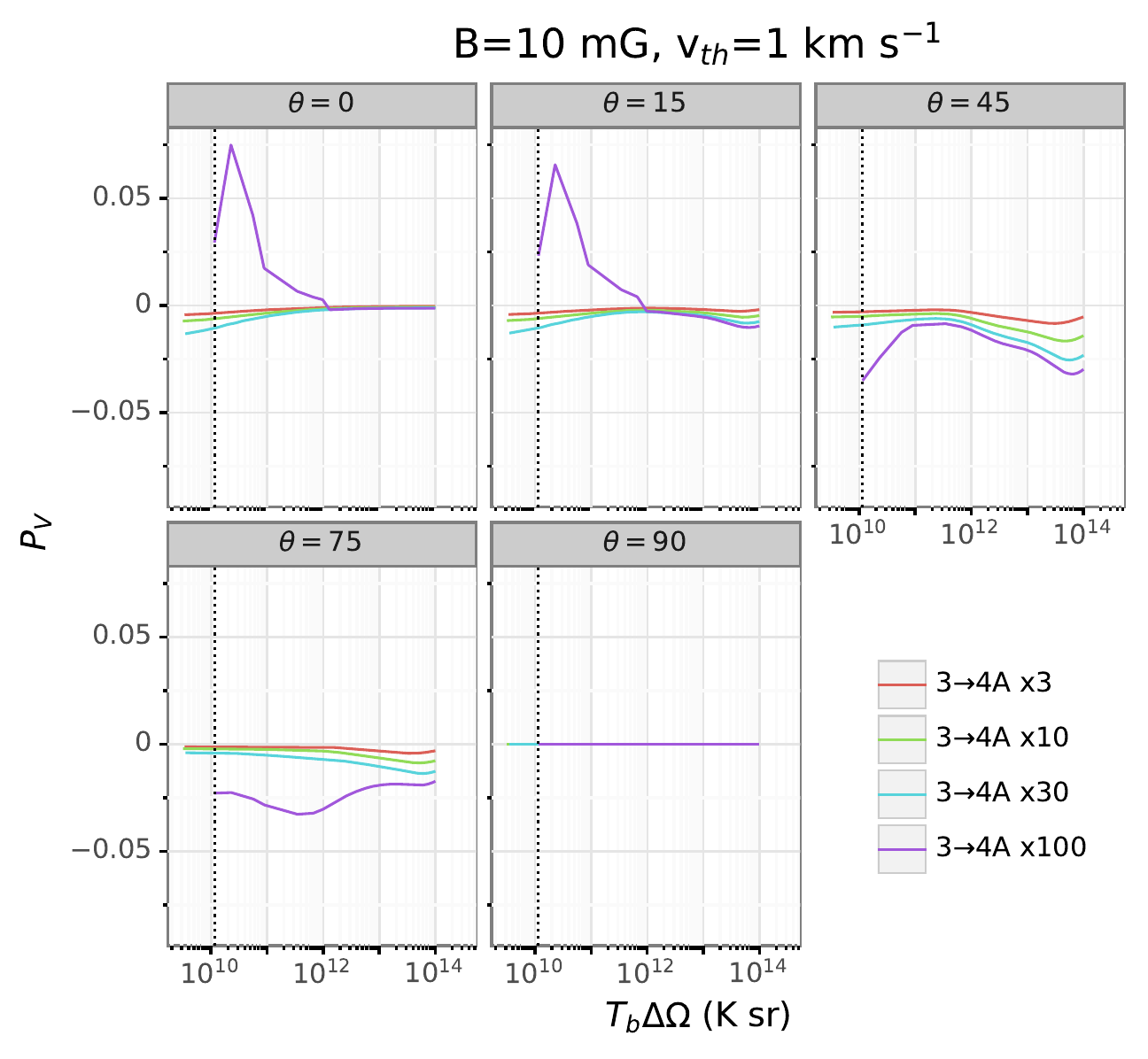}
  \caption{6.7 GHz methanol maser linear and circular
    polarization fraction as a function of the maser
    luminosity. Colours indicates different degrees of preferred
    pumping (3, 10, 30, 100 times) for the hyperfine transition
    $3\rightarrow4$~A. The vertical dotted lines indicate
    $g\Omega=10R$. Magnetic field strength, angle $\theta$, and
    thermal width are indicated in the plot.}
 \label{Fig:trans43}
\end{figure}
\begin{figure}
 \centering
  \includegraphics[width=\columnwidth, clip]{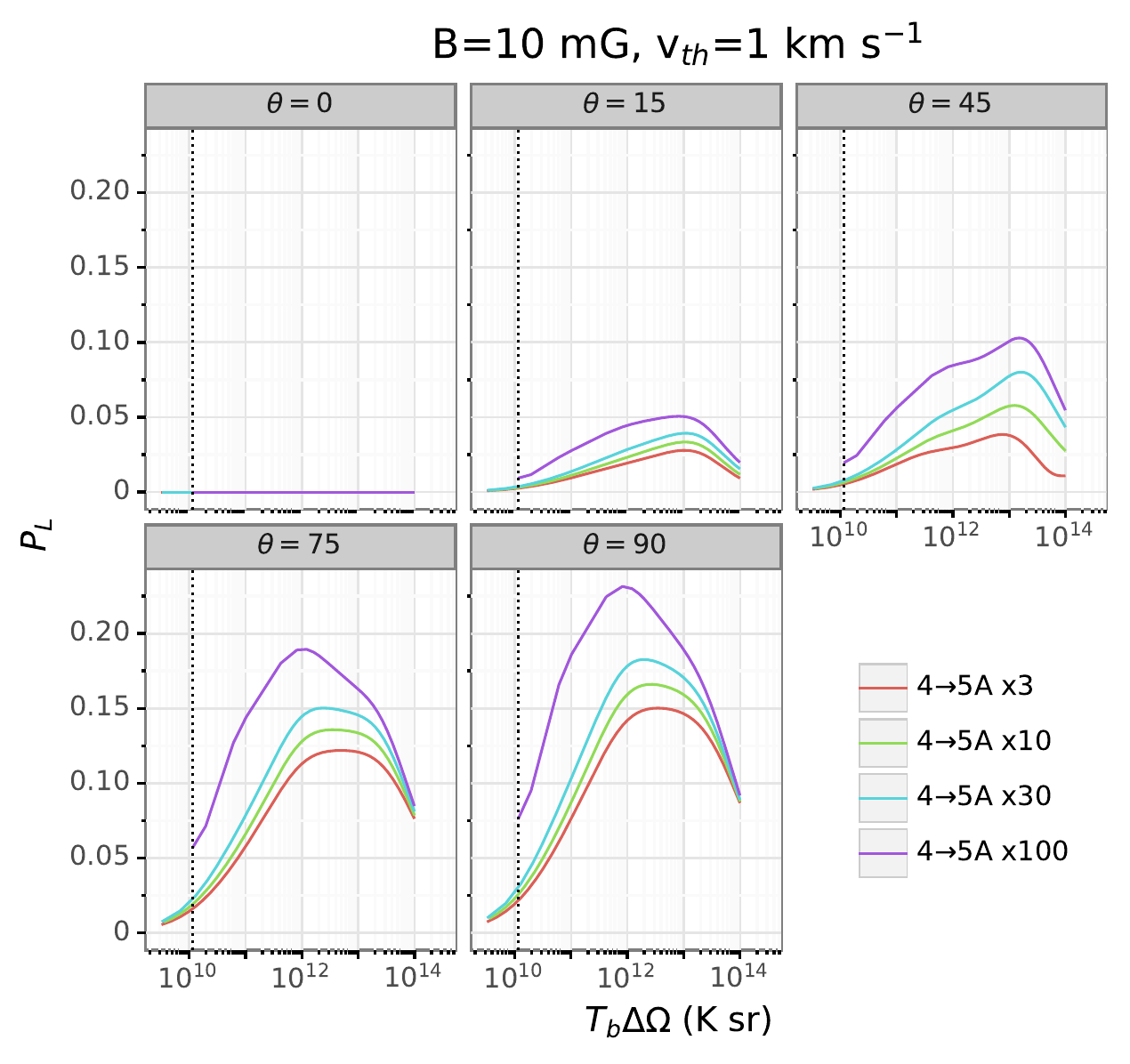}
  \includegraphics[width=\columnwidth, clip]{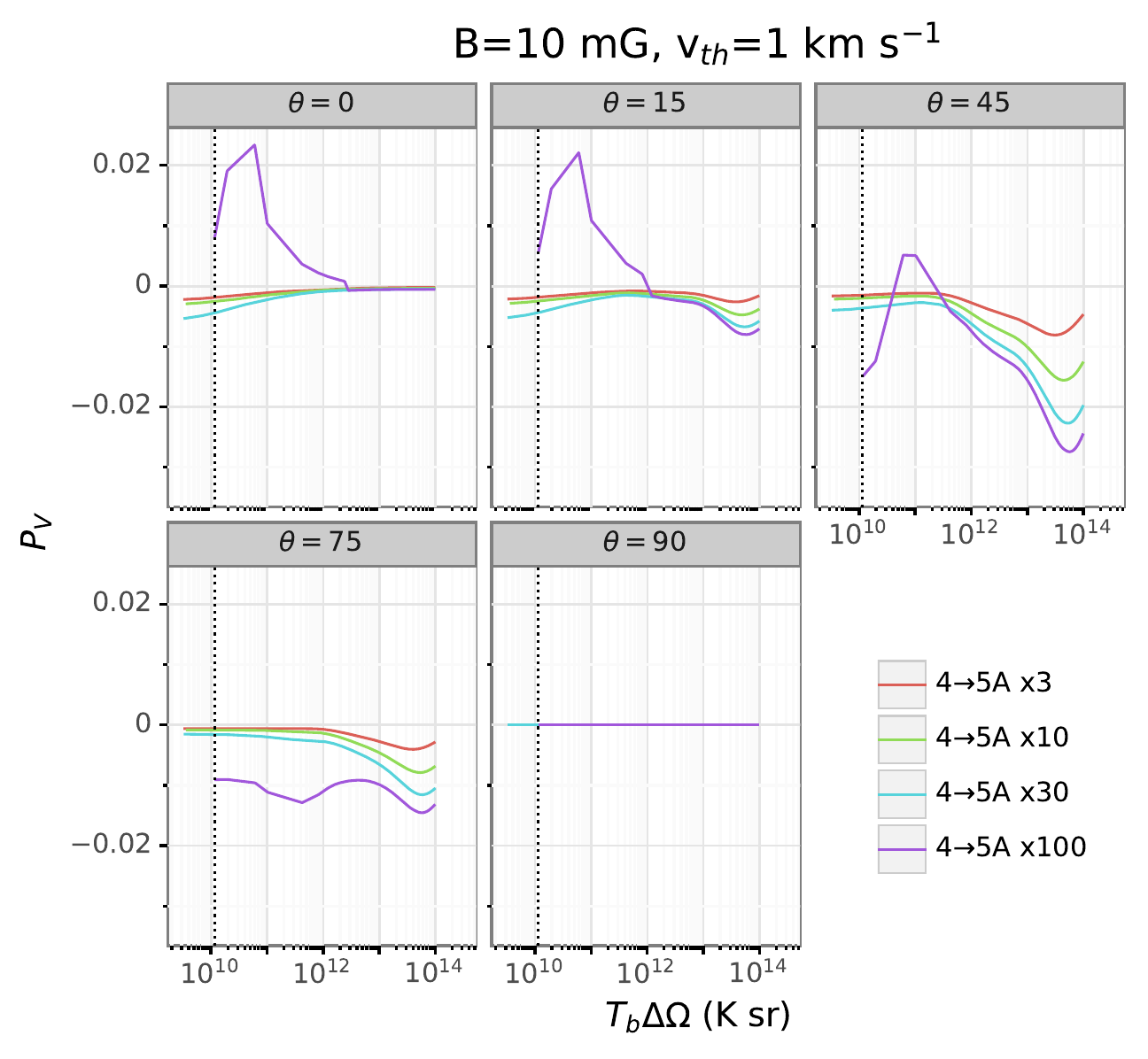}
  \caption{6.7 GHz methanol maser linear and circular
    polarization fraction as a function of the maser
    luminosity. Colours indicates different degrees of preferred
    pumping (3, 10, 30, 100 times) for the hyperfine transition
    $4\rightarrow 5$~A. The vertical dotted lines indicate
    $g\Omega=10R$. Magnetic field strength, angle $\theta$, and
    thermal width are indicated in the plot.}
 \label{Fig:trans54}
\end{figure}

\begin{figure*}[h!]
 \centering
 \includegraphics[width=\columnwidth,clip]{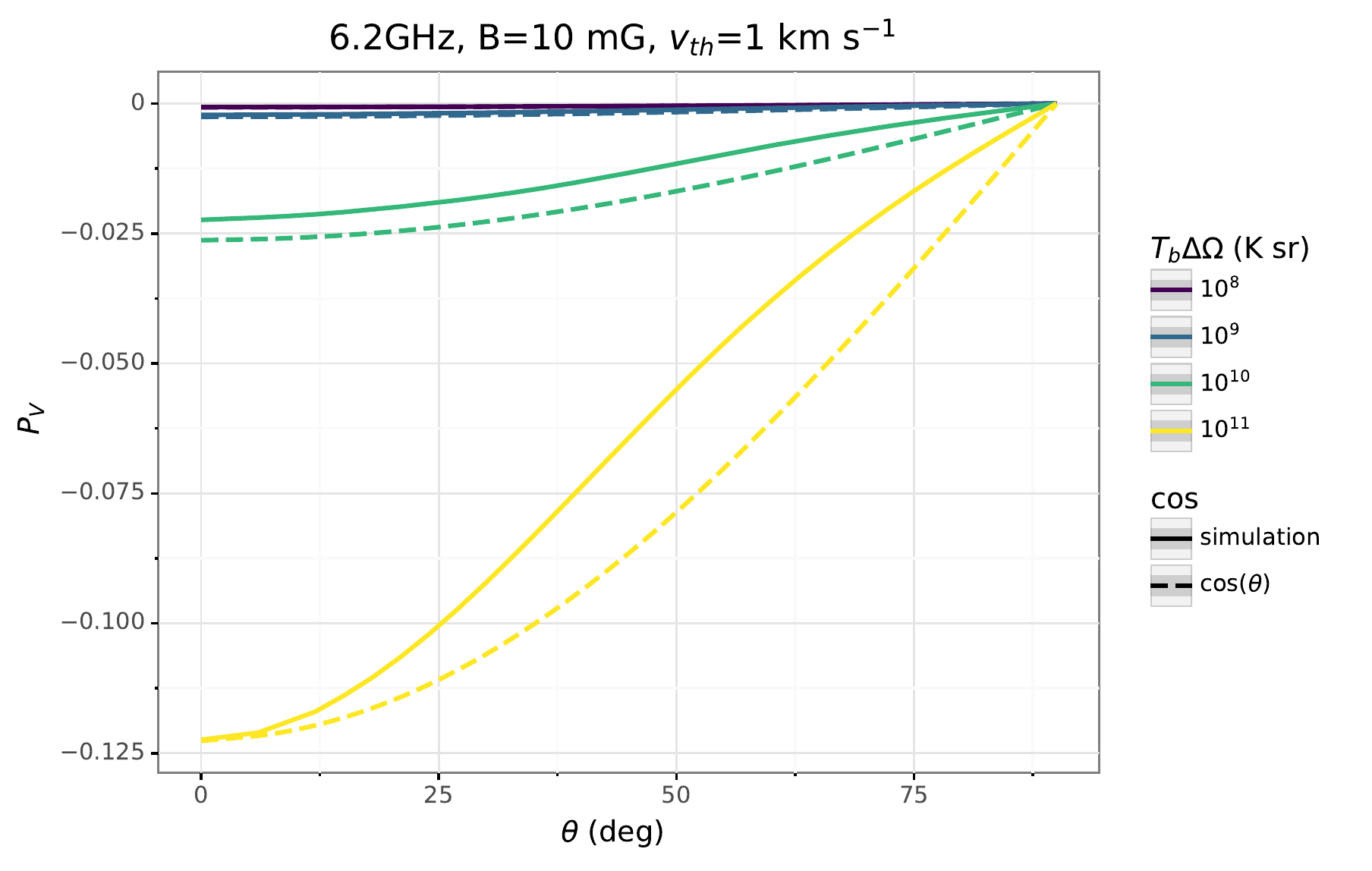}
 \includegraphics[width=\columnwidth,clip]{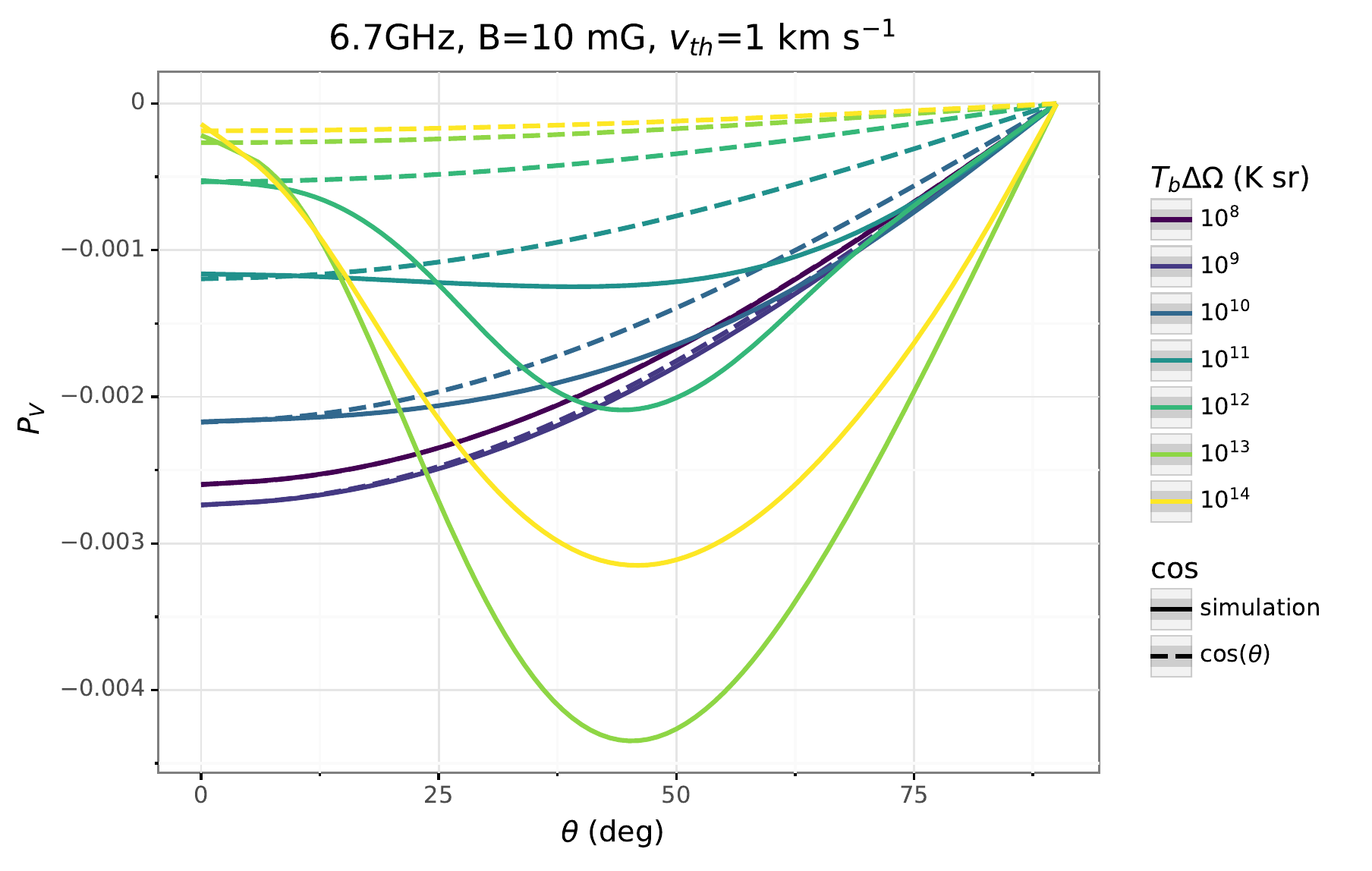}
 \includegraphics[width=\columnwidth,clip]{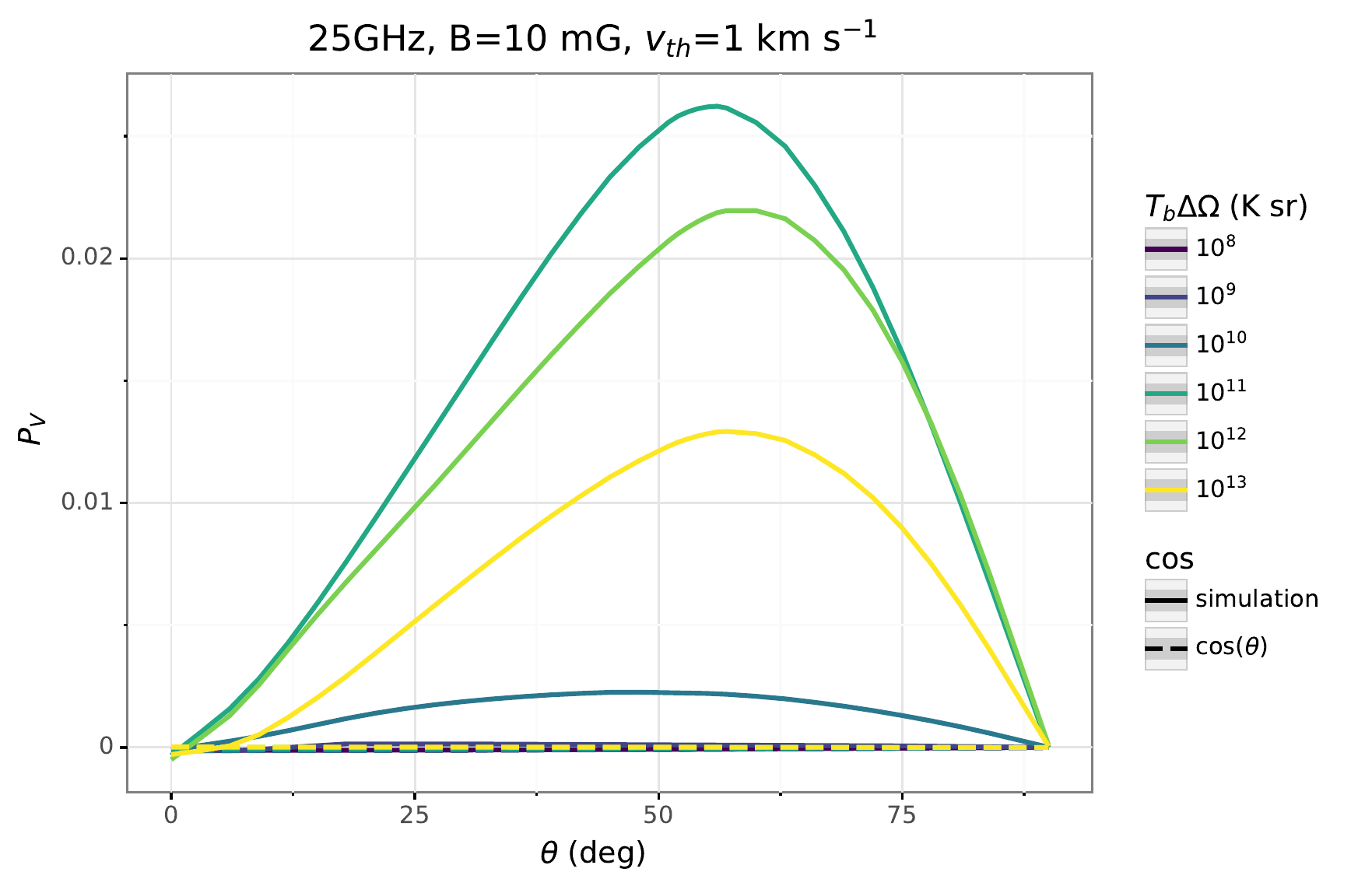}
 \includegraphics[width=\columnwidth,clip]{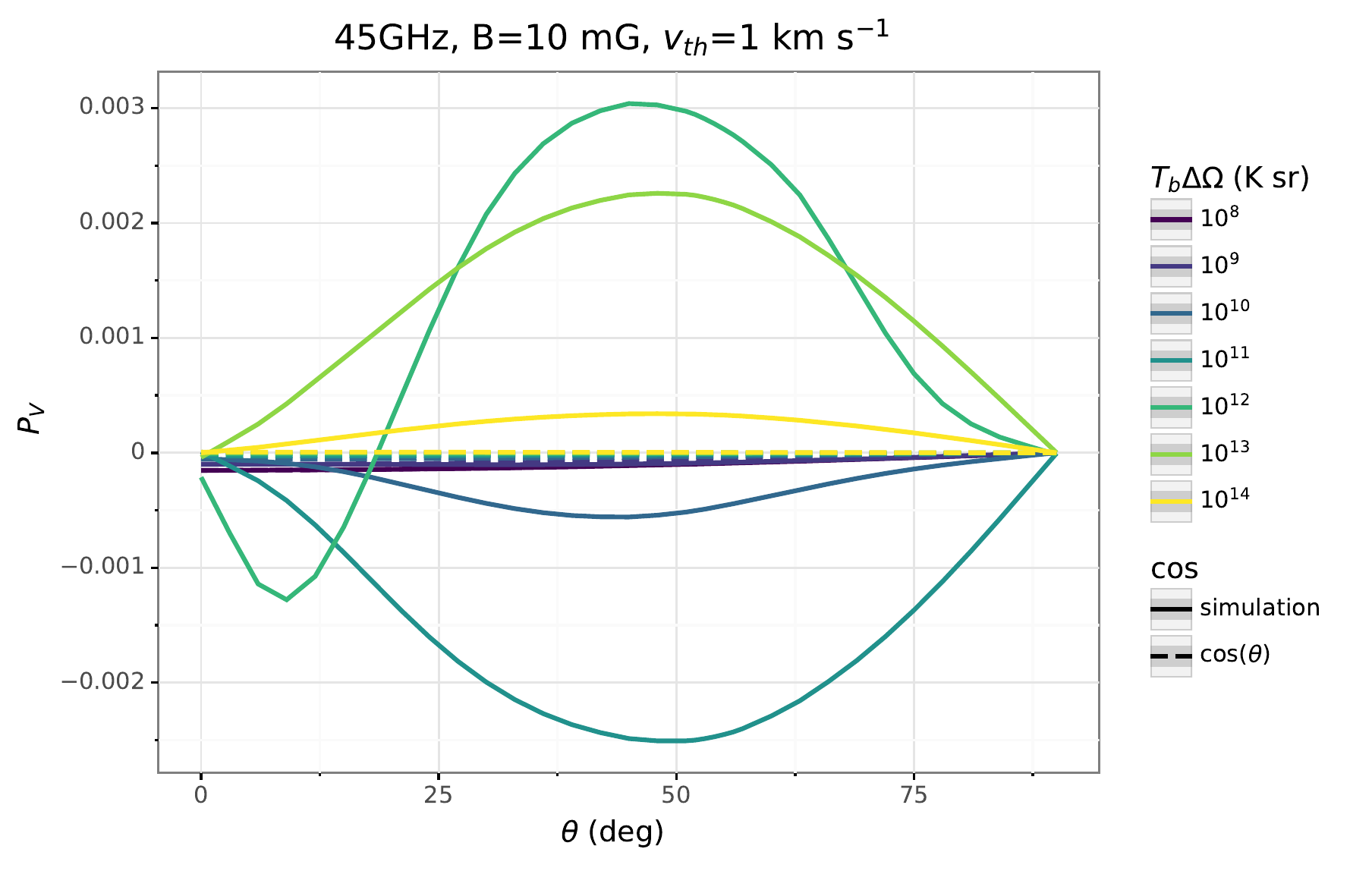}
 \caption{Circular polarization fraction as a function of $\theta$ for
   different $T_B\Delta\Omega$, for 6.2 GHz, 6.7 GHz, 25 GHz and 45
   GHz methanol maser baselines. Magnetic field strength is 10 mG,
   thermal velocity width is 1 km s$^{-1}$.  For the 6.2 GHz methanol
   maser $T_B\Delta\Omega$ corresponding to $g\Omega=10R$ limit is
   $\sim 2\times10^{10}$ K sr; for the 6.7 GHz methanol maser
   $T_B\Delta\Omega$ corresponding to $g\Omega=10R$ is $\sim10^{10}$ K
   sr ; for the 25 GHz methanol maser $T_B\Delta\Omega$ corresponding
   to $g\Omega=10R$ is $\sim10^{9}$ K sr; for the 45 GHz methanol
   maser $T_B\Delta\Omega$ corresponding to $g\Omega=10R$ is
   $\sim 9\times10^{8}$ K sr.}
 \label{fig:msersPV_TB}
\end{figure*}
  
   \begin{figure}[h!]
    \centering
    \begin{subfigure}[b]{.4\textwidth}
       \includegraphics[width=\textwidth]{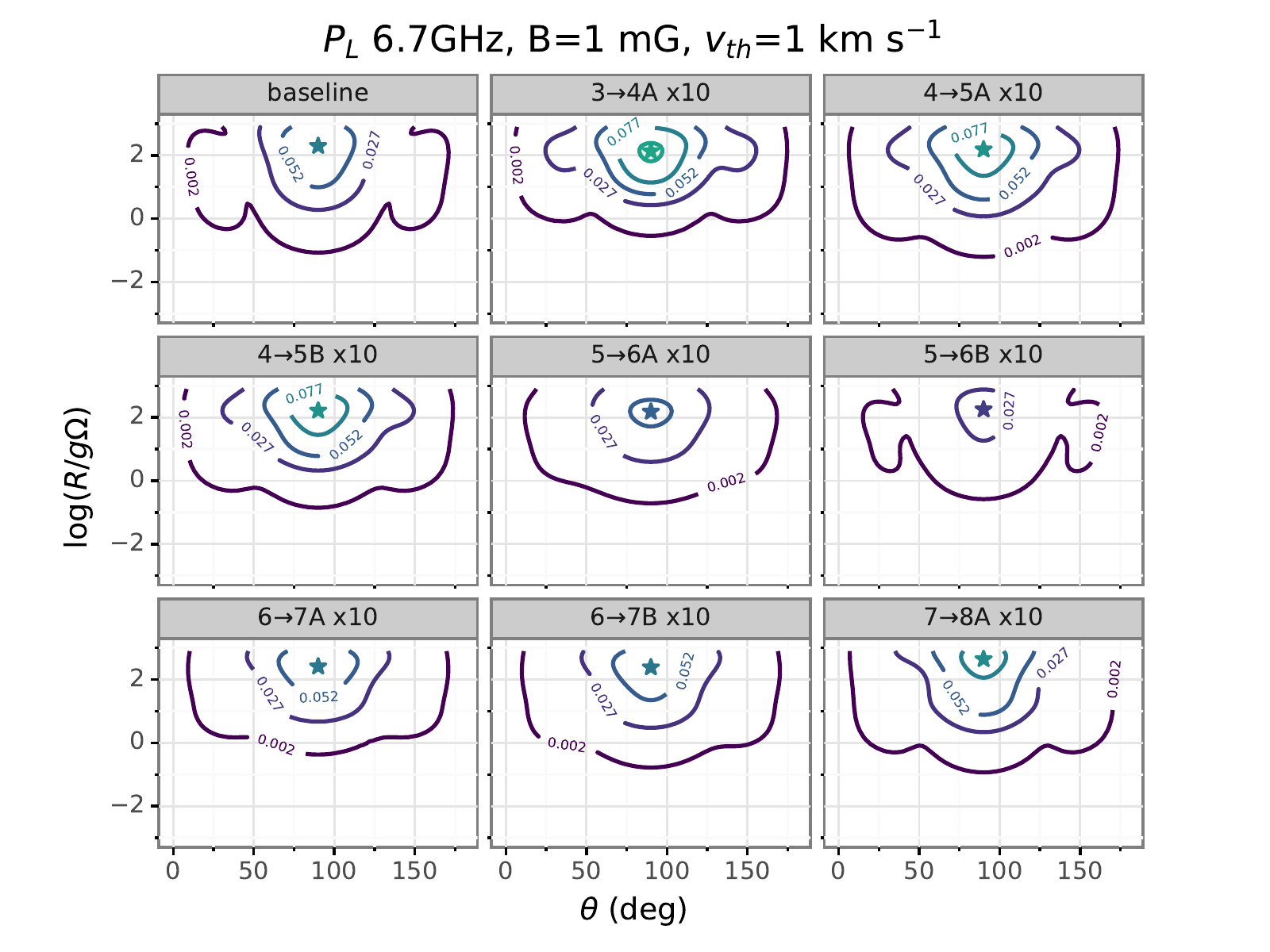}

    \end{subfigure}
    ~ 
    \begin{subfigure}[b]{0.4\textwidth}
       \includegraphics[width=\textwidth]{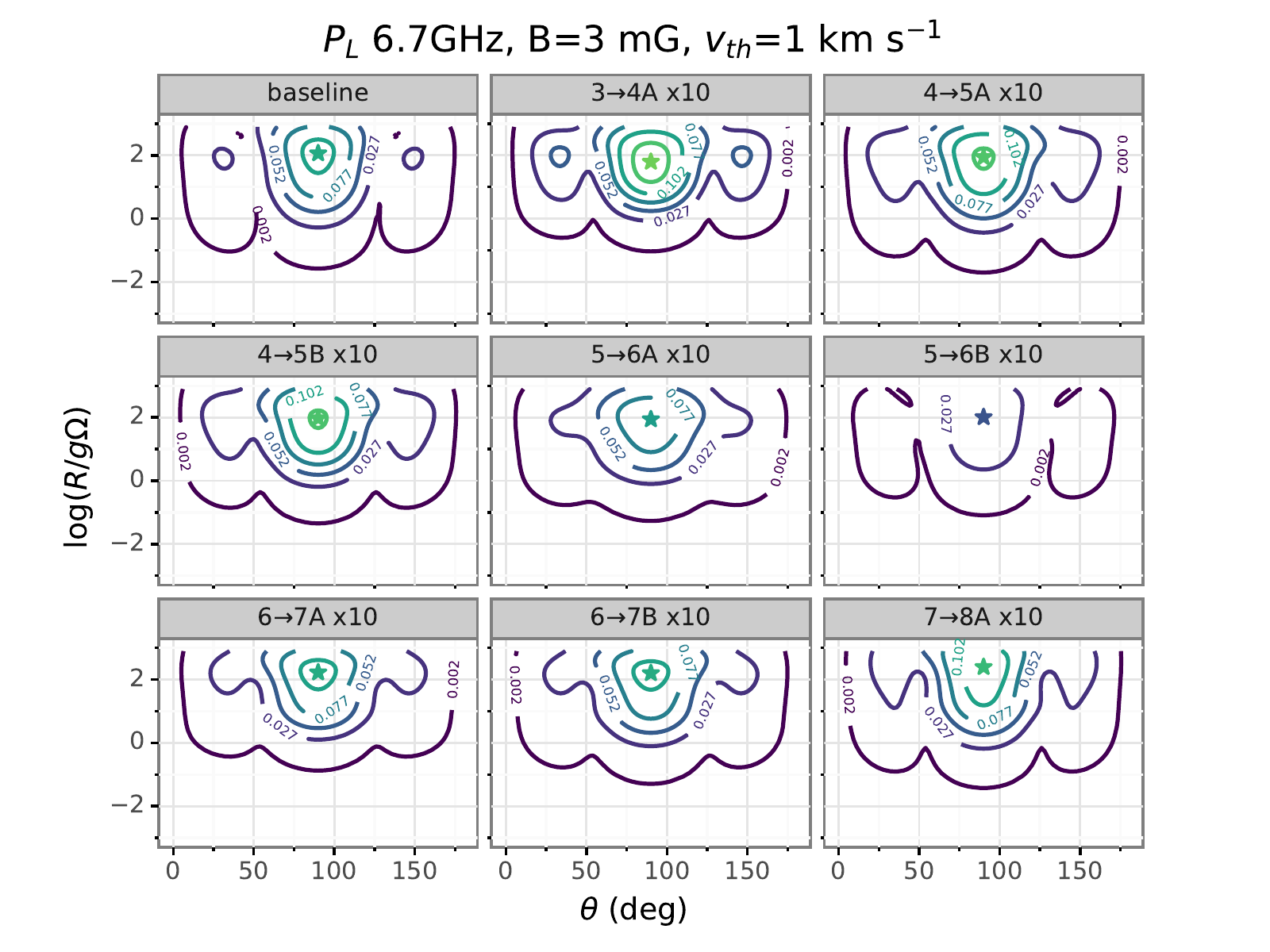} 

    \end{subfigure}
    ~ 
    \begin{subfigure}[b]{0.4\textwidth}
       \includegraphics[width=\textwidth]{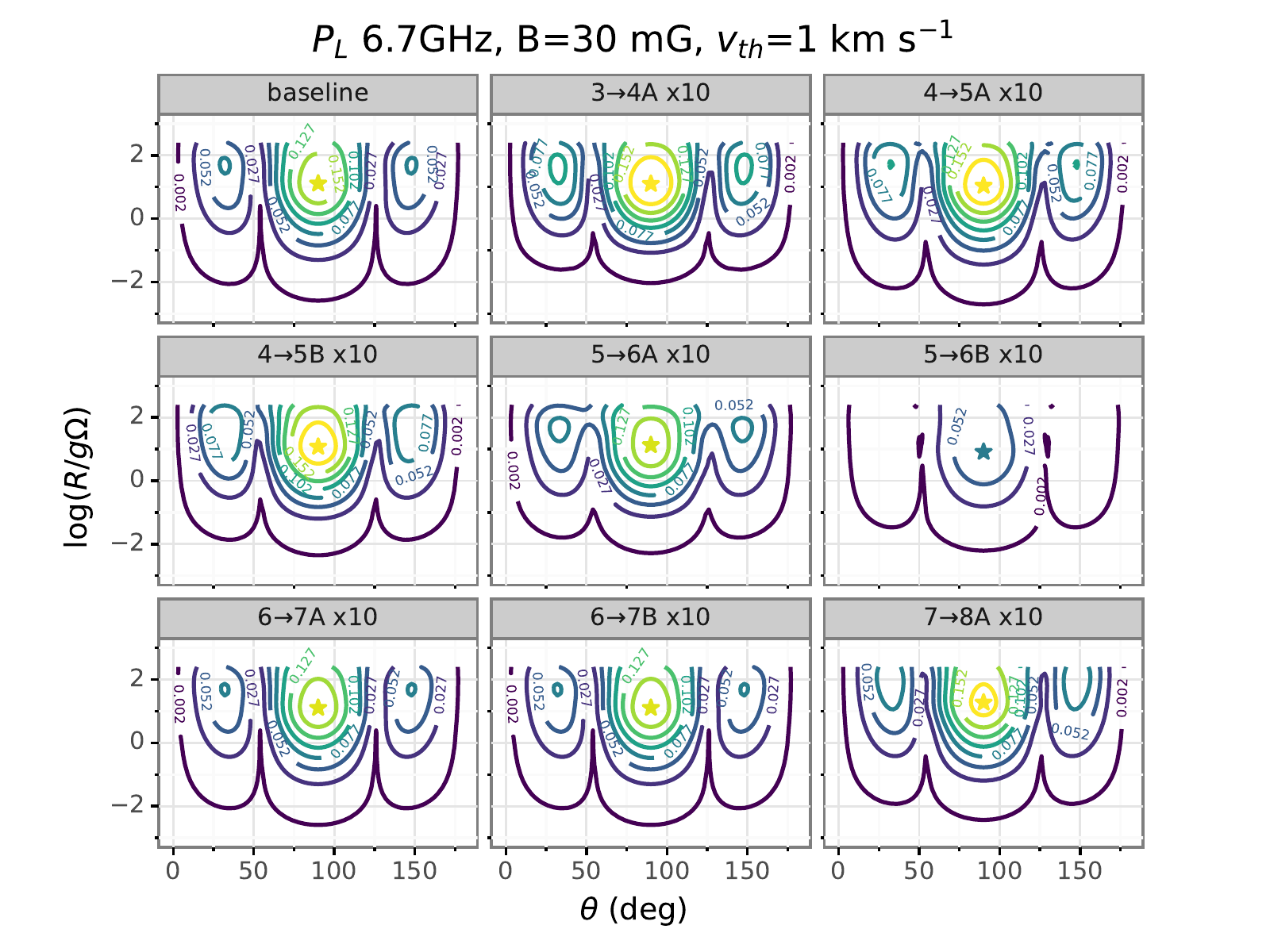} 

     \end{subfigure}
      ~ 
    \begin{subfigure}[b]{0.4\textwidth}
       \includegraphics[width=\textwidth]{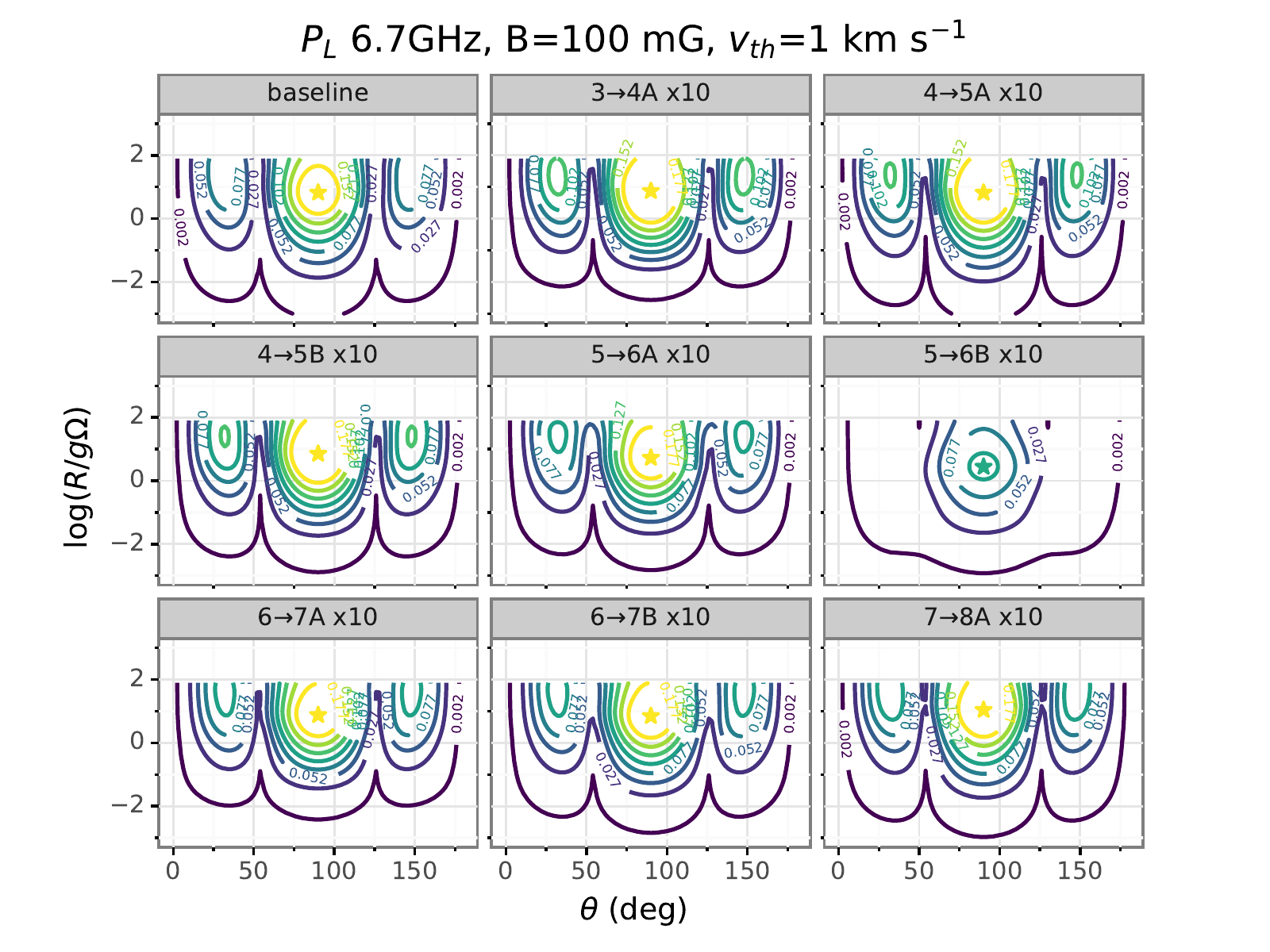} 

     \end{subfigure}
    \caption{6.7 GHz methanol maser linear polarization fraction P$_L$, plotted as a function
      of the propagation angle $\theta$ and the rate of stimulated
      emission. B strengths, v$_{th}$, and the preferred hyperfine
    transitions are indicated in each panel. The panel at the top left
    labelled ``baseline'' indicates a fixed pumping rate equal for all
    the hyperfine transitions, while all others assume a $10\times$
    preferred pumping for the indicated $i\rightarrow j$ transition.}
    \label{fig:PL_cont_altriB}
\end{figure}

\begin{figure}[h!]
    \centering
    \begin{subfigure}[b]{0.4\textwidth}
       \includegraphics[width=\textwidth]{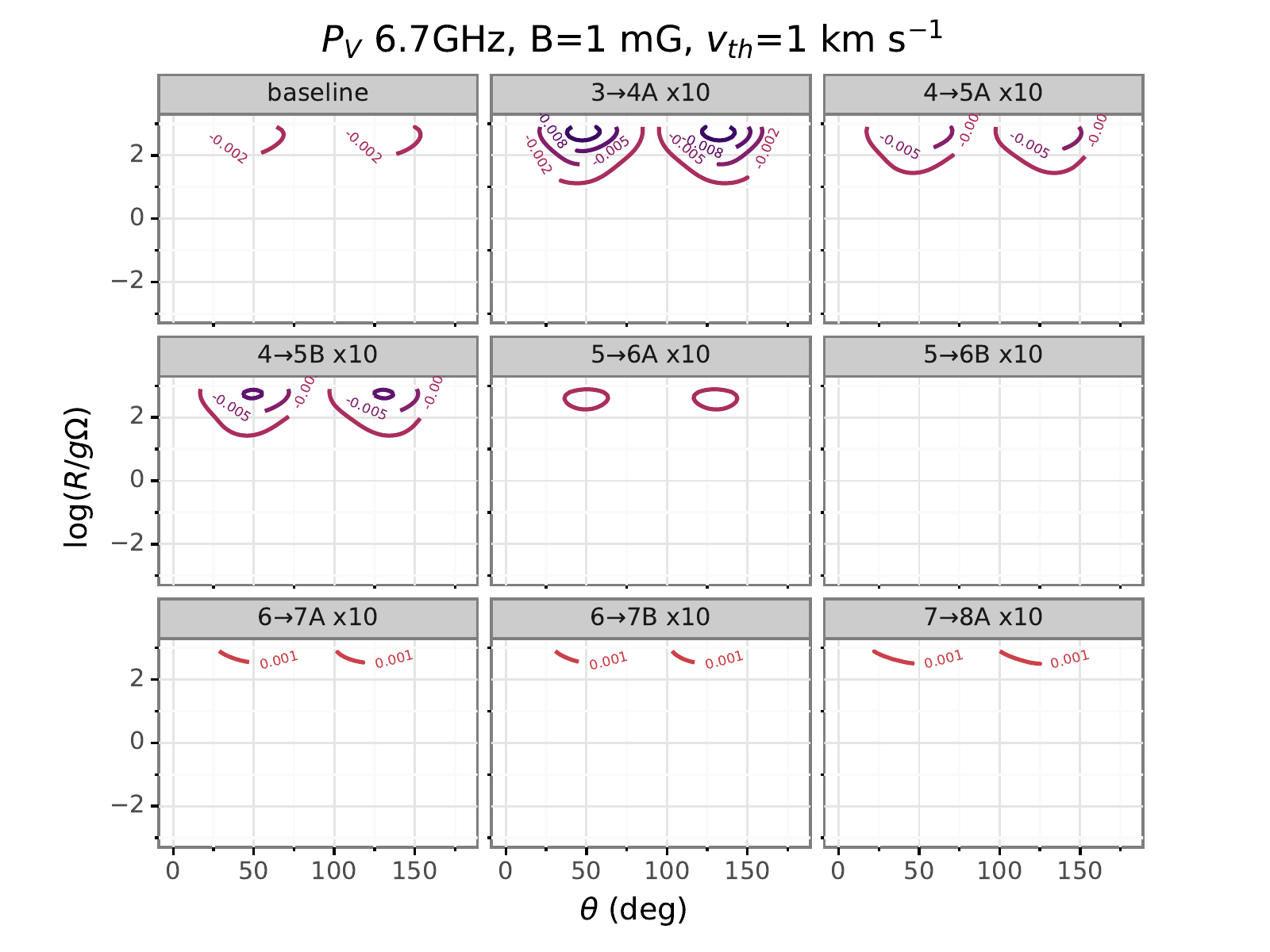}

    \end{subfigure}
    ~ 
    \begin{subfigure}[b]{0.4\textwidth}
       \includegraphics[width=\textwidth]{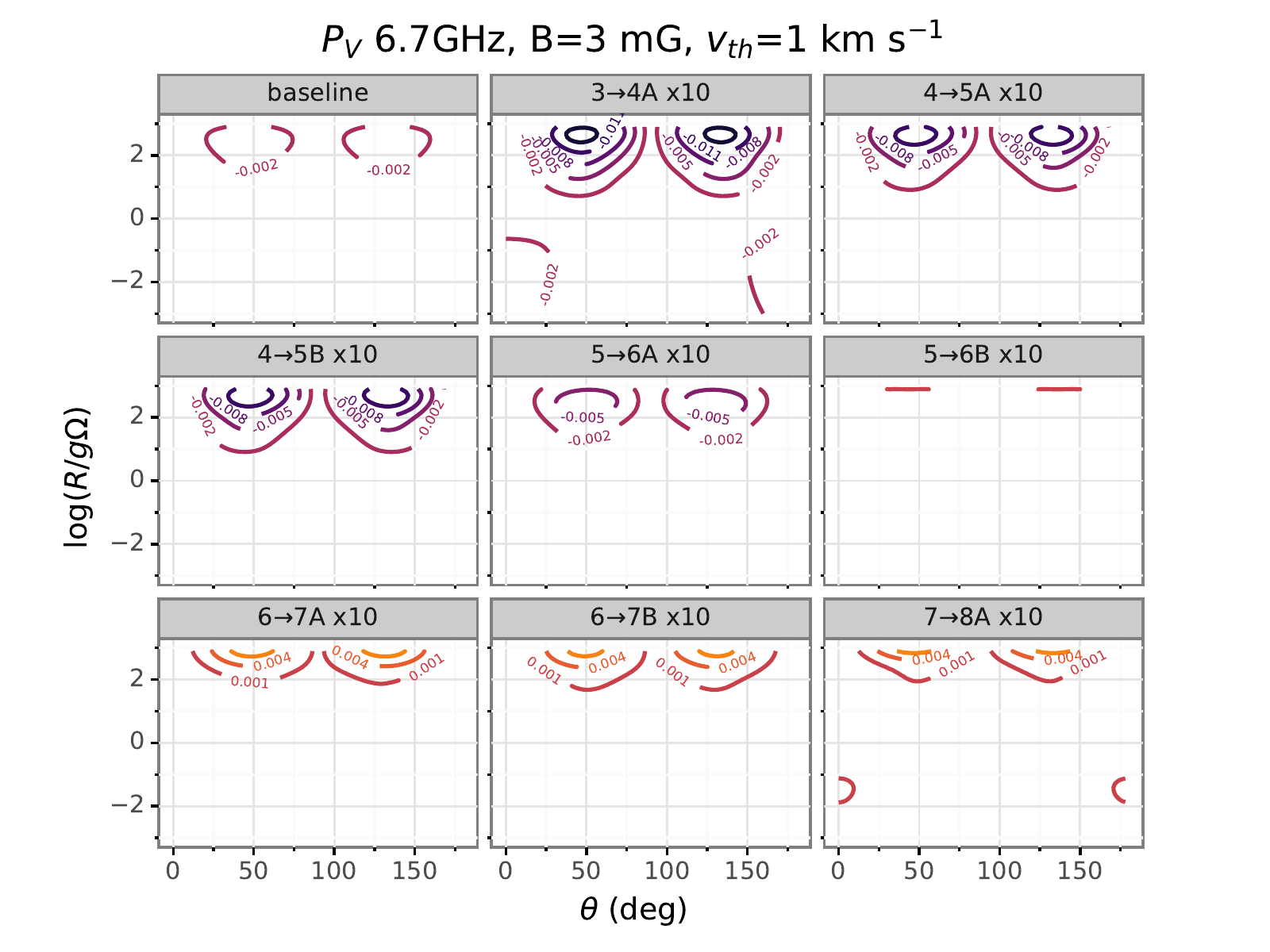} 

    \end{subfigure}
    ~ 
    \begin{subfigure}[b]{0.4\textwidth}
       \includegraphics[width=\textwidth]{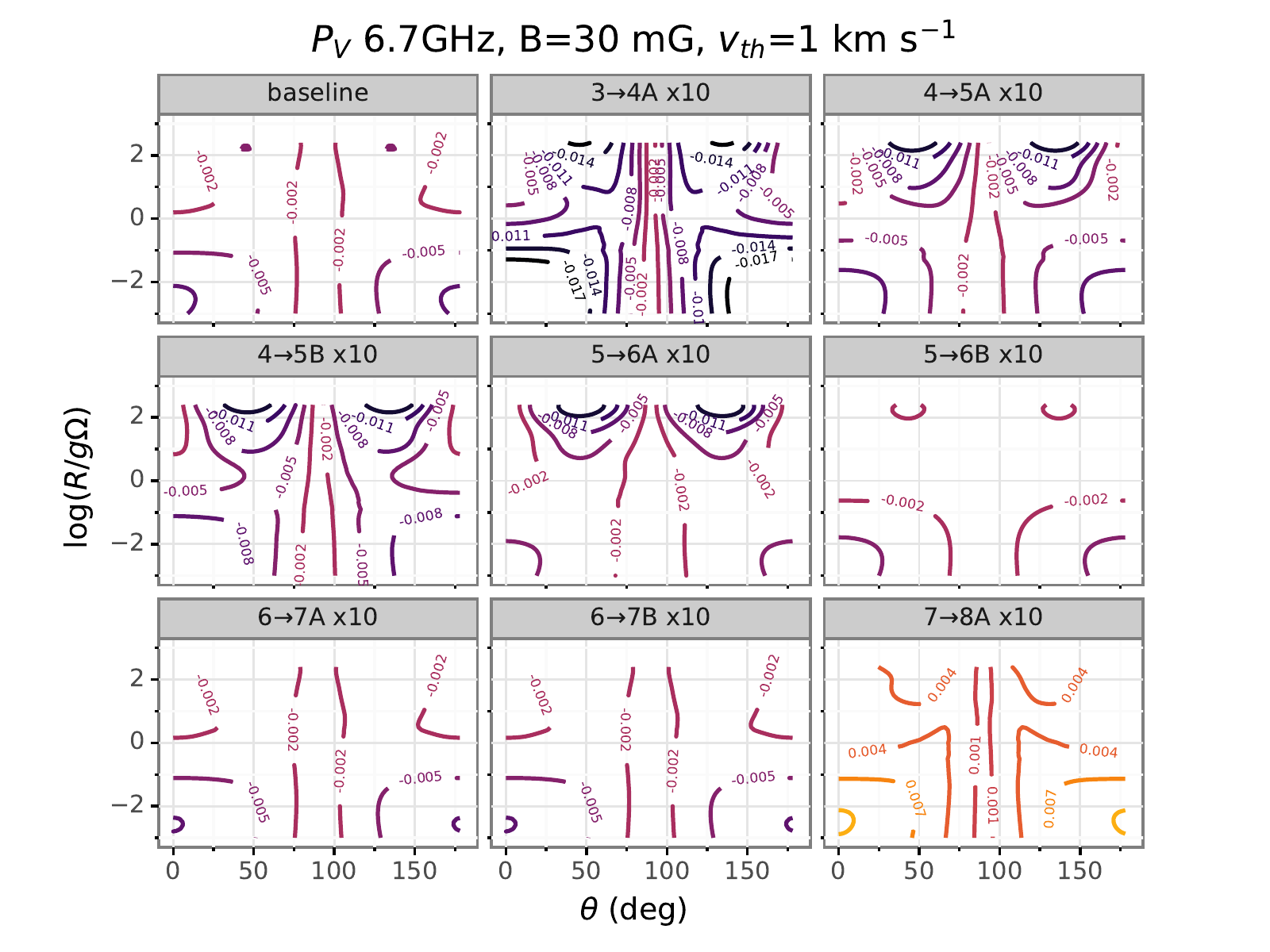} 

     \end{subfigure}
      ~ 
    \begin{subfigure}[b]{0.4\textwidth}
       \includegraphics[width=\textwidth]{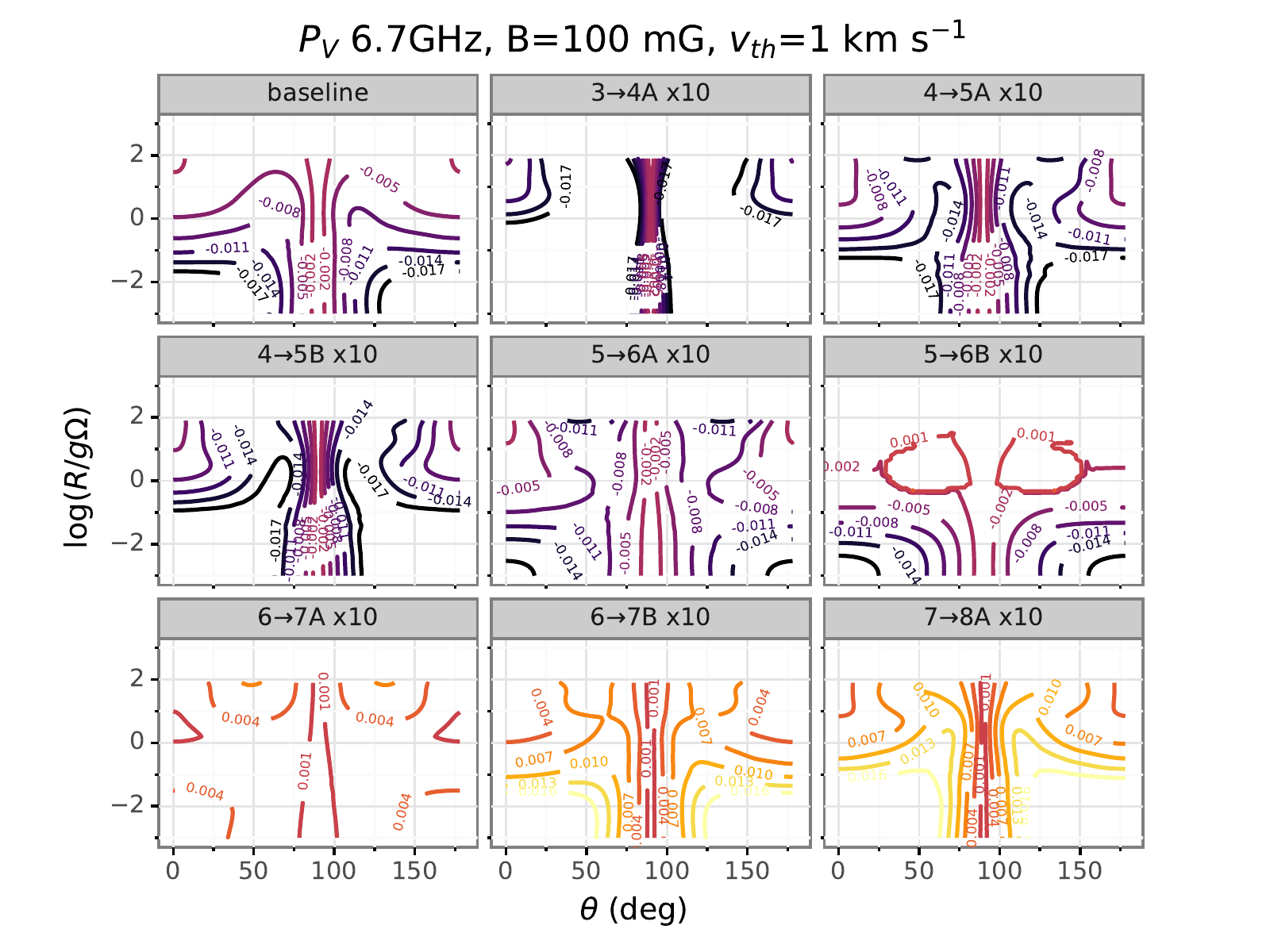} 

     \end{subfigure}
    \caption{6.7 GHz methanol maser circular polarization fraction P$_V$, plotted as a function
      of the propagation angle $\theta$ and the rate of stimulated
      emission.  Panels as in Fig.~\ref{fig:PL_cont_altriB}}
    \label{fig:PV_cont_altriB}
  \end{figure}

\begin{figure}[h!]
    \centering
    \begin{subfigure}[b]{0.4\textwidth}
       \includegraphics[width=\textwidth]{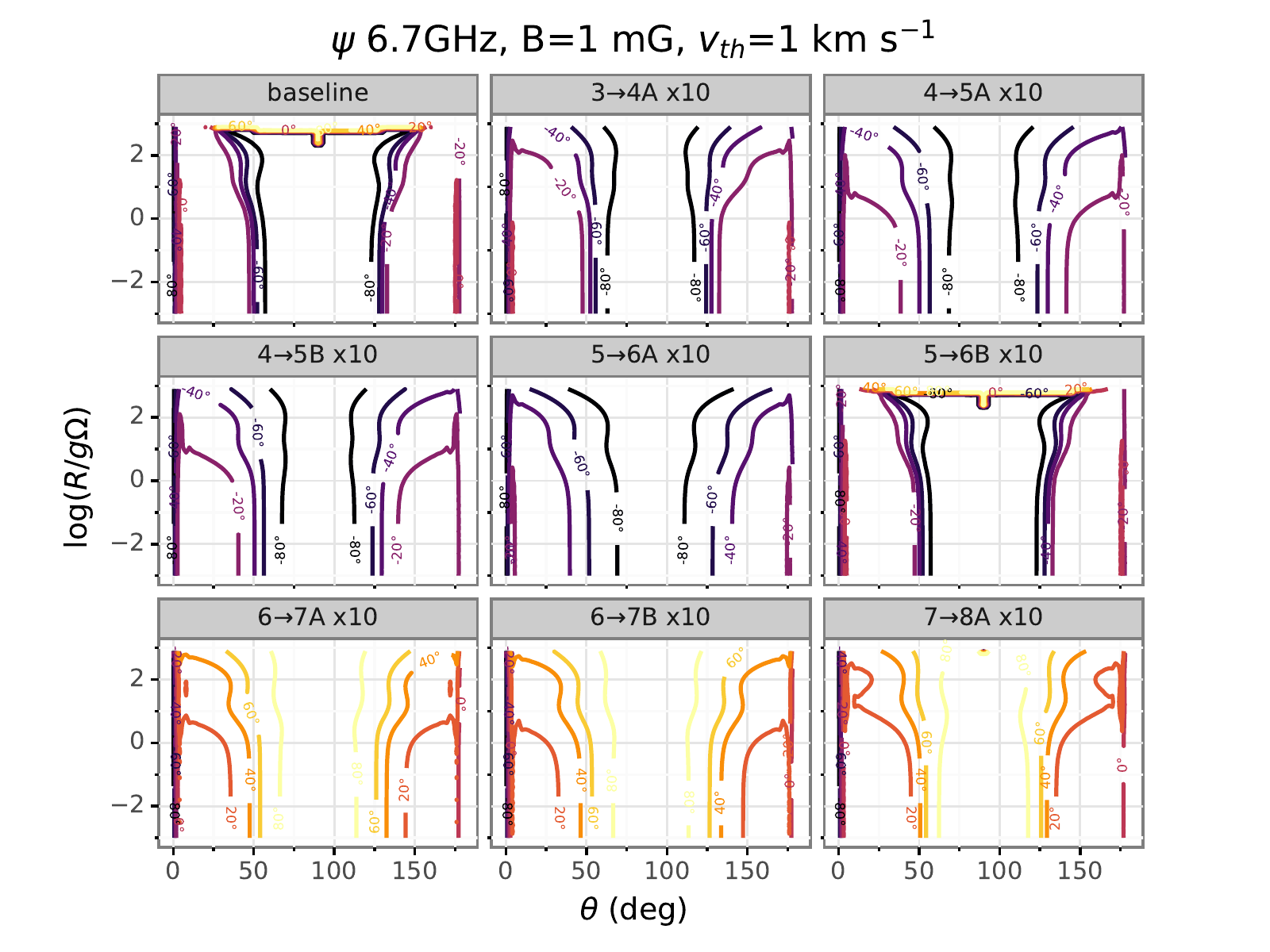}

    \end{subfigure}
    ~ 
    \begin{subfigure}[b]{0.4\textwidth}
       \includegraphics[width=\textwidth]{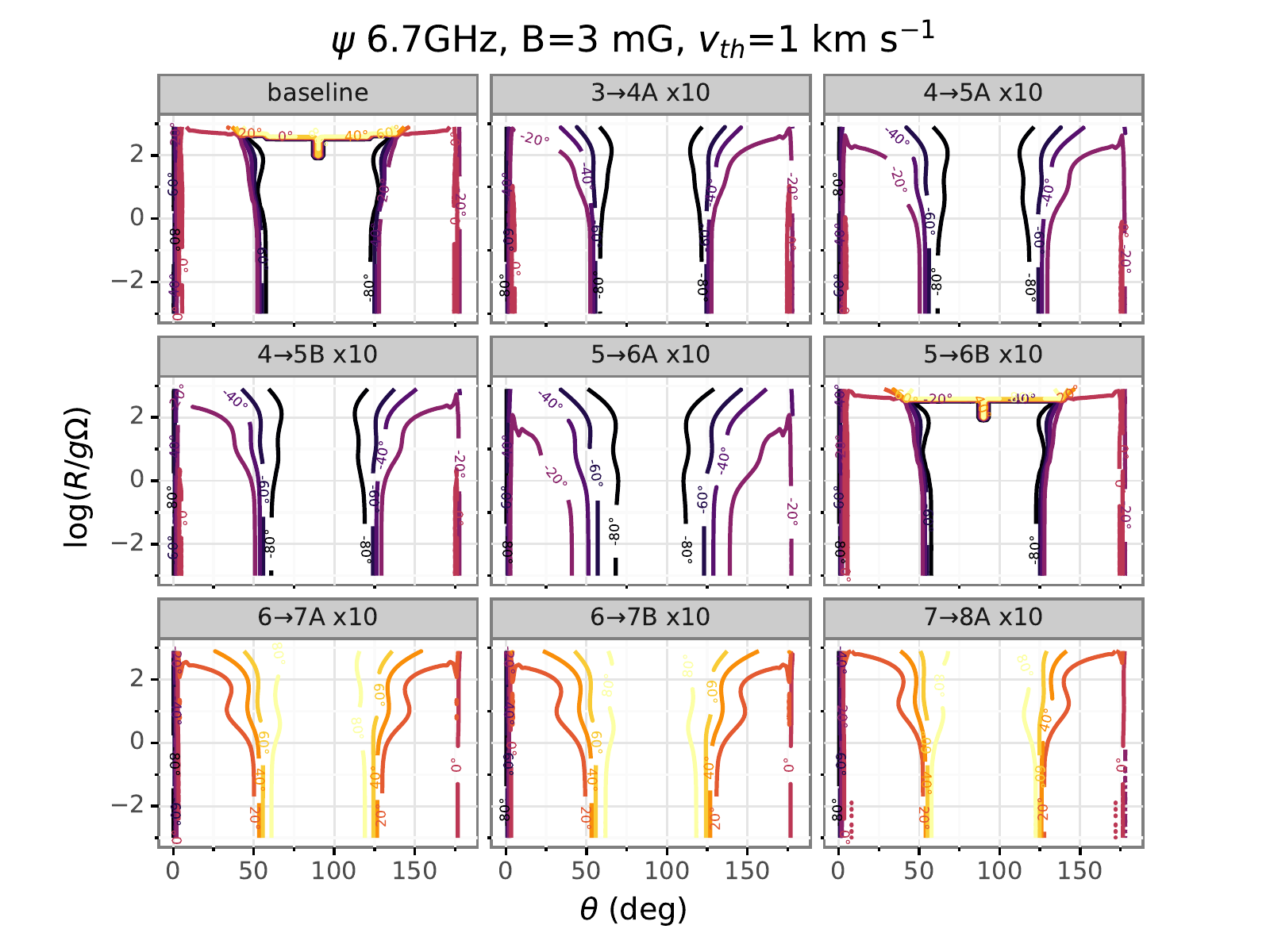} 

    \end{subfigure}
    ~ 
    \begin{subfigure}[b]{0.4\textwidth}
       \includegraphics[width=\textwidth]{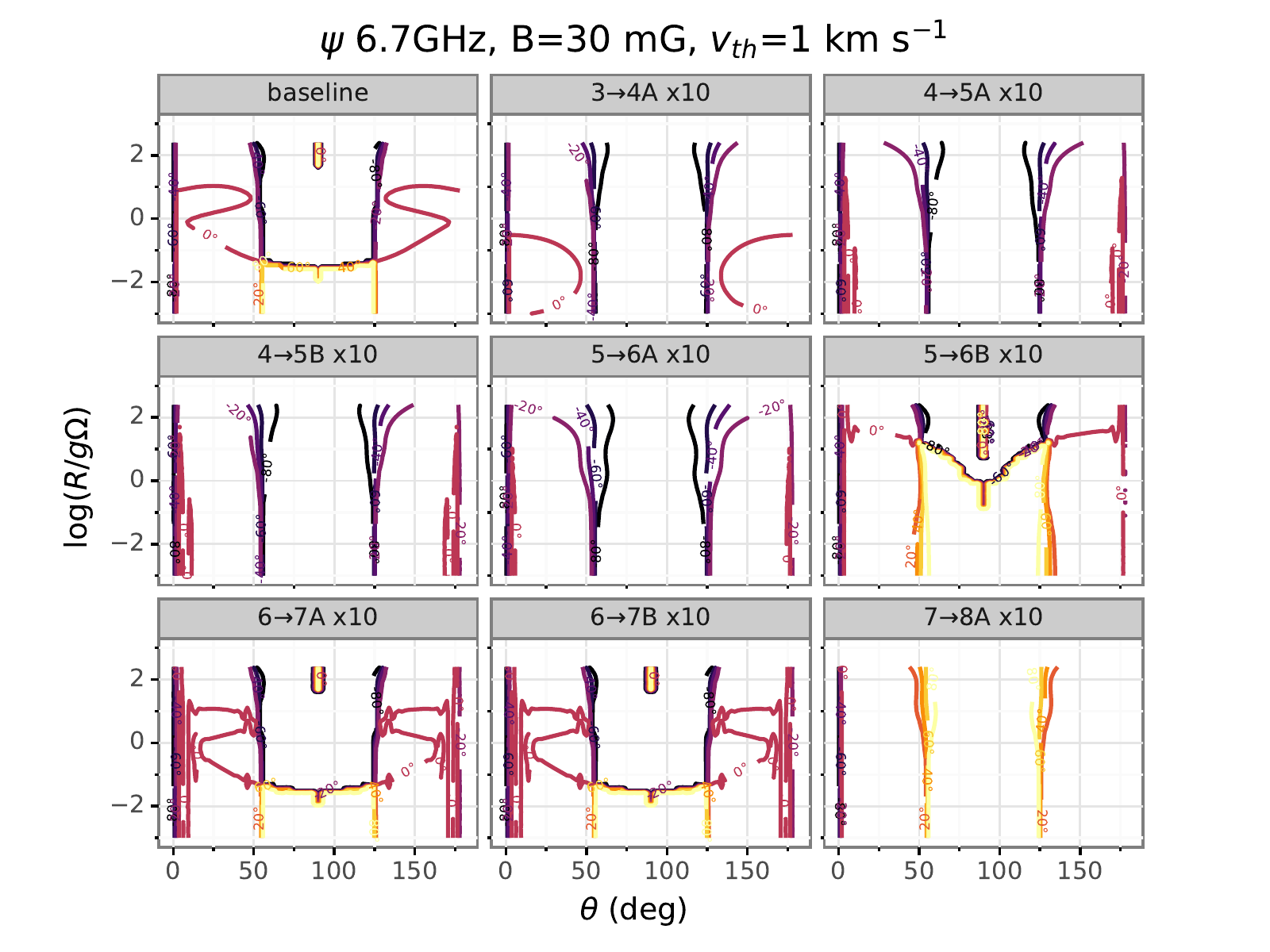} 

     \end{subfigure}
      ~ 
    \begin{subfigure}[b]{0.4\textwidth}
       \includegraphics[width=\textwidth]{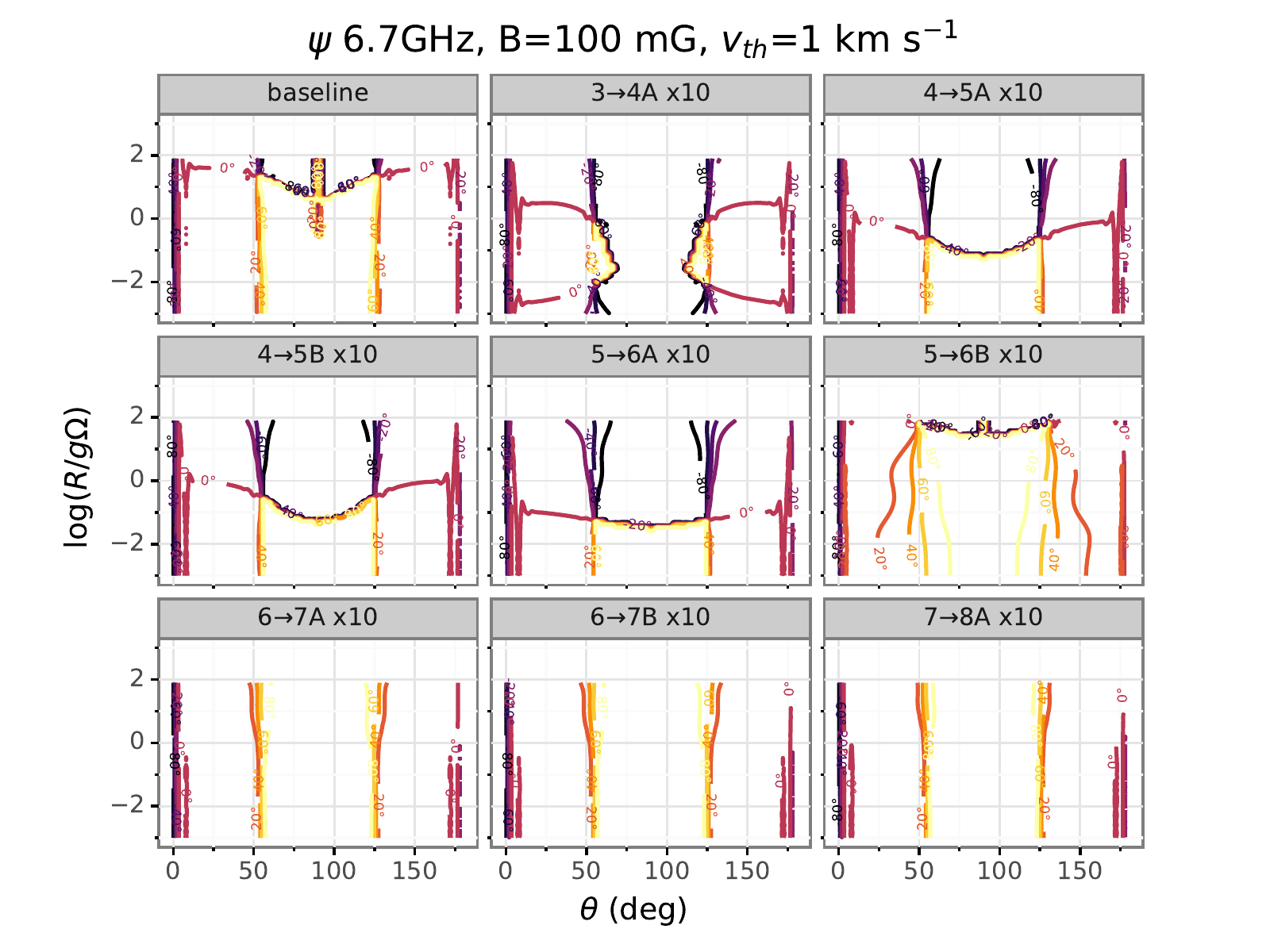} 

     \end{subfigure}
    \caption{6.7 GHz methanol maser linear polarization angle
      $\psi$, plotted as a function of the propagation angle $\theta$
      and the rate of stimulated emission. Contours are plotted every
      $20^\circ$. Panels as in Fig.~\ref{fig:PL_cont_altriB}}
    \label{fig:Pa_cont_altriB}
\end{figure}

 \begin{figure}[h!]
    \centering
    \begin{subfigure}[b]{0.4\textwidth}
       \includegraphics[width=\textwidth]{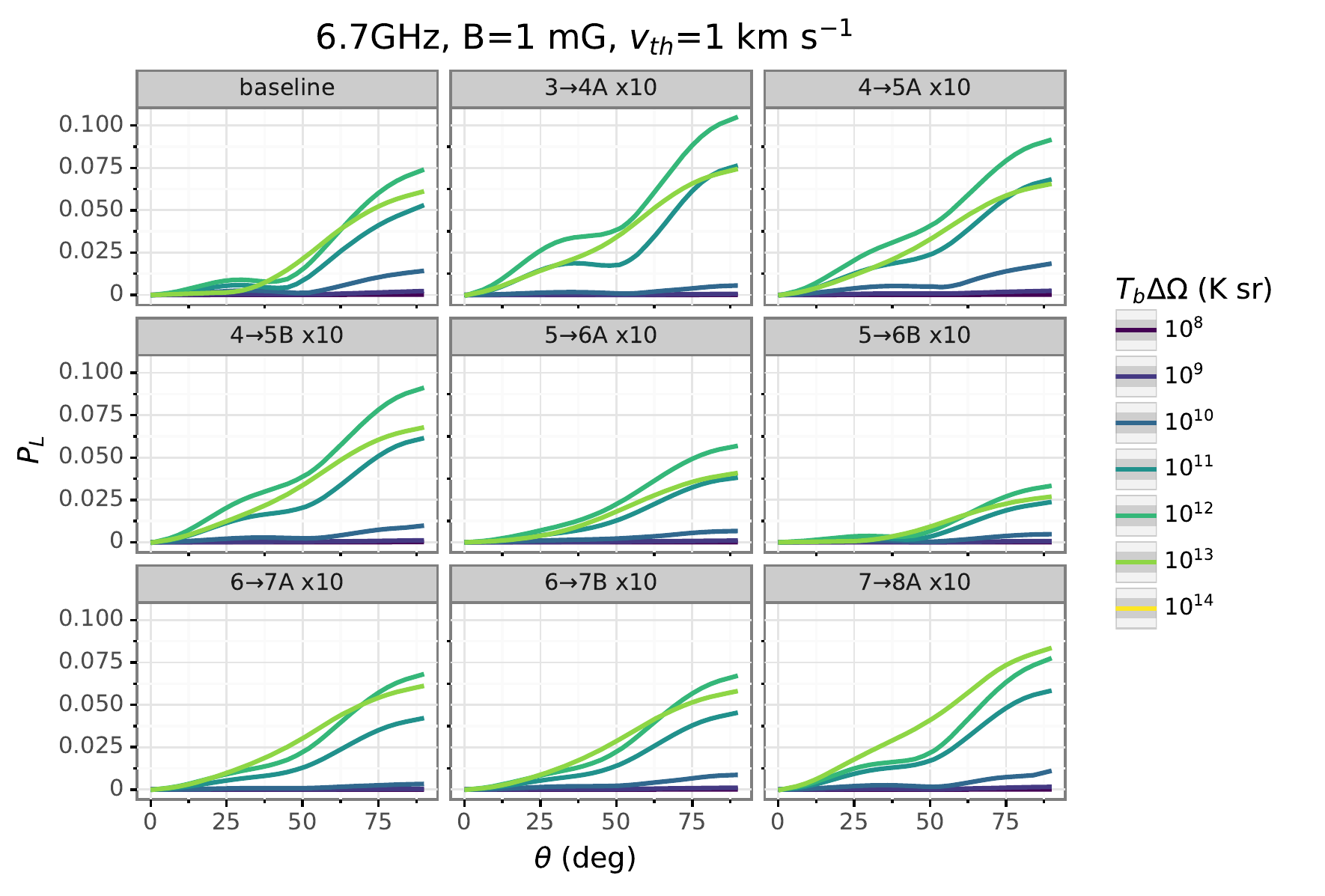}

    \end{subfigure}
    ~ 
    \begin{subfigure}[b]{0.4\textwidth}
       \includegraphics[width=\textwidth]{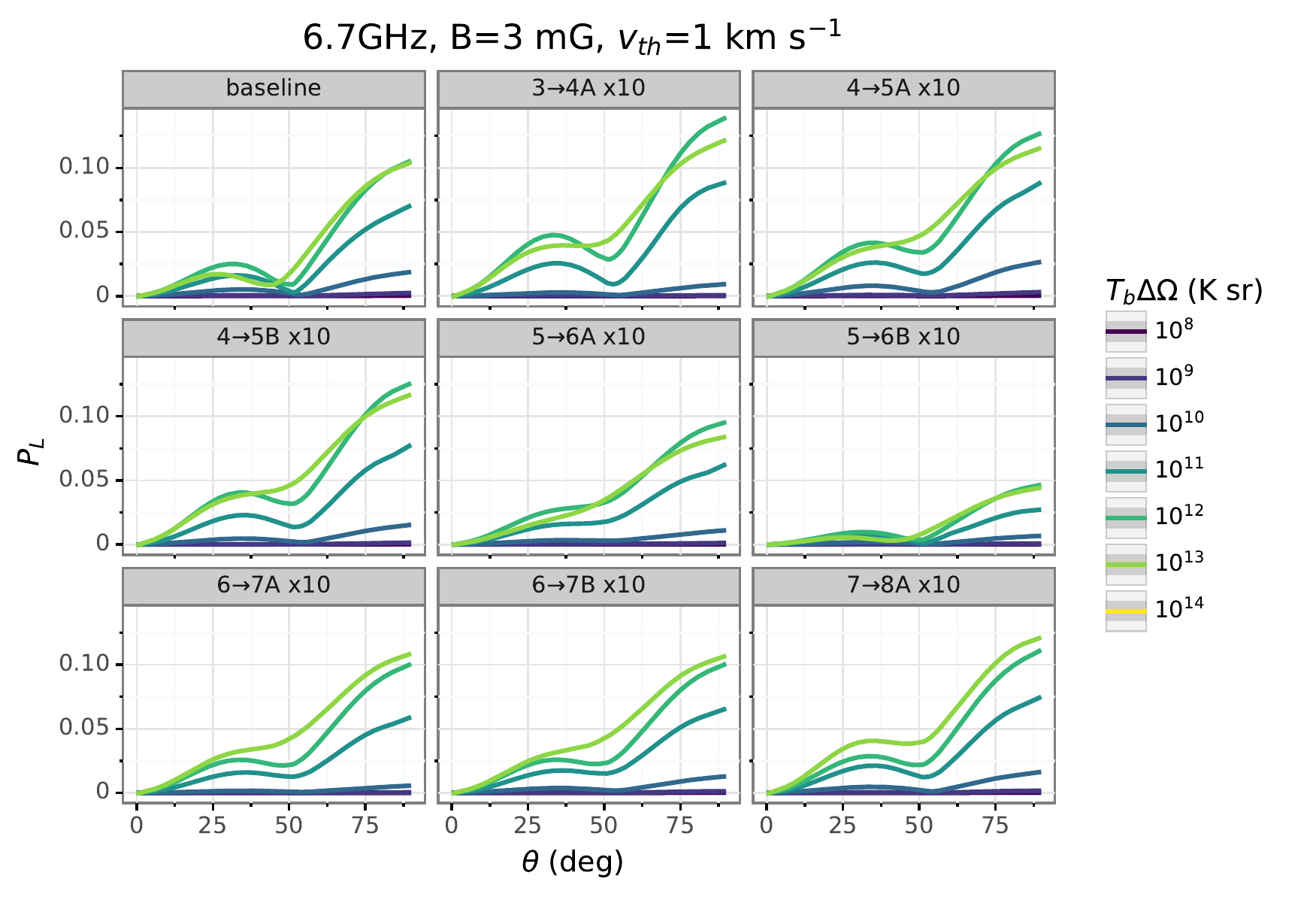} 

    \end{subfigure}
    ~ 
    \begin{subfigure}[b]{0.4\textwidth}
       \includegraphics[width=\textwidth]{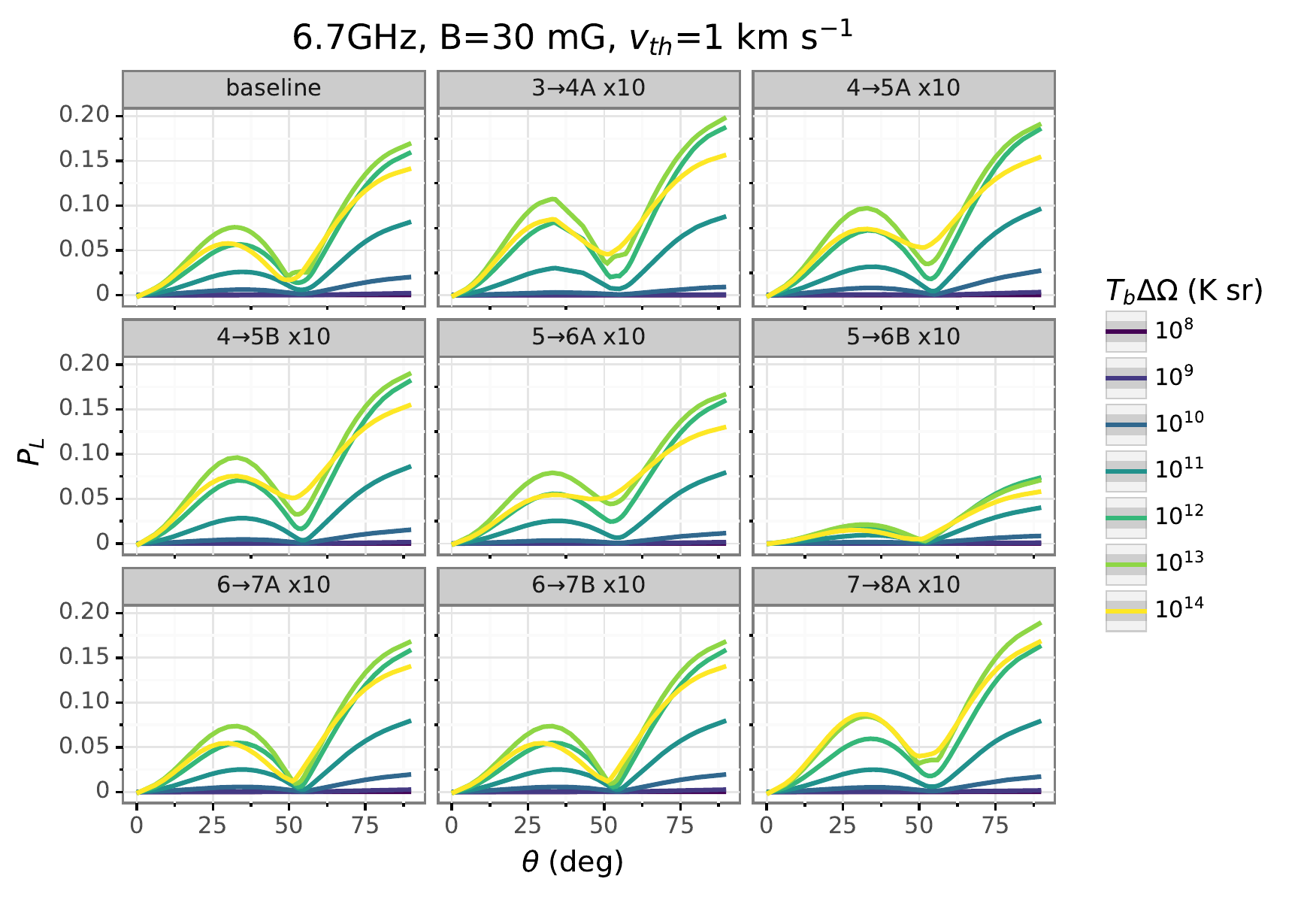} 

     \end{subfigure}
      ~ 
    \begin{subfigure}[b]{0.4\textwidth}
       \includegraphics[width=\textwidth]{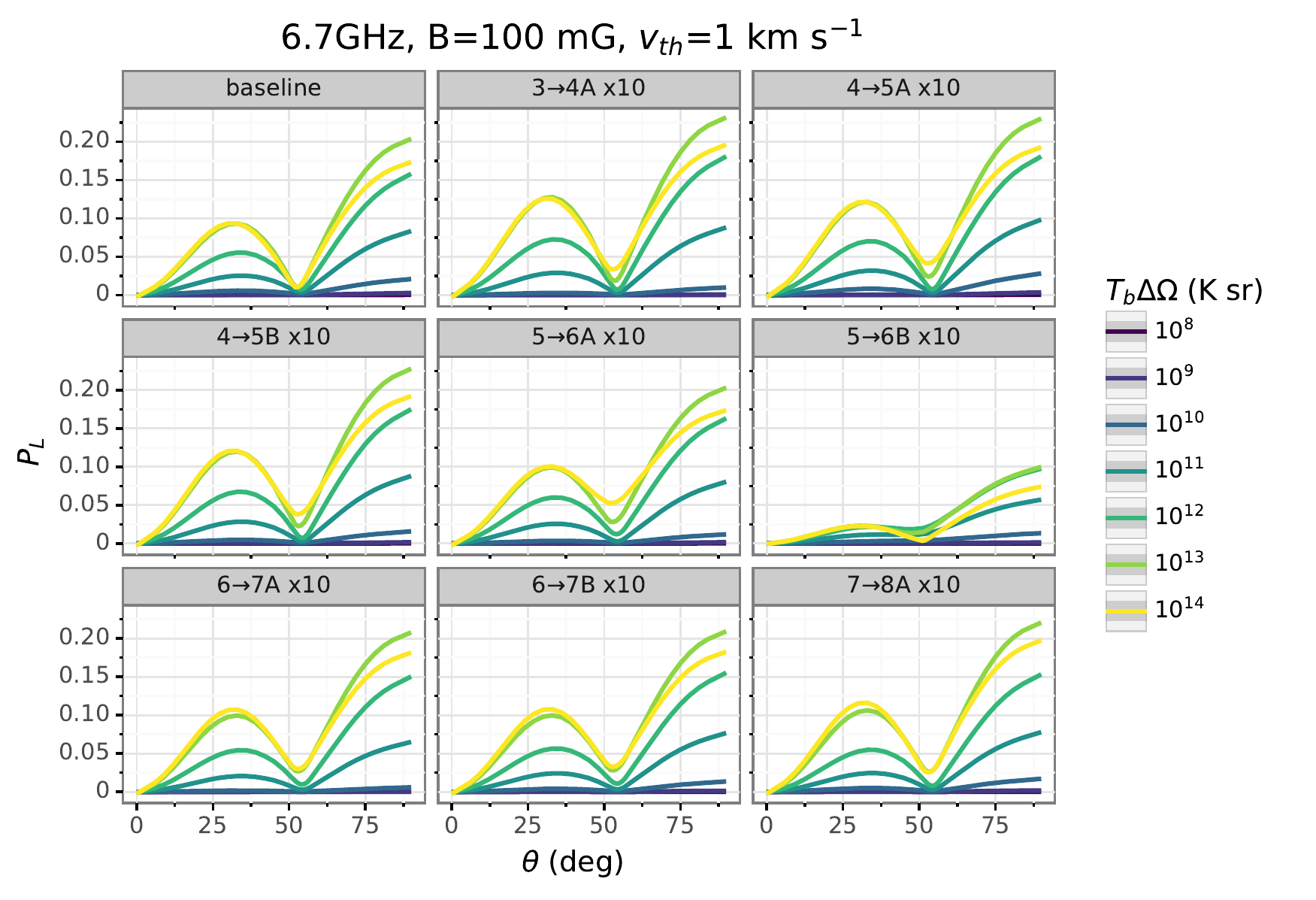} 

     \end{subfigure}
     \caption{6.7 GHz methanol maser linear polarization
       fraction, plotted as a function of the propagation angle
       $\theta$ for different brightness temperatures. B strength,
       v$_{th}$ and the preferred hyperfine transitions are indicated
       in each panel. The panel at the top left labelled ``baseline''
       indicates a fixed pumping rate equal for all the hyperfine
       transitions, while all others assume a $10\times$ preferred
       pumping for the indicated $i\rightarrow j$ transition.}
  \label{fig:PL_profile_iso_altriB}
  \end{figure}

\begin{figure}[h!]
    \centering
    \begin{subfigure}[b]{0.4\textwidth}
       \includegraphics[width=\textwidth]{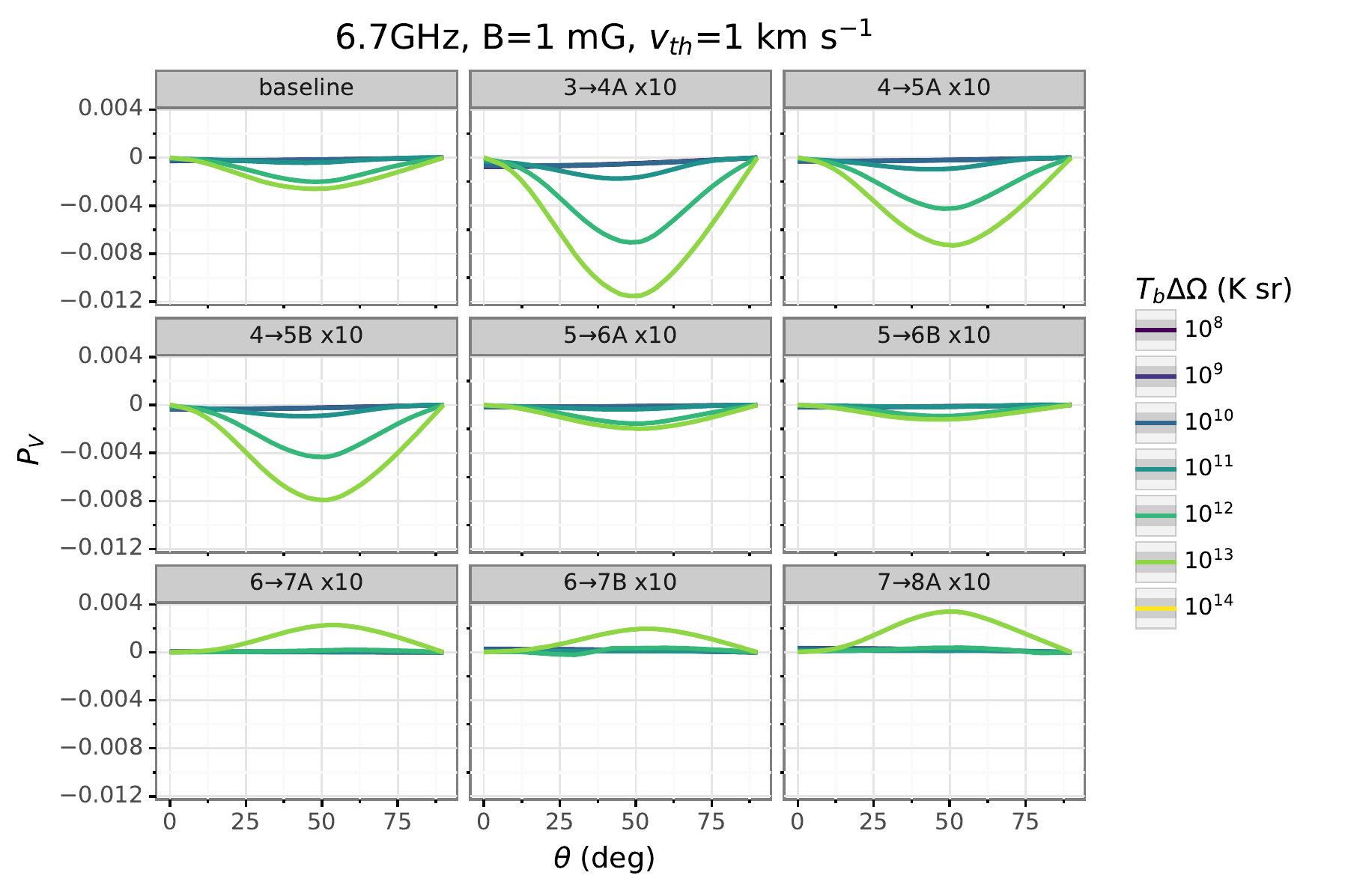}

    \end{subfigure}
    ~ 
    \begin{subfigure}[b]{0.4\textwidth}
       \includegraphics[width=\textwidth]{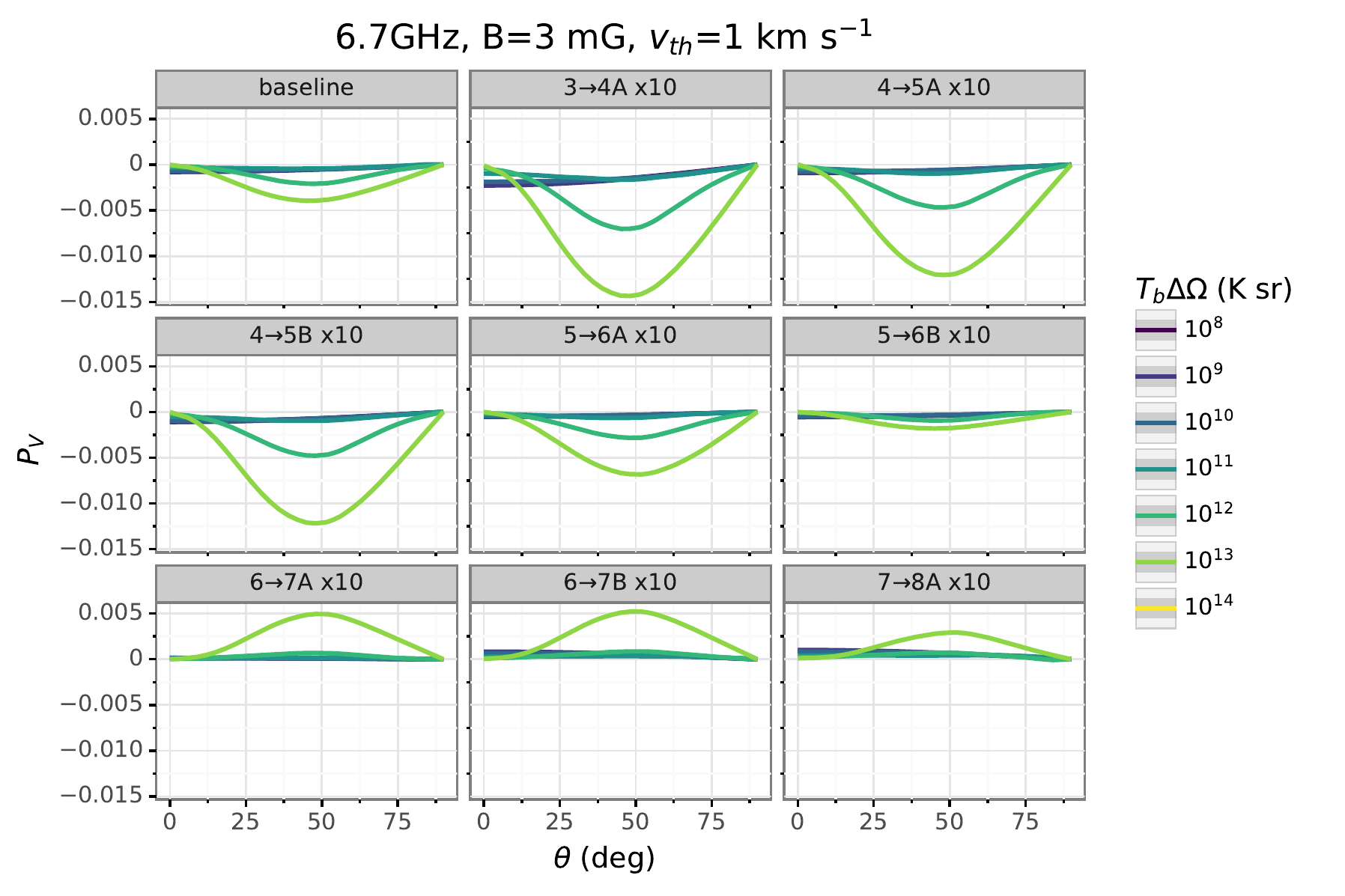} 

    \end{subfigure}
    ~ 
    \begin{subfigure}[b]{0.4\textwidth}
       \includegraphics[width=\textwidth]{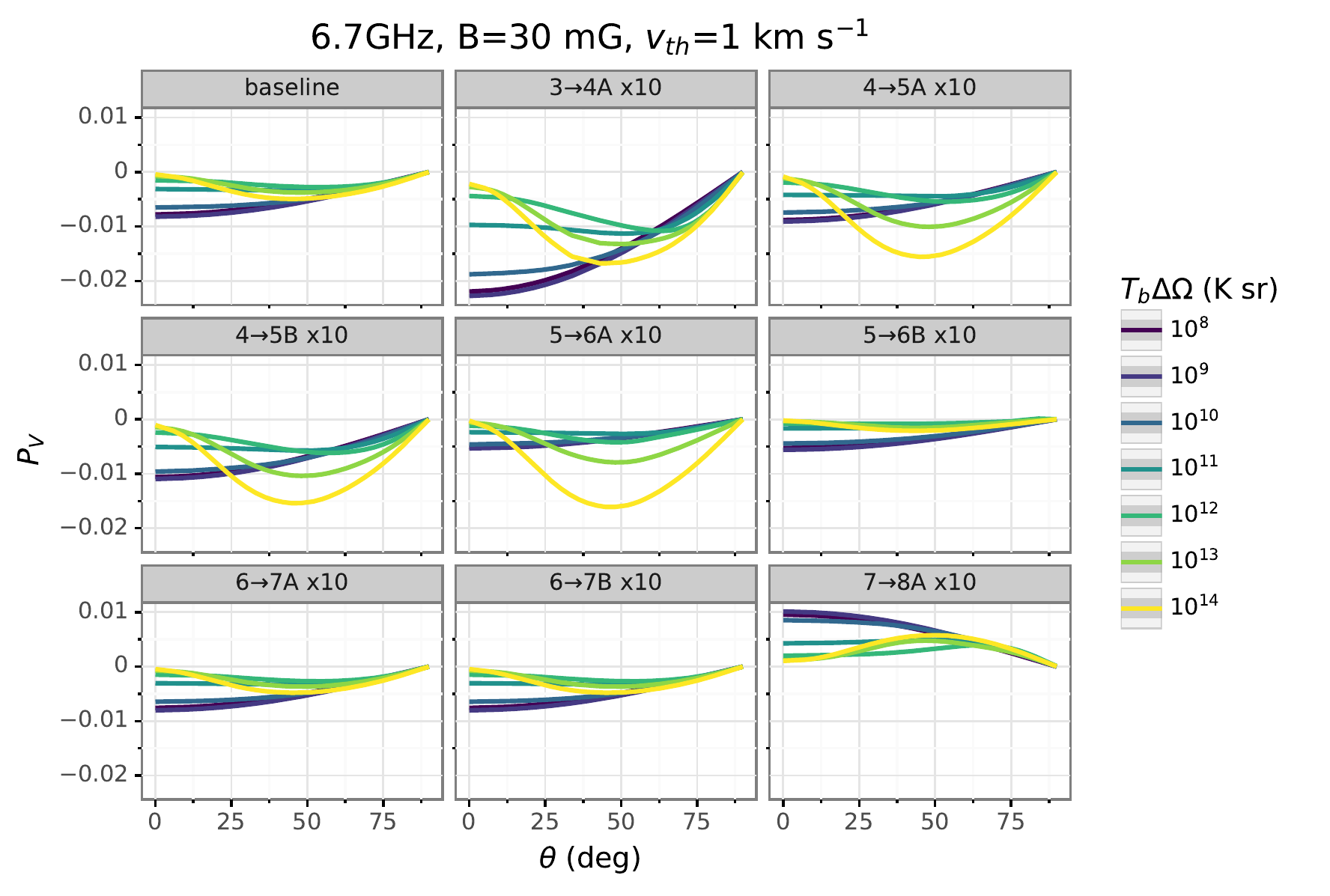} 

     \end{subfigure}
      ~ 
    \begin{subfigure}[b]{0.4\textwidth}
       \includegraphics[width=\textwidth]{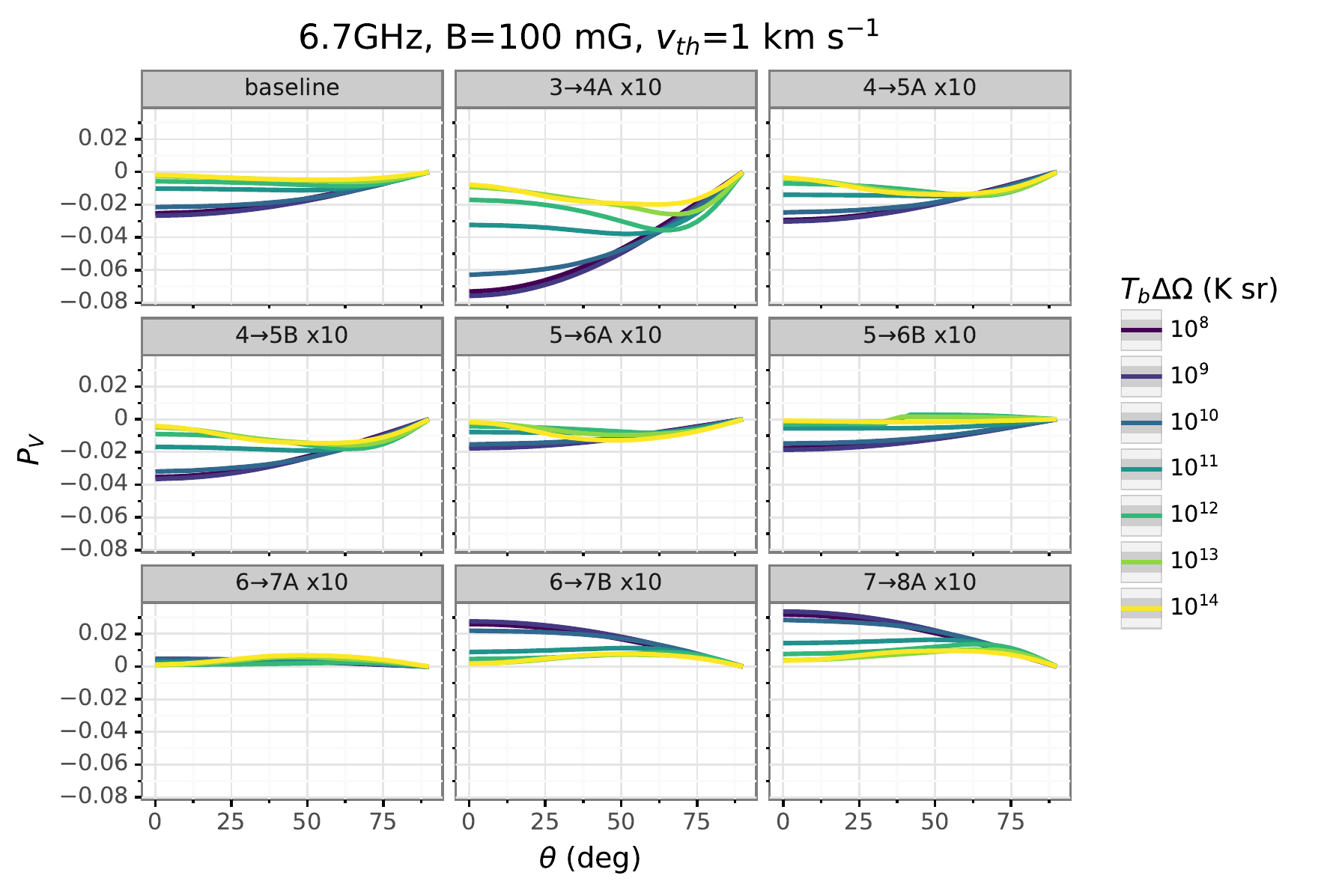} 

     \end{subfigure}
 \caption{6.7 GHz methanol maser circular polarization
    fraction $P_V$, plotted as a function of the propagation angle
    $\theta$ for different brightness temperatures. Panels as in
    Fig.~\ref{fig:PL_profile_iso_altriB}. }
  \label{fig:pv_profile_iso_altriB}
\end{figure}
  
  \begin{figure}[h!]
    \centering
    \begin{subfigure}[b]{0.4\textwidth}
       \includegraphics[width=\textwidth]{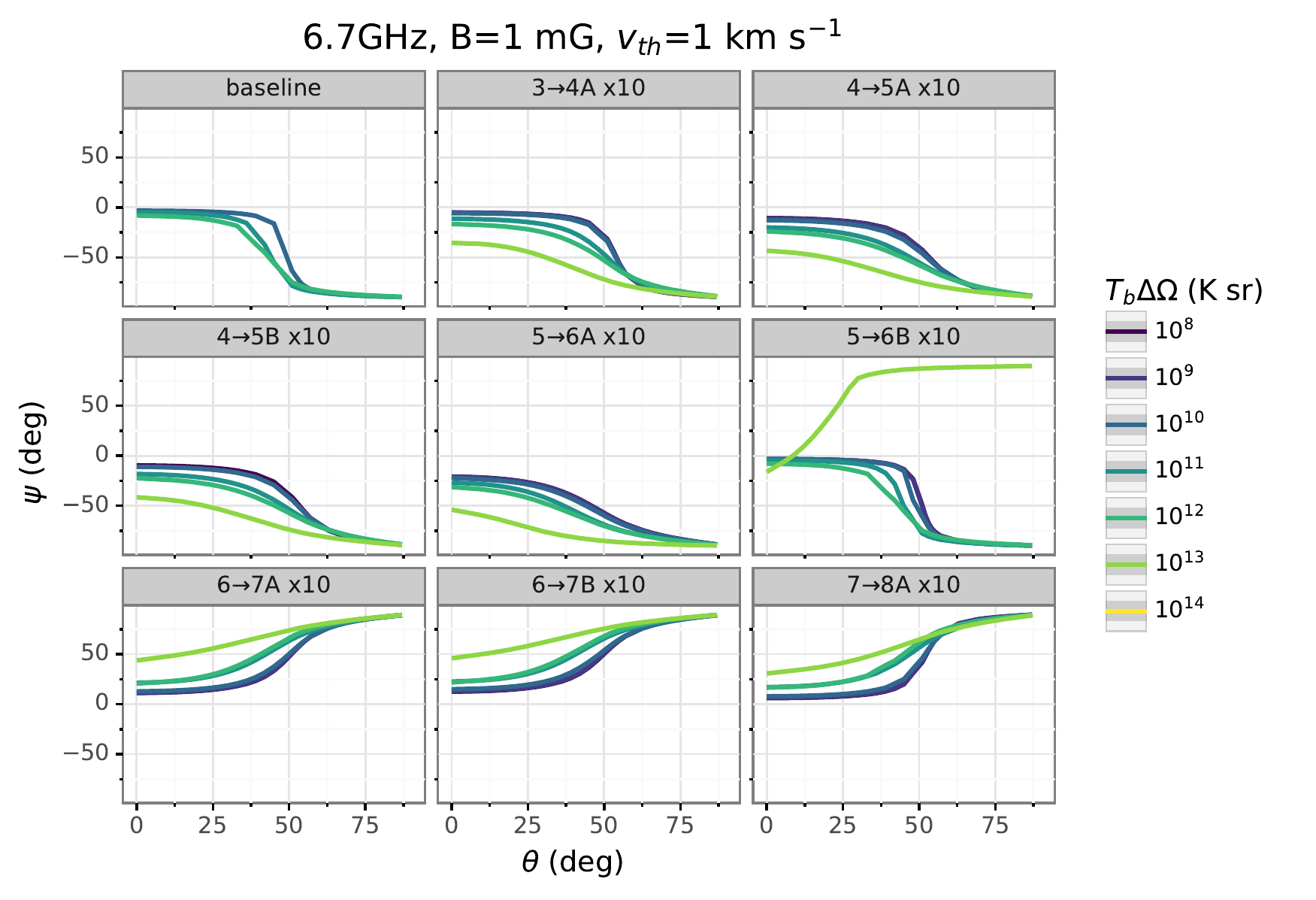}

    \end{subfigure}
    ~ 
    \begin{subfigure}[b]{0.4\textwidth}
       \includegraphics[width=\textwidth]{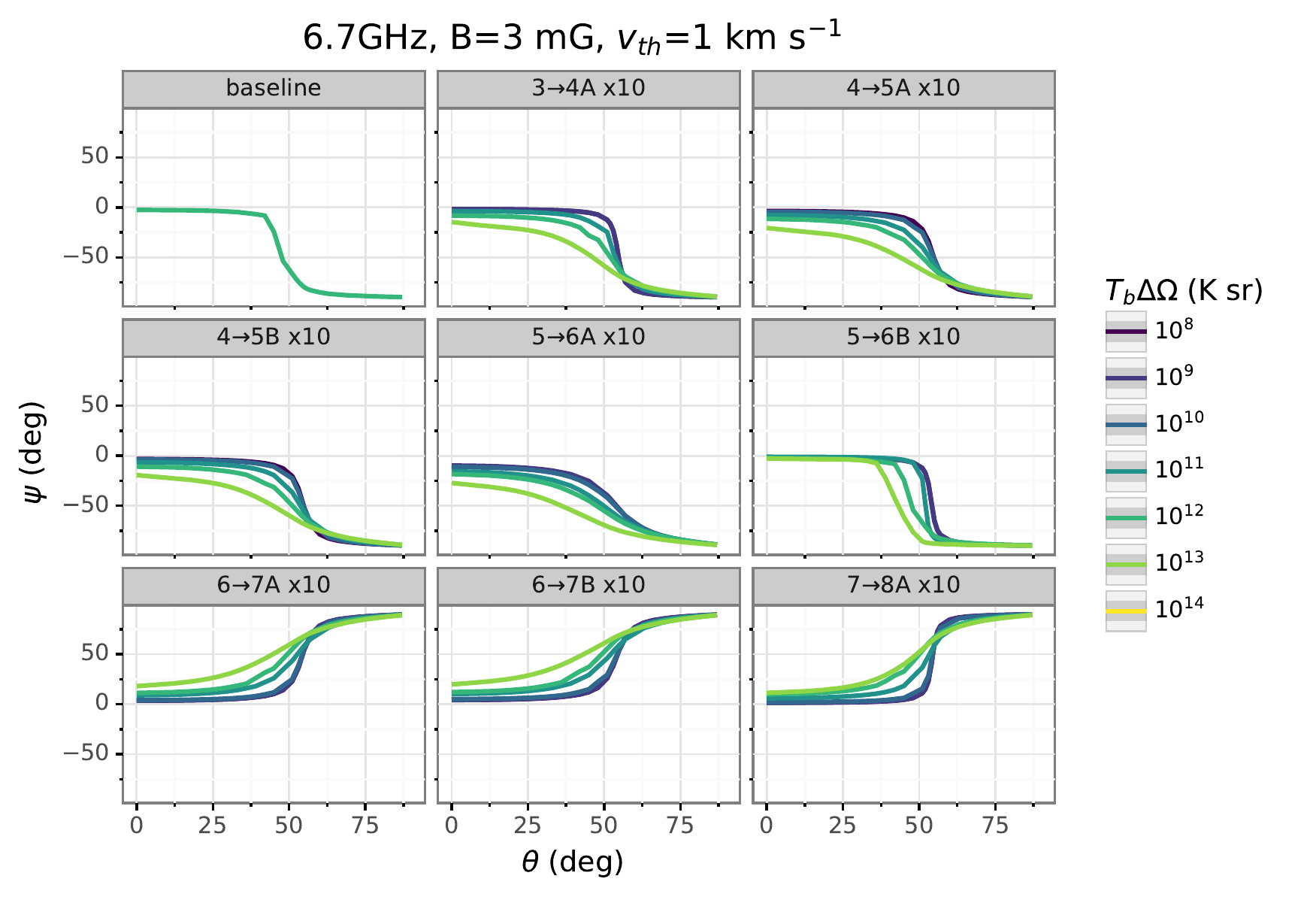} 

    \end{subfigure}
    ~ 
    \begin{subfigure}[b]{0.4\textwidth}
       \includegraphics[width=\textwidth]{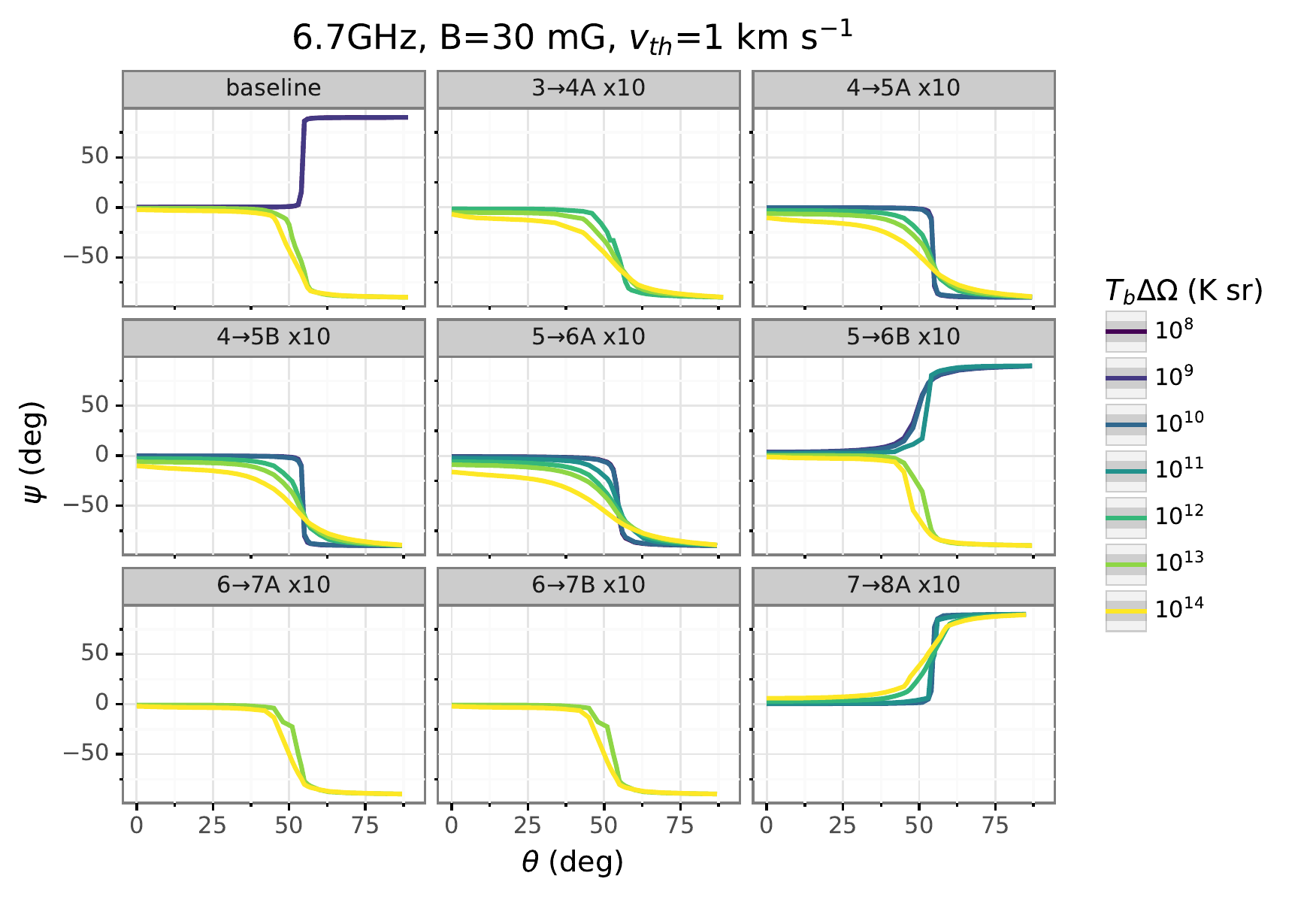} 

     \end{subfigure}
      ~ 
    \begin{subfigure}[b]{0.4\textwidth}
       \includegraphics[width=\textwidth]{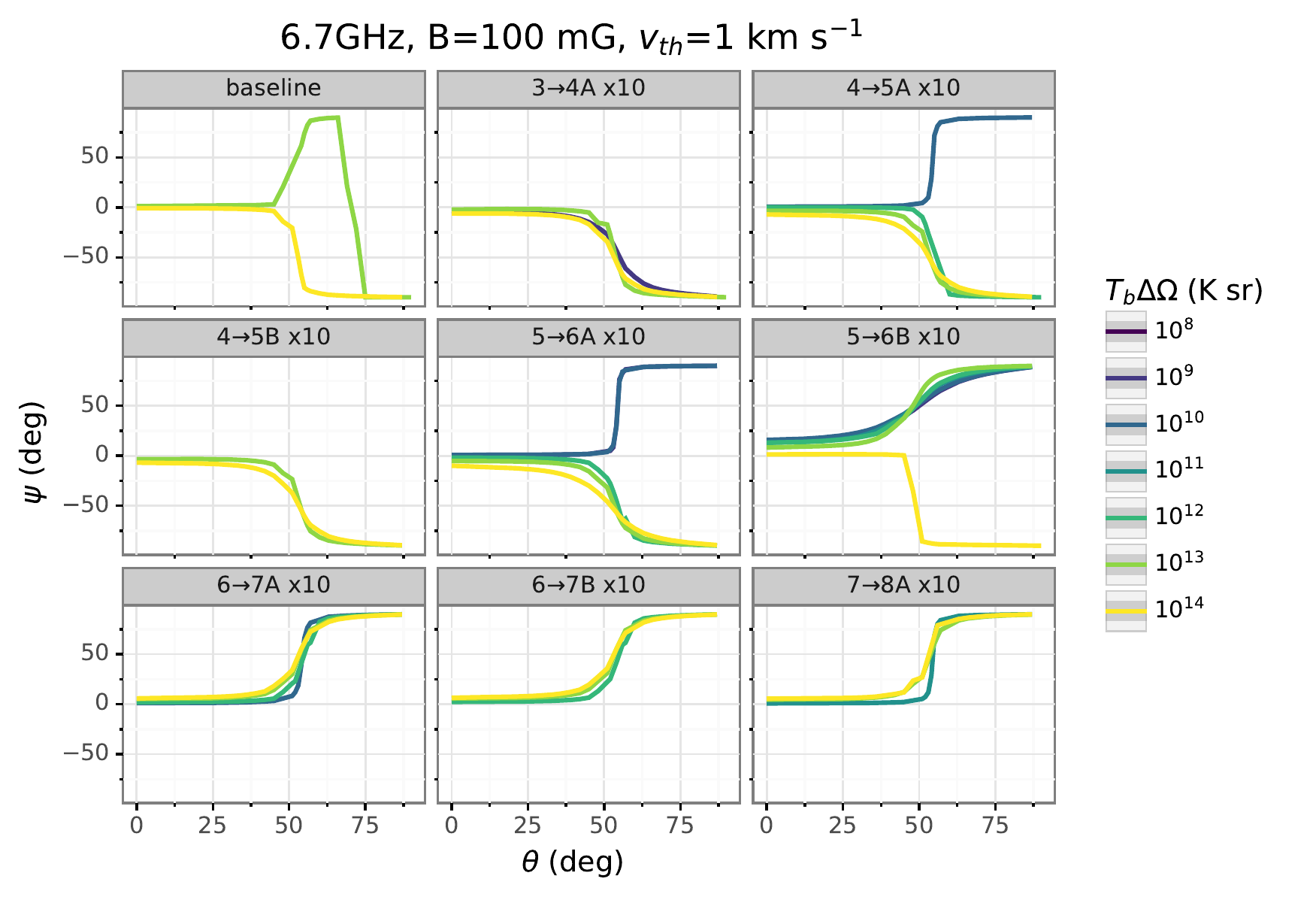} 

     \end{subfigure}
 \caption{6.7 GHz methanol maser linear polarization
    angle $\psi$, plotted as a function of the propagation angle
    $\theta$ for different brightness temperatures. Panels as in
    Fig.~\ref{fig:PL_profile_iso_altriB}. }
  \label{fig:pa_profile_iso_altriB}
\end{figure}

\begin{figure}
  \centering
  \includegraphics[width=0.9\columnwidth, clip]{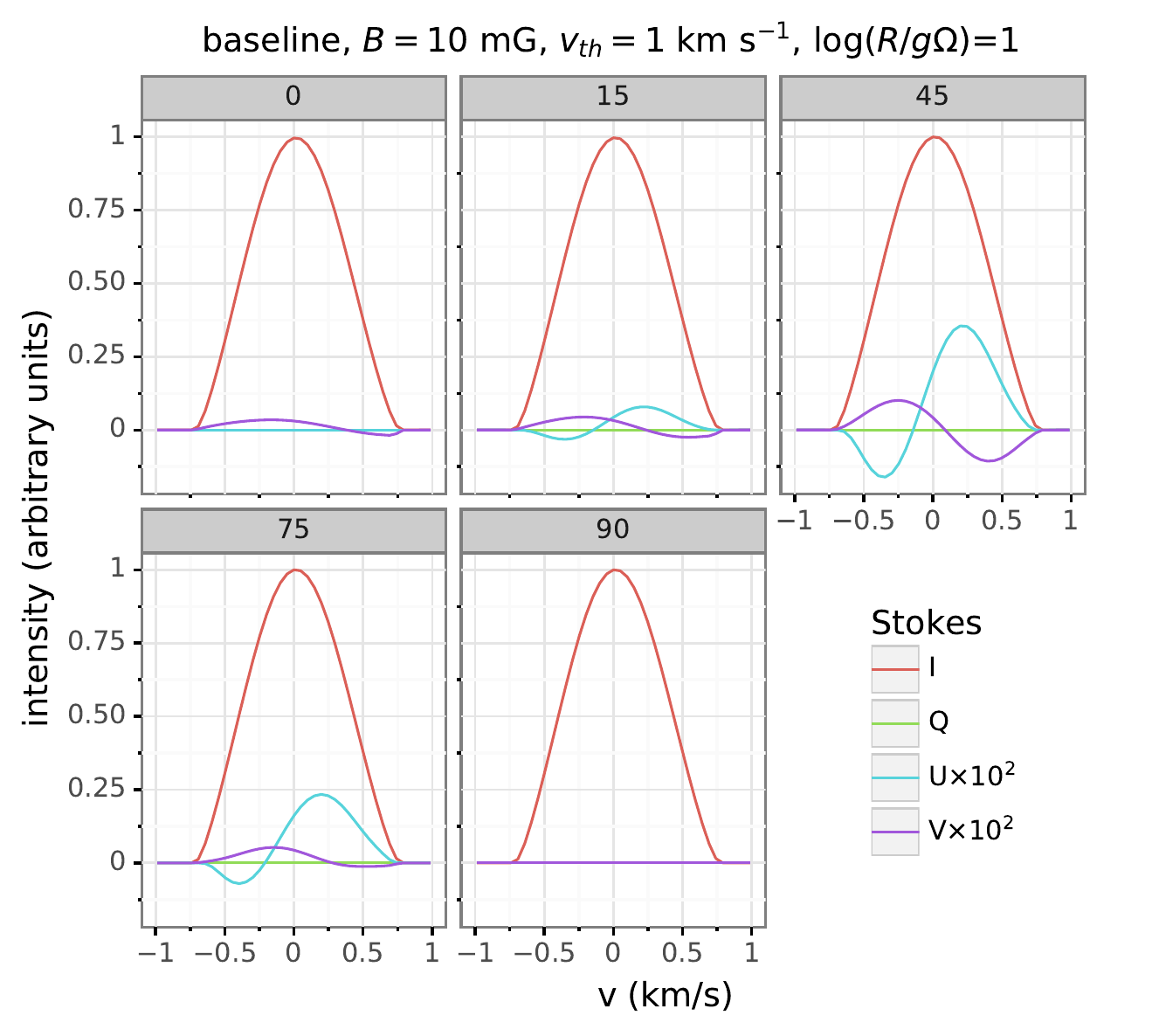}
  \includegraphics[width=0.9\columnwidth, clip]{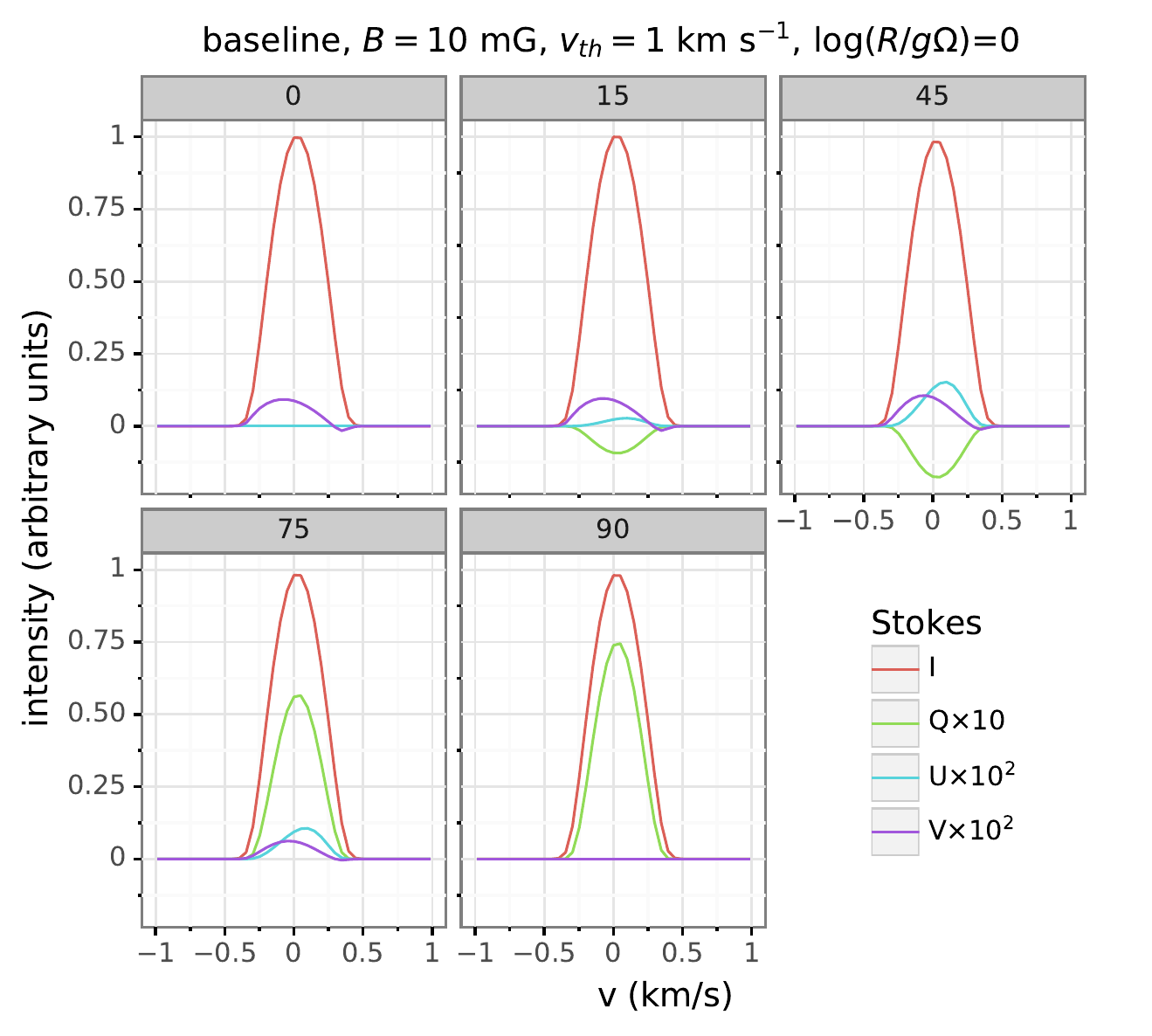}
  \includegraphics[width=0.9\columnwidth, clip]{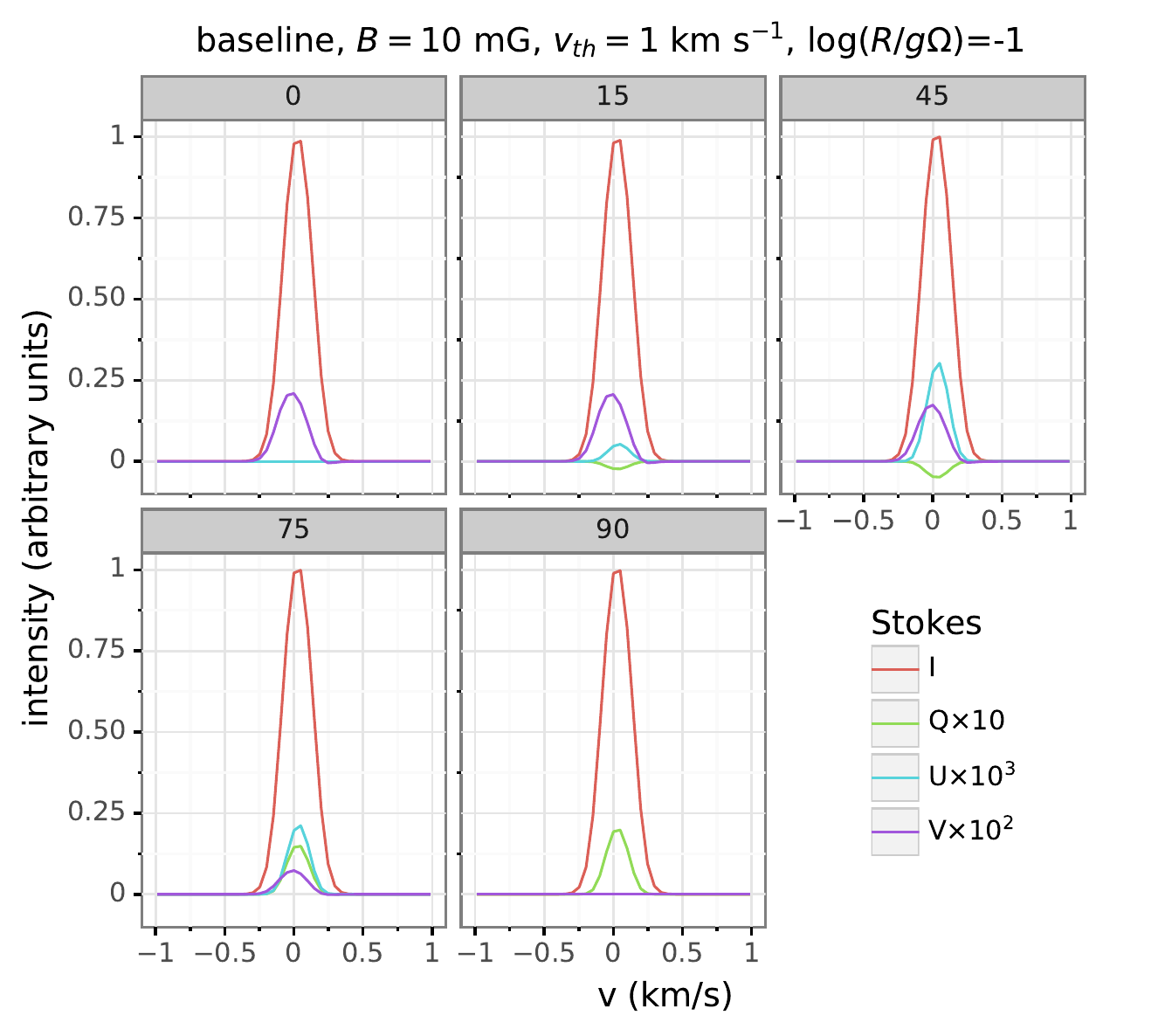}
  
  \caption{6.7 GHz methanol maser spectra for different
    levels of saturation, considering all the hyperfine transitions
    equally pumped. Simulations were performed for stokes I, Q, U,
    and V, and propagation angles $\theta$ of 0, 15, 45, 75, and
    90. }
 \label{Fig:spectra6.7_baseline}
\end{figure}

\begin{figure}
  \centering
  \includegraphics[width=0.9\columnwidth, clip]{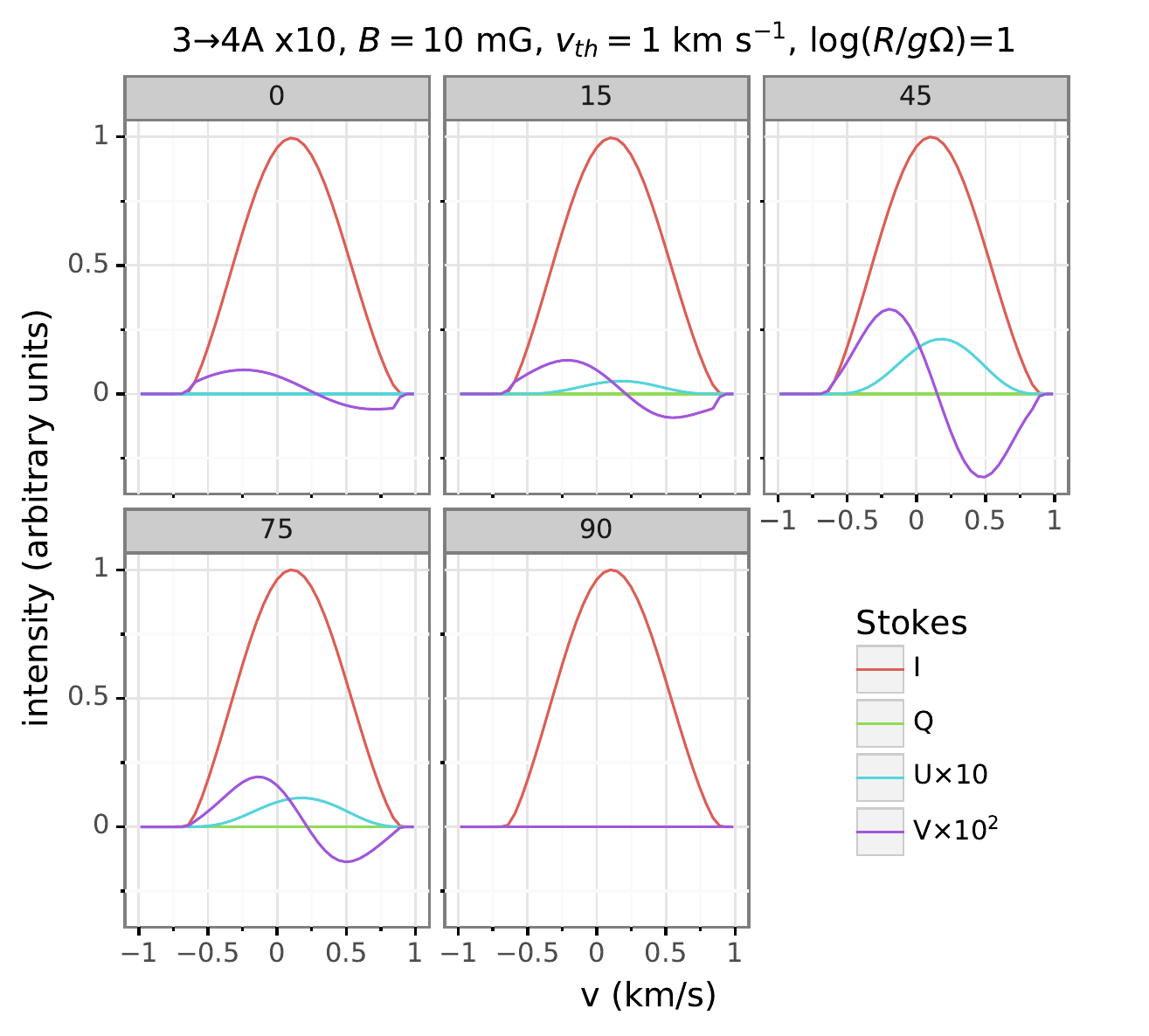}
  \includegraphics[width=0.9\columnwidth, clip]{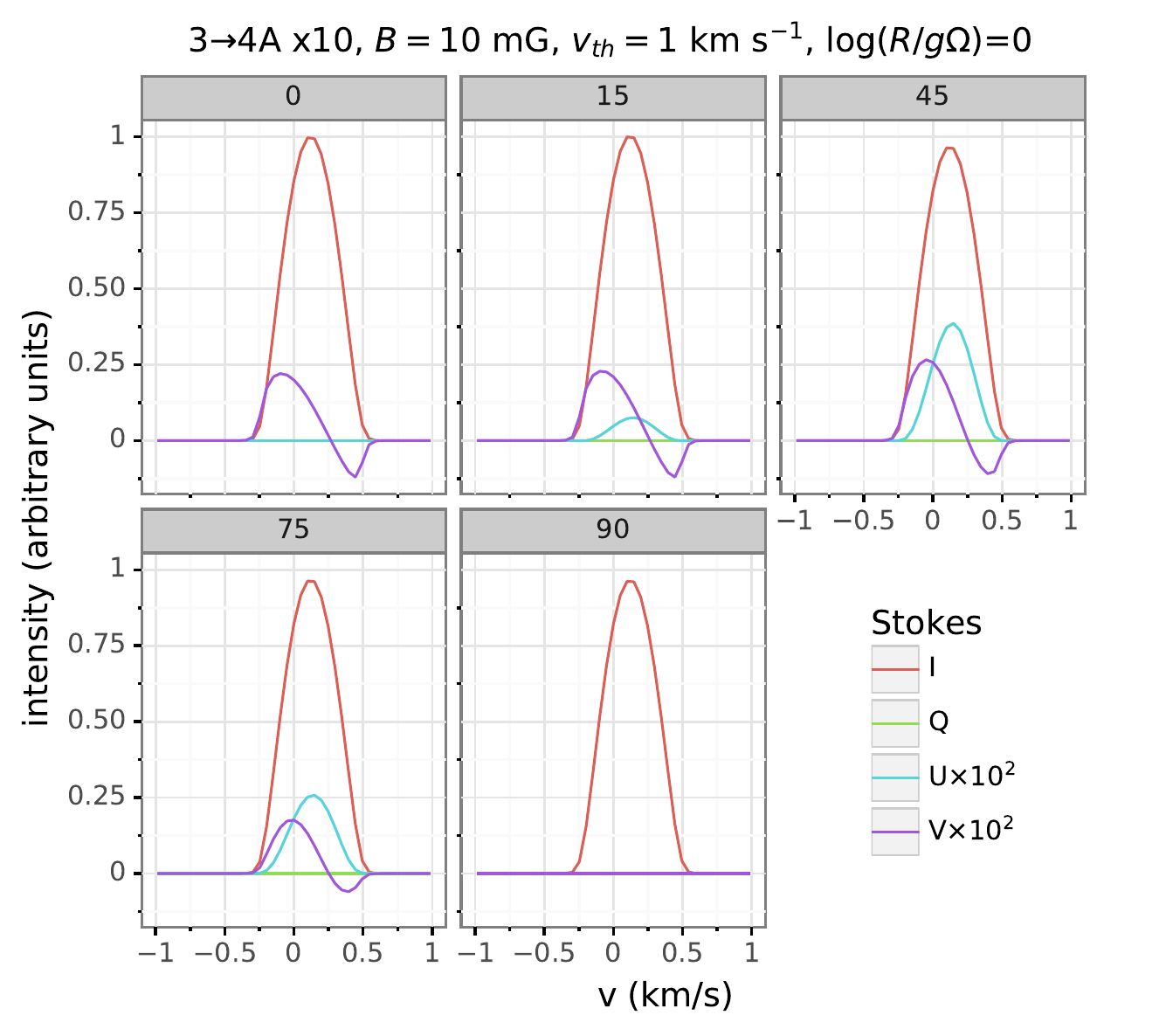}
  \includegraphics[width=0.9\columnwidth, clip]{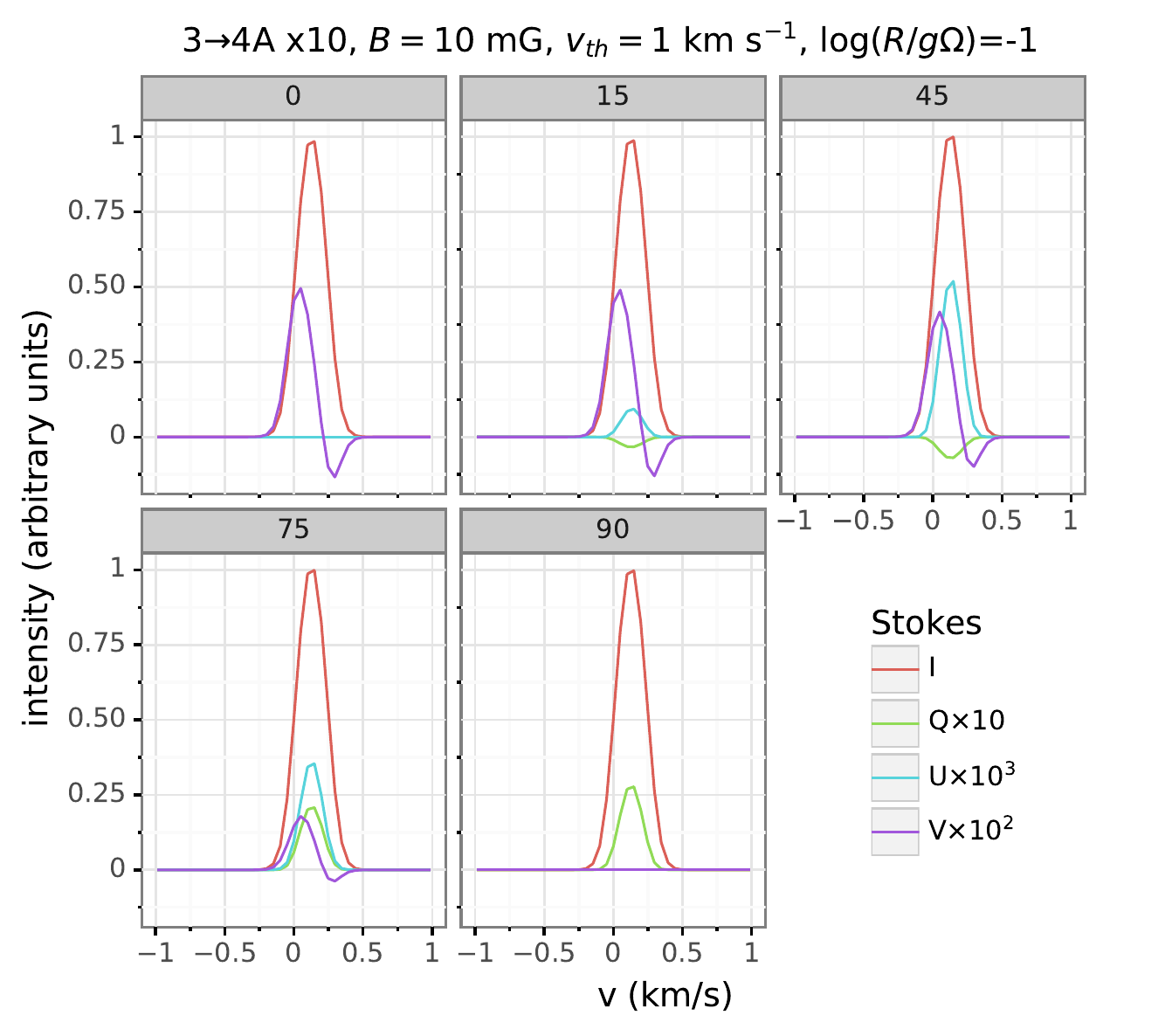}
  
  \caption{6.7 GHz methanol maser spectra for different
    levels of saturation. Simulations were performed for stokes I, Q, U,
    and V, and propagation angles $\theta$ of 0, 15, 45, 75, and
    90. Preferred pumping on the $3\rightarrow4$A hyperfine transition was
    applied.}
 \label{Fig:spectra6.7_1}
\end{figure}

\begin{figure}
  \centering
  \includegraphics[width=0.9\columnwidth, clip]{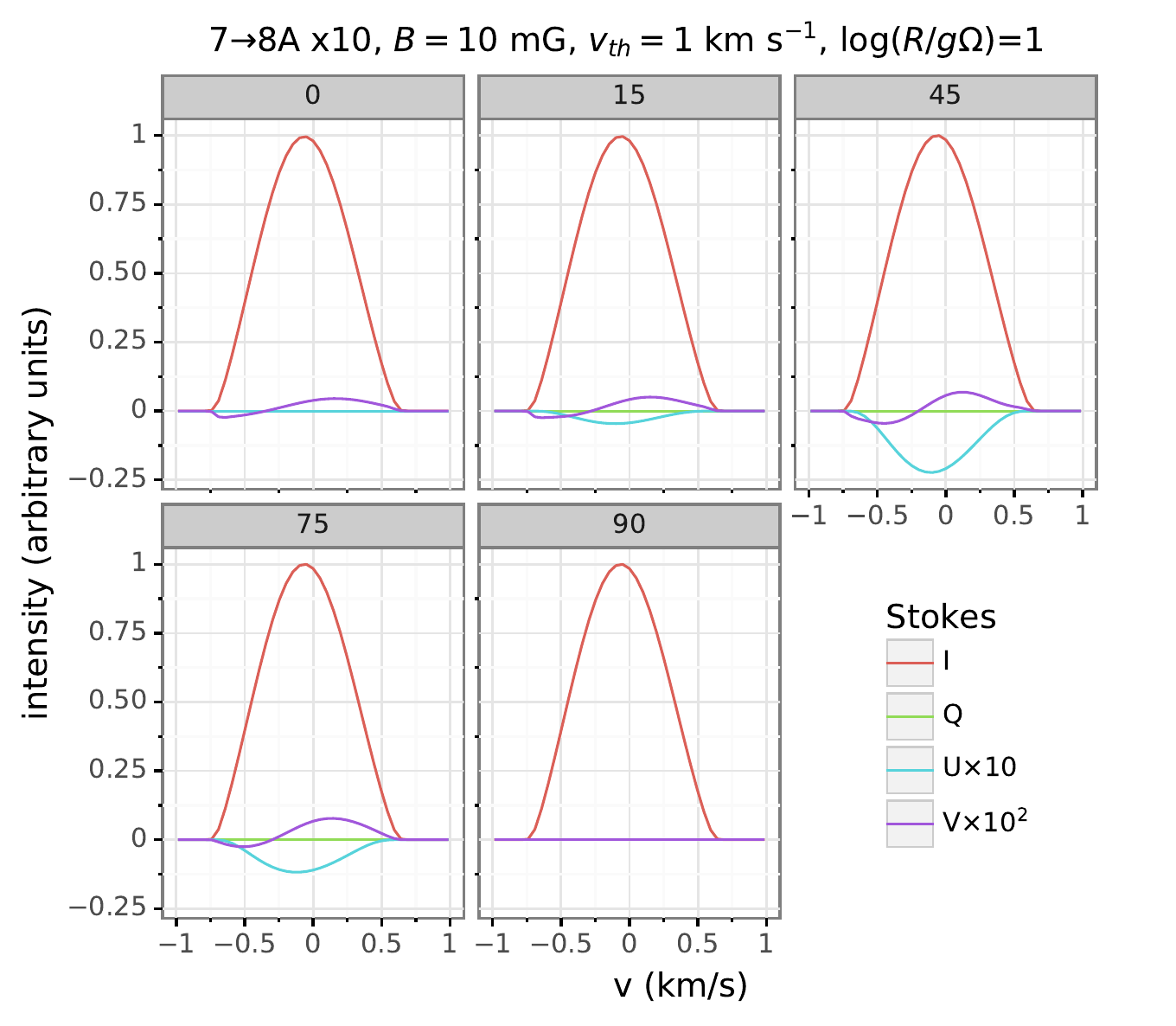}
  \includegraphics[width=0.9\columnwidth, clip]{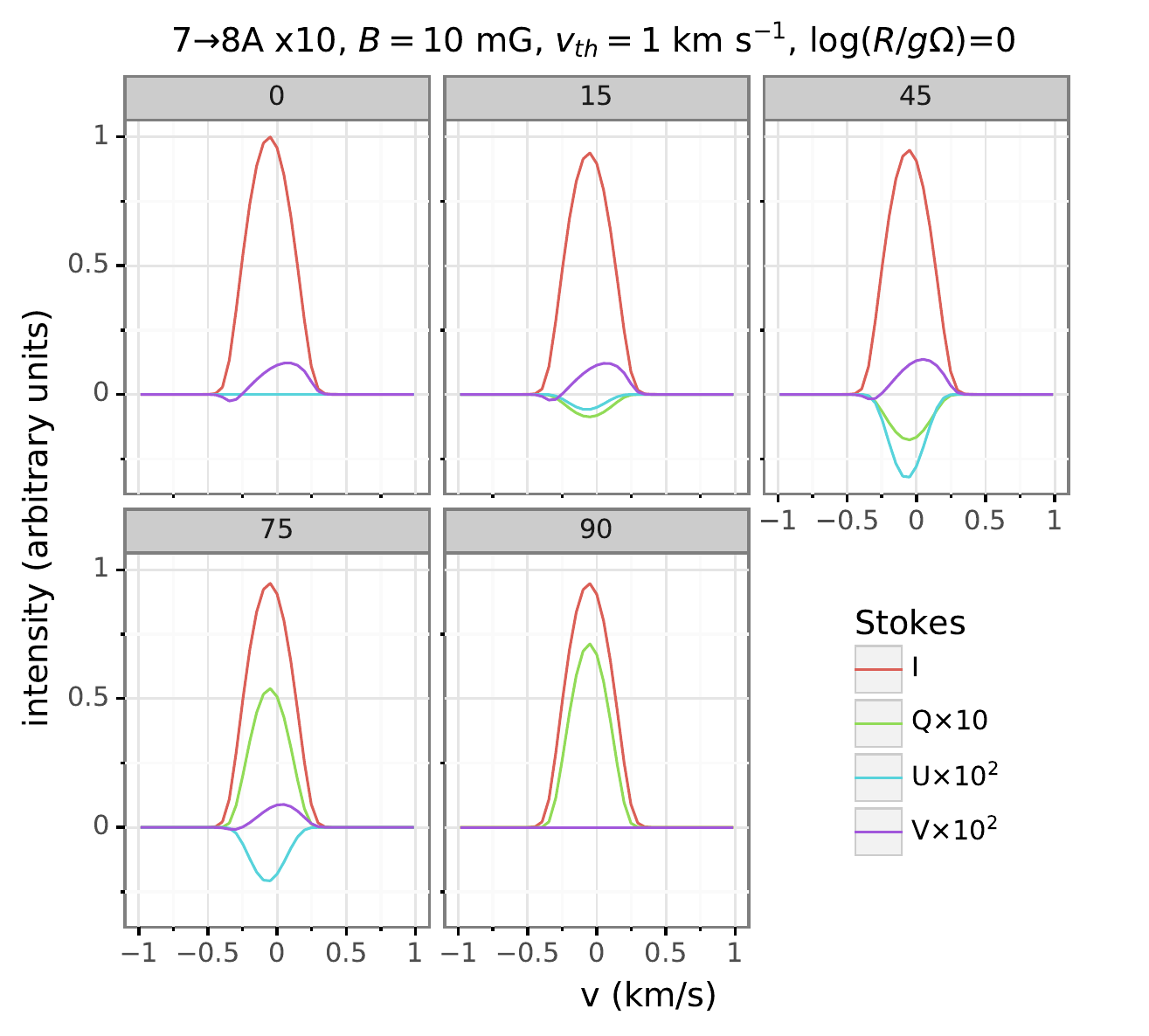}
  \includegraphics[width=0.9\columnwidth, clip]{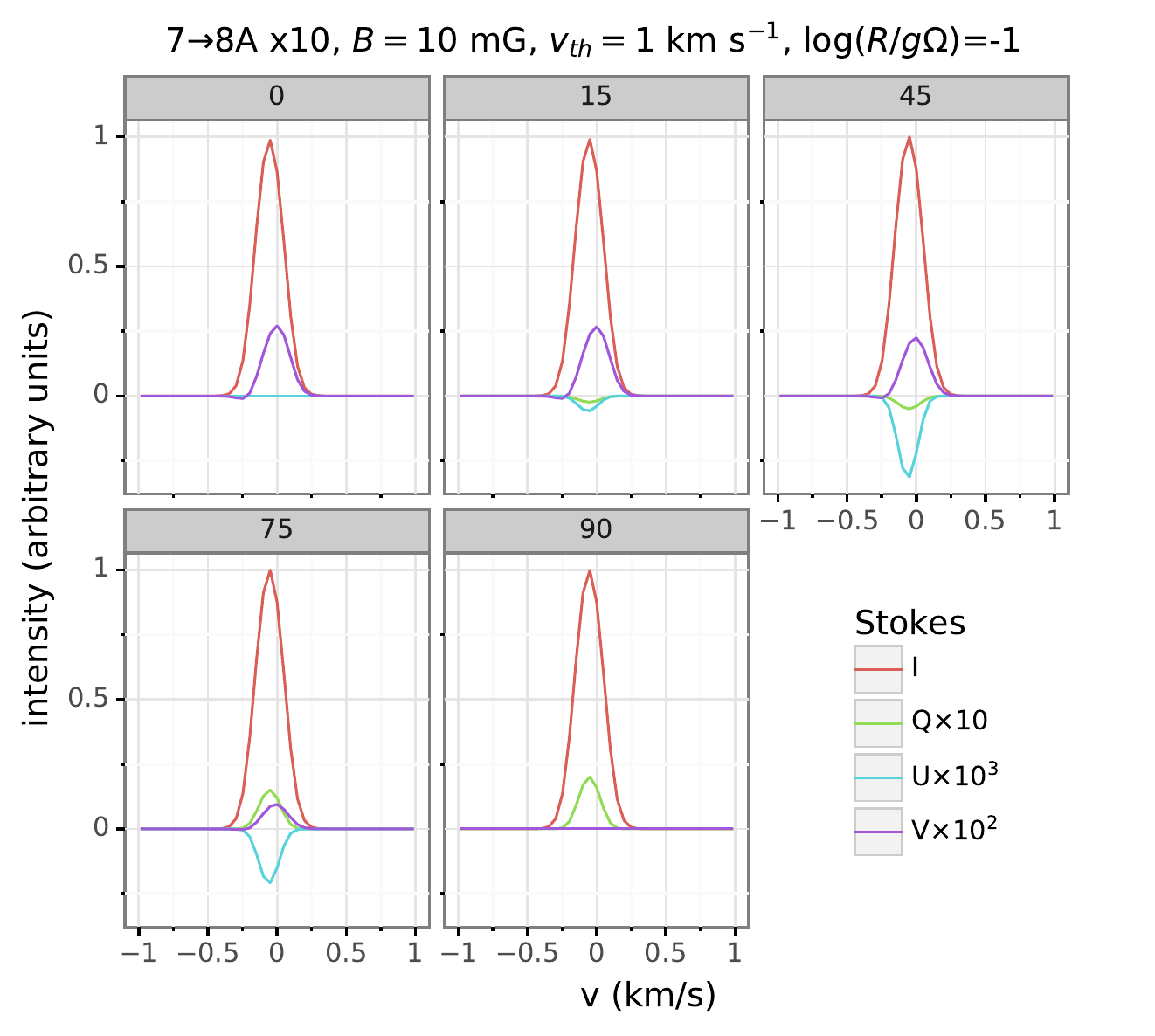}
  \caption{6.7 GHz methanol maser spectra for different
    levels of saturation. Simulations were performed for stokes I, Q, U,
    and V, and for propagation angles $\theta$ of 0, 15, 45, 75 and
    90. Preferred pumping on the $7\rightarrow8$A hyperfine transition was
    applied.}
 \label{Fig:spectra6.7_8}
\end{figure}

\begin{figure}
  \centering
  \includegraphics[width=0.9\columnwidth, clip]{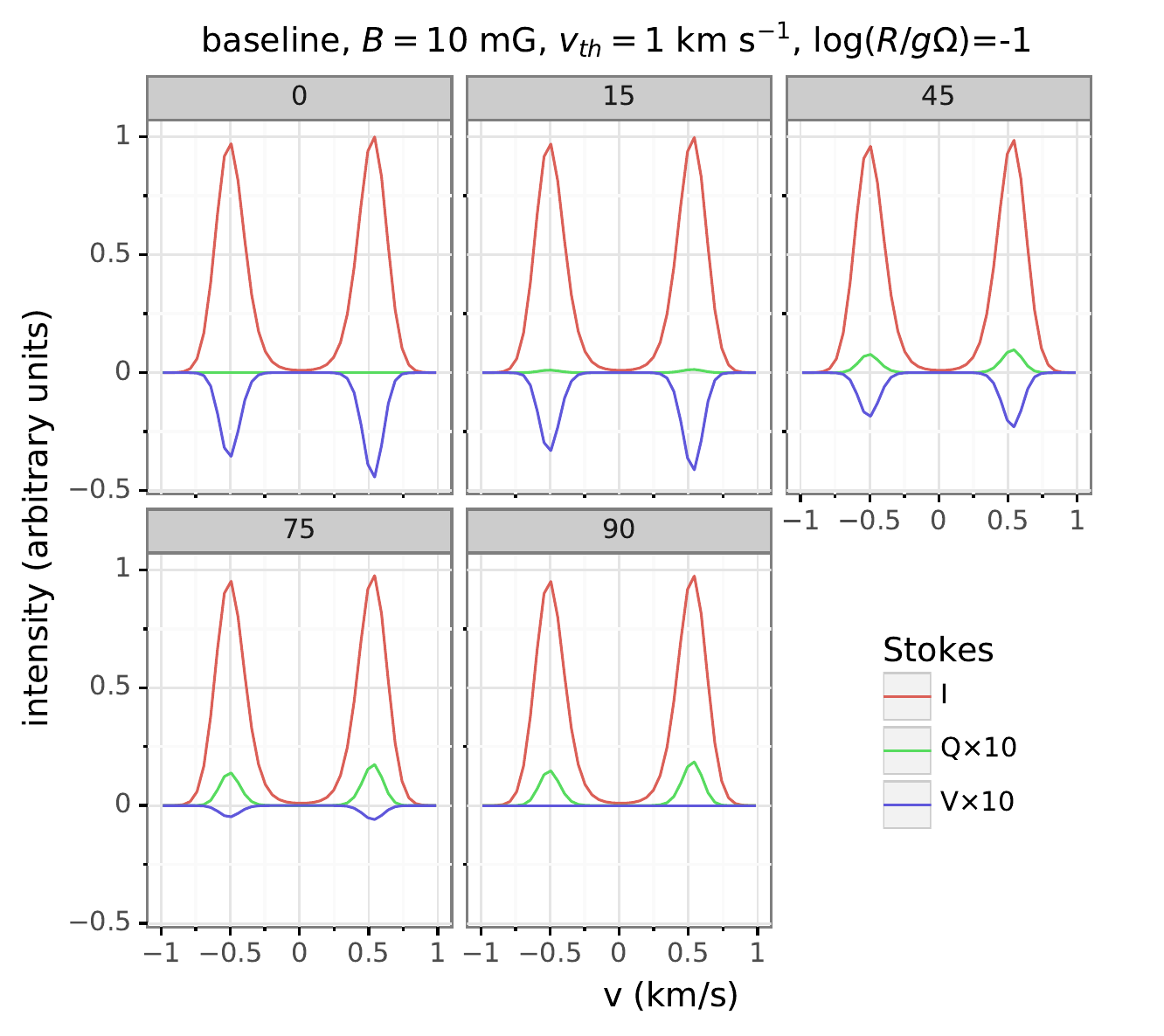}
  \includegraphics[width=0.9\columnwidth, clip]{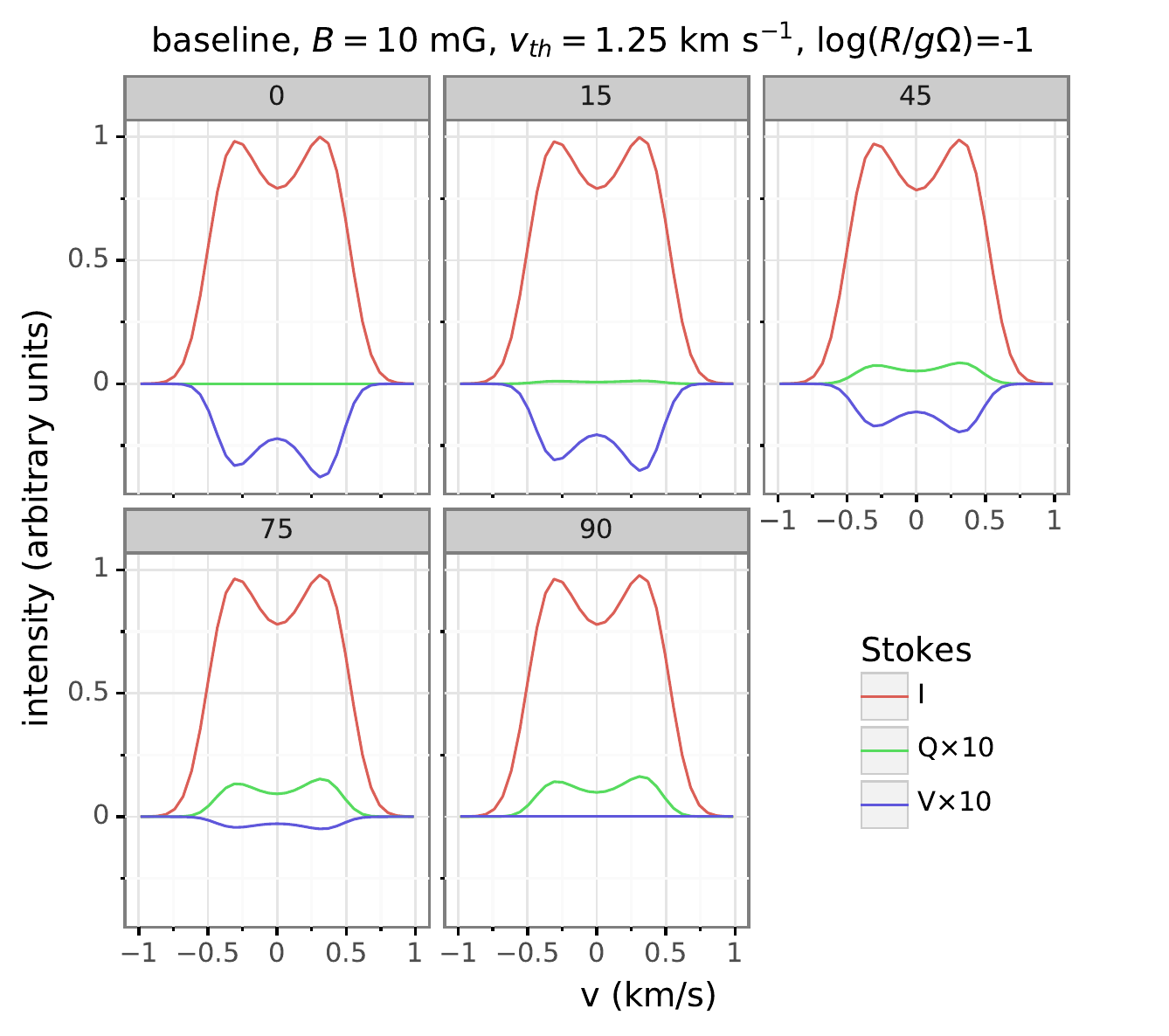}
  \includegraphics[width=0.9\columnwidth, clip]{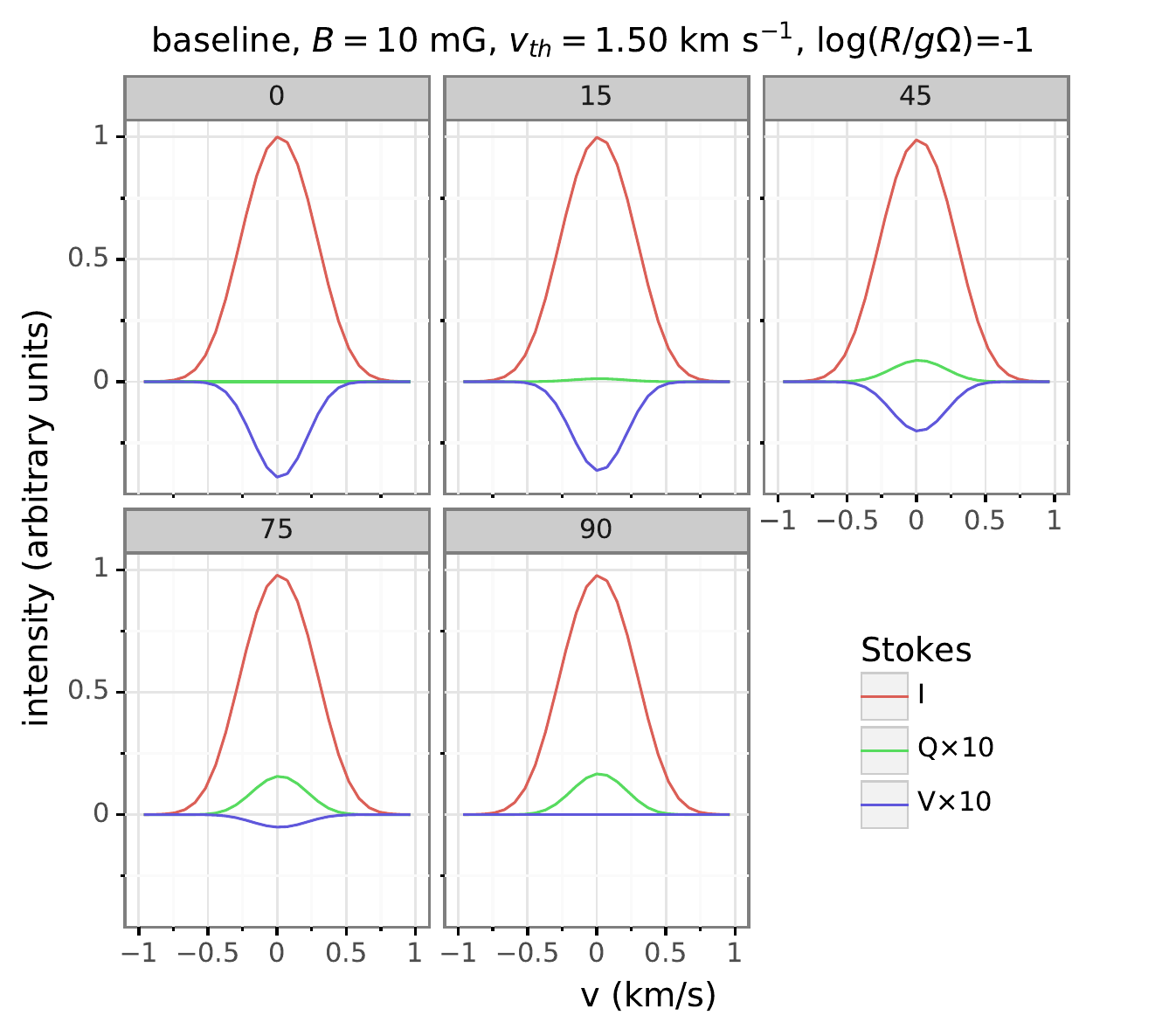}
  \caption{6.2 GHz methanol maser spectra for different
    intrinsic thermal velocity width. Simulations were performed for stokes I, Q, U,
    and V, and for propagation angles $\theta$ of 0, 15, 45, 75 and
    90.}
 \label{Fig:spectra6.2}
\end{figure}

\end{appendix}

\bibliographystyle{aa}
\bibliography{../bibliografia}

\end{document}